\documentclass[a4paper, 10pt, notitlepage]{article}
\usepackage{graphicx}
\usepackage{a4wide}
\usepackage{amsmath}
\usepackage{amssymb}
\usepackage{url}
\usepackage{bm}
\usepackage{subfig}

\begin{document}
\newcommand{\bra}[1]{\langle #1|}
\newcommand{\ket}[1]{|#1\rangle}
\newcommand{\sembrack}[1]{\left[\!\left[#1\right]\!\right]}
\renewcommand{\thefootnote}{\roman{footnote}}

\begin{titlepage}
 
\begin{center}


\begin{minipage}{0.7\textwidth}
\begin{flushleft} \large
{\large \today} \\
\large Lausanne, Switzerland
\end{flushleft}
\end{minipage}
\begin{minipage}{0.2\textwidth}
\begin{flushright} 
\includegraphics[width=\textwidth]{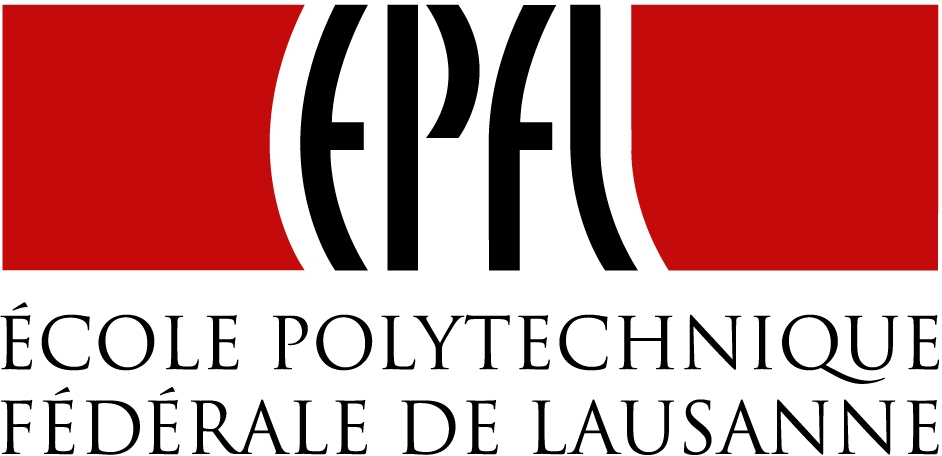}
\end{flushright}
\end{minipage}\\[1 cm]

\textsc{\LARGE Numerical treatment of disorder \\
\vspace{0.2cm}
in PHC slabs}\\[1 cm]

\includegraphics[width=\textwidth]{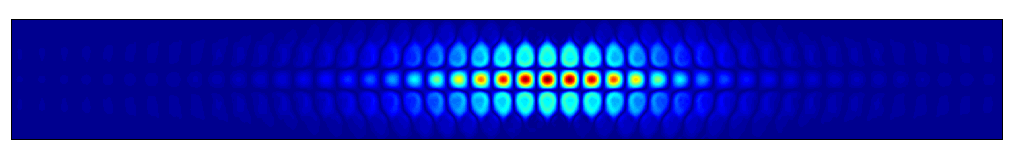}
 
\begin{center}
\large Master Thesis\\[0.3 cm]
\textbf{\large by}\\
\textsc{\Large Momchil Minkov}\\[0.5 cm]
\textbf{\large Academic supervisor}\\
\textsc{\large Prof. Vincenzo Savona}
\end{center}

\vspace{1 cm}
 

\begin{abstract}
 This work concentrates on numerical simulations of Photonic Crystal structures using basis-expansion methods, with a main focus on simulating disorder. The plane-wave and guided-mode expansions are outlined as tools to compute the Bloch modes of a structure, on the basis of which the Bloch-mode expansion formalism is outlined - the latter allowing for simulations of large structures in presence of disorder. As a first illustration of the method, we apply it to three gentle-confinement cavities, to obtain results for their quality factors similar to the theoretically predicted in the literature. Furthermore, we compute that random disorder can drive those factors down to the experimentally measured values. As a second application, we study the effect of irregular hole shapes in a PHC waveguide, and find that the correlation length of the irregularity (i.e. the typical scale of the roughness of the features) matters: for higher correlation lengths, the computed modes show both higher band broadening and higher loss rates.
\end{abstract}

\end{center}


\newpage
\thispagestyle{empty}
\tableofcontents
\end{titlepage}

\setcounter{page}{1}
\clearpage

\section{Introduction}
\subsection{What is a photonic crystal?}

It has been shortly less than a quarter of a century since what are now considered some of the seminal works on photonic crystals, \cite{john}, \cite{yablonovitch} first appeared, but the significance of those structures was so quickly realized that the field is currently in the spotlight of photonics research and development. A photonic crystal, in essence, is any form of a medium with periodically varying dielectric constant, but more generally it can be a structure with such an underlying periodicity, which is however broken by e.g. unwanted disorder or deliberately placed defects for the achievement of particular tasks. Such crystals also appear in nature, but it is the ability to design and manufacture them in the lab which shows a potential for numerous applications in optical devices.  

\begin{figure}[h!]
\begin{center}
 \subfloat[]{\includegraphics[width = 0.3\textwidth]{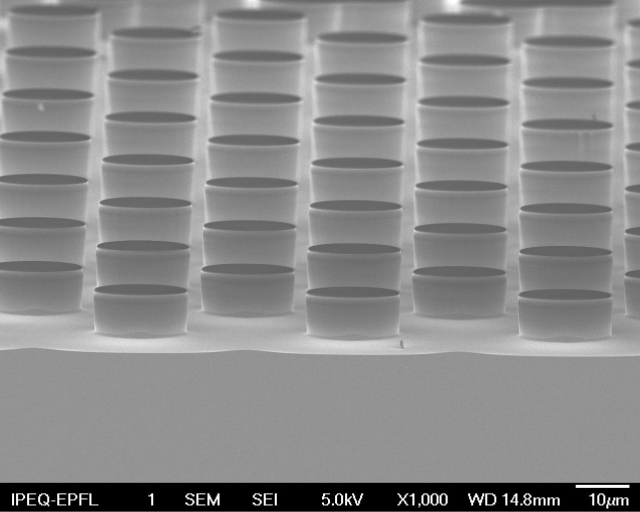}} 
 \subfloat[]{\includegraphics[width = 0.37\textwidth]{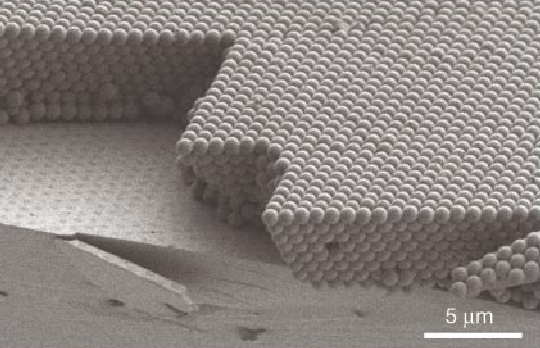}}
 \subfloat[]{\includegraphics[width = 0.33\textwidth]{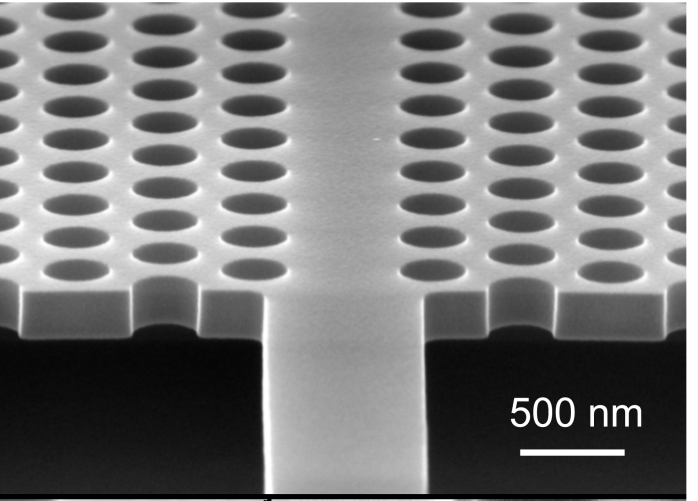}}
 \caption{(a): GaAs/AlGaAs pillars in air, image from \cite{zabelin}. (b): 3D photonic crystal with the ``opal structure'' of dielectric spheres in air, image from \cite{vlasov}. (c): waveguide created by a ``missing'' row of air holes in a dielectric slab, image from \cite{ferrini}.}
 \label{crystals}
 \end{center}
\end{figure}

A general classification that can be made between different photonic crystals is the dimensionality in which the periodicity occurs; namely, there can be 1D photonic crystals, where the periodicity is in one dimension only, and the crystal is homogeneous in the other two dimensions - an example for this is a collection of alternating layers of two different materials. Similarly, a 2D crystal would be periodic in two spacial directions and homogeneous in the third - an example can be periodically occurring dielectric rods in air (fig. \ref{crystals} (a)). And finally, a 3D crystal is described by some periodic dielectric constant in all directions - for example tangential dielectric spheres in air (fig. \ref{crystals} (b)).

Photonic crystals (PHCs) have many similarities to the standard crystals studied in Solid State Physics. Generally, the periodicity of the dielectric function can be represented in some form as a lattice of primitive cells. Photons in a PHC thus behave in a similar way to electrons in a crystal lattice, both particles being governed by an underlying wave equation. We can thus construct band diagrams for a PHC, just as we do for a standard crystal. The periodicity of the structure of the PHC often provides a band-gap over a given frequency range which provides a basis for interesting properties and applications. The main distinction between the two types of crystals is that, first, electrons are fermions, while photons are bosons, so they essentially obey different statistics, and second, that photons do not carry electric charge, so we do not need to consider Coulomb interaction, as opposed to when considering electrons and holes. 

\subsection{Current developments and applications}

For a detailed outline of photonic crystal development and application possibilities, refer to Chapter 1.4 of \cite{igor}, and references therein; here we will briefly highlight the areas of science where PHCs can prove useful.

\begin{itemize}
\item Spontaneous radiation management: the main topic of \cite{yablonovitch} and more recently of a detailed review paper by S. Noda \cite{noda}, this is arguably the area where PHCs will prove most useful. In a 3D PHC, there can be a band gap in all spatial directions such that control as strong as complete inhibition of spontaneous emission of a material introduced into such a structure can be achieved. However, even 2D structures present interesting opportunities for control of spontaneous emission, e.g. when the emission is inhibited in two of the spatial dimensions so that all radiation energy is redirected to only exit through the third. An overview of the applications of the use of PHCs for spontaneous emission control includes improving semiconductor lasers or solar cell efficiency, and optical cavities based on PHCs can be a basis for new types of radiation sources for illumination, display and optical communication purposes.
\item Introducing defects in a PHC provides numerous possibilities for constructing waveguides (fig. \ref{crystals} (c)), microcavities, splitters, couplers and combiners. The waveguides are usually produced by introducing a line defect, and they work due to the presence of the photonic band gap, rather than due to internal reflection. This allows them to form very sharp bends, e.g. at an angle $\pi/2$. A splitter is essentially a number of such waveguides connected at a single point. Since we have great freedom in varying the waveguide parameters, we can finely control the operation of the splitter.
\item PHCs can also be used as super-prisms, super-lenses, multiplexers and demultiplexers for dispersion management.
\item Nonlinear optics applications: for example for optical storage elements or logical elements.
\item Slow light applications: the group velocity at a specific wavelength in a PHC can be very low.
\item Optical fibers: PHCs provide many possibilities to improve existing technology, or to provide a basis for a completely new type of fibers, where light is localized due to the photonic band gap and an introduced defect.
\item Quantum computing: PHCs can be potentially valuable for manufacturing quantum computation elements e.g. for linear optics quantum computation as proposed in \cite{knill}. In addition, at least theoretically, quantum dots used as qubits can be embedded in a PHC structure, and coupling between individual dots (e.g. in order to perform two-qubit operations) can be achieved through the electromagnetic modes of the underlying structure, which can be very finely tuned. As is the case in virtually every area of development related to quantum computers, there is a myriad of technical challenges in realizing such a setup in practice, but it is definitely worth the effort as it has certain advantages over some of the other qubit and qubit gates realizations. 
\end{itemize}

The core of this thesis is contained in sections 2 and 3: in section 2, we outline some of the possibilities to numerically compute the properties of certain PHC structures. Section 3 introduces the Bloch-mode expansion method, a tool which is well-known and established in the field of Solid State Physics, but has only recently been applied in the PHC case \cite{savona}. We then use this method to obtain some interesting results regarding disorder (both designed and random) in the crystals. 

\newpage

\section{Simulating a regular structure}

The starting point for a theoretical treatment of a Photonic Crystal, when no quantum mechanics effects need be taken into account, are the Maxwell equations of electrodynamics. As long as we are interested only in the free light propagation in the crystal, we can assume zero external charges and electrical currents, which allows us to write the Maxwell equations for the electric field $\mathbf{E}$ and the magnetic field $\mathbf{H}$ as

\begin{align}
 \bm{\nabla} \cdot (\varepsilon \mathbf{E}) &= 0,  \\ \nonumber
 \bm{\nabla} \cdot \mathbf{H} &= 0, \\ \nonumber
 \bm{\nabla} \times \mathbf{E} &= - \frac{\mu}{c} \frac{\partial\mathbf{H}}{\partial t}, \\ \nonumber
 \bm{\nabla} \times \mathbf{H} &= \frac{\varepsilon}{c}\frac{\partial\mathbf{E}}{\partial t}, 
\label{max}
\end{align}

where $\varepsilon = \varepsilon(\mathbf{r})$ is the dielectric permittivity and $\mu = \mu(\mathbf{r})$ is the magnetic permeability of the medium, which in general depend on $\mathbf{r}$. In fact, it is their periodic dependence with $\mathbf{r}$ which is in the very definition of a PHC structure. Everywhere from now on, we will assume $\mu(\mathbf{r}) = 1$, which is true for a wide range of materials. If we also restrict the investigation to monochromatic light of frequency $\omega$, the time dependence will simply be given by a phase shift of $\omega t$, and we are left with the following equations for the electric and magnetic fields:

\begin{align}
 \frac{1}{\varepsilon(\mathbf{r})} \bm{\nabla} \times \bm{\nabla} \times \mathbf{E}(\mathbf{r}) = \frac{\omega^2}{c^2} \mathbf{E}(\mathbf{r}), \\ \nonumber
 \bm{\nabla} \times \frac{1}{\varepsilon(\mathbf{r})} \bm{\nabla} \times \mathbf{H}(\mathbf{r}) = \frac{\omega^2}{c^2} \mathbf{H}(\mathbf{r}).
\label{helmh}
\end{align}

In general, there are several popular ways to approach numerically the problem of finding the properties of a certain PHC structure. The Finite-Difference Time-Domain method (FDTD) is a brute-force solution of Maxwell's equations over a discretized space- and time-grid; while it is a very general method based on a first-principle computation, it is unfortunately often too demanding in computational resources. This is partly due to the fact that the discretization of space and time can introduce unwanted artifacts, and it is only in the limit of infinitesimal intervals that the solution is expected to correspond to the full Maxwell equations solution. In the ``scattering-matrix'' method, the modulation of the dielectric constant is represented as a system of scattering sources, repeated periodically in agreement with the PHC periodic structure. The scattering-matrix formalism (encountered frequently in quantum mechanics) is then used to investigate the propagation of light through the medium. This method, while being a bit more involved mathematically than the FDTD, is still essentially a first-principle simulation and so is expected to produce very good results, but it can also easily become computationally demanding to apply. 

The class of methods this thesis specializes in is the ``mode expansion'' methods, in which the light mode of an unknown system is expanded over the basis of known functions. Such a method is not very ubiquitous in the sense that one should choose a proper basis (which is not always easy to find) for every structure at hand, and that it involves an infinite summation which is not guaranteed to converge for any finite number of terms taken into account. Still, in cases where such a method works (and there are some in which it works perfectly), it proves to be very advantageous to use it, because it would usually have a lower computational complexity, and furthermore it could provide insight into the physics of the system by showing the degree of mixing between the actual mode of the system and the basis modes. 

In the rest of this thesis, we will concentrate on a PHC formed by cutting circular holes of radius $R$, on a lattice (triangular or square) with periodicity $a$ in a dielectric slab of thickness $d$ with dielectric constant $\varepsilon_{2}$. In general, the outside medium can be thought of as having some dielectric constant $\varepsilon_{1}$, but in all the actual computations we stick to air and so set $\varepsilon_{1} = 1$. Such structures are currently easy to produce in the lab with standard lithographic techniques, and thus provide a good research focus.

\subsection{Plane-wave expansion}

In standard solid state structures, the most common basis over which to expand the electron modes is the set of plane-wave functions of momentum $\mathbf{k}$, $\psi_{\mathbf{k}}(\mathbf{r}) = e^{i\mathbf{k}\mathbf{r}}$. A straightforward analogy that can help us obtain the band diagram of a regular PHC is by plane-wave expansion (PWE) of the electromagnetic modes. To do that, we first invoke the Bloch theorem, which can be proven to apply to the classical electromagnetic wave-equations in just the same way it applies for particles governed by the Schr\"{o}dinger wave-equation. Its main result is that the modes of the electric and the magnetic fields of a periodic structure can be written as 

\begin{align}
\mathbf{E}(\mathbf{r}) = \mathbf{E}_{\mathbf{k}n}(\mathbf{r}) \cdot e^{\mathrm{i}\mathbf{k}\cdot \mathbf{r}},\\
\nonumber \mathbf{H}(\mathbf{r}) = \mathbf{H}_{\mathbf{k}n}(\mathbf{r}) \cdot e^{\mathrm{i}\mathbf{k}\cdot \mathbf{r}},
\end{align}

where the functions $\mathbf{E}_{\mathbf{k}n}(\mathbf{r})$ and $\mathbf{H}_{\mathbf{k}n}(\mathbf{r})$ are called the Bloch modes of the structure. They have periodicity of the underlying lattice, and are labeled by their momentum $\mathbf{k}$ and a band index $n$. Based on the periodicity of the lattice, one can also define inverse lattice vectors $\mathbf{G}$ such that the Bloch modes corresponding to momenta $\mathbf{k}$ and $\mathbf{k} + \mathbf{G}$ are equivalent; thus, generally, only the modes within an area of the $k$-space where they can be uniquely defined (e.g. the first Brillouin zone) are considered. Finally, those modes can be represented in Fourier space by a summation over all reciprocal lattice vectors $\mathbf{G}$, or in other words they can be expanded on the basis of plane waves with momenta $\mathbf{k} + \mathbf{G}$:

\begin{equation}
\begin{split}
\mathbf{E}_{\mathbf{k}n}(\mathbf{r}) = \sum_{\mathbf{G}} \mathbf{E}_{\mathbf{k}n}(\mathbf{G})e^{\mathrm{i}(\mathbf{k} + \mathbf{G})\cdot \mathbf{r}},\\ 
 \mathbf{H}_{\mathbf{k}n}(\mathbf{r}) = \sum_{\mathbf{G}} \mathbf{H}_{\mathbf{k}n}(\mathbf{G})e^{\mathrm{i}(\mathbf{k} + \mathbf{G})\cdot \mathbf{r}}.
\end{split}
\label{bloch}
\end{equation}

\subsubsection{Numerical procedure}
\label{secpwe}

Here and in the rest of this thesis, we will consider only 2D-crystals. If in the third dimension translation symmetry is kept, i.e. if $\varepsilon(\bm{\rho}, z) = \varepsilon(\bm{\rho}, 0) \; \forall z$ and so $d = \infty$, and we restrict the discussion to light with in-plane wave-vector, there are two polarization states possible, namely Transverse Electric (TE), in which the z-component of the electric field vanishes as do the in-plane components of the magnetic field, and Transverse Magnetic (TM), in which the magnetic field is in-plane while the electric field is in the z-direction only. For TE-polarization, one obtains from eq. (2) the so-called Helmholtz equation for the perpendicular magnetic field component (we assume the periodicity of the crystal is in the x-y-direction) \cite{sakoda}:

\begin{equation}
- \left(\frac{\partial}{\partial x} \frac{1}{\varepsilon(\mathbf{r})} \frac{\partial}{\partial x} + \frac{\partial}{\partial y} \frac{1}{\varepsilon(\mathbf{r})} \frac{\partial}{\partial y}\right)H_z(\mathbf{r}) = \left(\frac{\omega}{c}\right)^2 H_z(\mathbf{r}).
\label{helmholtz}
\end{equation}

This equation holds for any mode supported by the system, so in particular - for the Bloch modes $\mathbf{H}_{\mathbf{k}n}$. It can be handled more conveniently in Fourier space, where eq. (\ref{helmholtz}) can be rewritten as the eigenvalue problem (often called the ``master equation''),

\begin{equation}
\sum_{\mathbf{G}'} \varepsilon^{-1}(\mathbf{G} - \mathbf{G}')(\mathbf{k} + \mathbf{G})(\mathbf{k} + \mathbf{G}') H_{z, \mathbf{k}n}(\mathbf{G}') = \left(\frac{\omega_{\mathbf{k}n}}{c}\right)^2 H_{z, \mathbf{k}n}(\mathbf{G}),
\label{master}
\end{equation} 

where $H_{z, \mathbf{k}n}(\mathbf{G})$ simply denotes the z-component of $\mathbf{H}_{\mathbf{k}n}(\mathbf{G})$, which for TE-polarization is the only non-vanishing one, and we have used the Fourier expansion of the inverse of the permittivity function, $\varepsilon^{-1}(\mathbf{G})$. Mathematically, this is simply 

\begin{equation}
\varepsilon^{-1}(\mathbf{G}) = \frac{1}{A} \int_{A} \frac{1}{\varepsilon(\mathbf{r})}\exp(-i\mathbf{G}\mathbf{r}) \mathrm{d^2}\mathbf{r},
\label{chi}
\end{equation}

with $A$ being the area of the unit cell. In the case of circular holes in a dielectric medium of permittivity $\varepsilon_2$, the integral in eq. (\ref{chi}) can be computed analytically and represented using a Bessel function of the first kind, namely 

\begin{align}
\varepsilon^{-1}(\mathbf{G} \neq 0) &= 2f\left(\frac{1}{\varepsilon_1} - \frac{1}{\varepsilon_2}\right) \frac{J_1(Gr)}{Gr}, \nonumber \\ 
\varepsilon^{-1}(0) &= f \frac{1}{\varepsilon_1} + (f-1) \frac{1}{\varepsilon_2},
\label{epsdirect}
\end{align}

where $\varepsilon_1$ is the permittivity of the surrounding medium (usually air), and $f$ is the so-called "filling factor", defined as the ratio of the hole area to the total elementary cell area, so in this case, $f = \pi R^2/A$. However, due to the discontinuities of the permittivity at the edges of the holes, it turns out that a method to compute $\varepsilon^{-1}(\mathbf{G} - \mathbf{G}')$ which was first used by Ho \cite{ho} gives much better numerical convergence. Namely, the method involves computing the Fourier transform $\varepsilon(\mathbf{G} - \mathbf{G}')$ as a matrix over all possible combinations of differences of $\mathbf{G}$-vectors taken into account,

\begin{equation}
 \begin{pmatrix}
  \varepsilon(\mathbf{G}_1 - \mathbf{G}_1) & \varepsilon(\mathbf{G}_1 - \mathbf{G}_2) & \cdots & \varepsilon(\mathbf{G}_1 - \mathbf{G}_n) \\
  \varepsilon(\mathbf{G}_2 - \mathbf{G}_1) & \varepsilon(\mathbf{G}_2 - \mathbf{G}_2) & \cdots & \varepsilon(\mathbf{G}_2 - \mathbf{G}_{n})\\
  \vdots  & \vdots  & \ddots & \vdots  \\
  \varepsilon(\mathbf{G}_{n} - \mathbf{G}_1)& \varepsilon(\mathbf{G}_{n} - \mathbf{G}_2) & \cdots & \varepsilon(\mathbf{G}_{n} - \mathbf{G}_n)
 \end{pmatrix},
\label{inveps}
\end{equation}

and then using matrix inversion to obtain $\varepsilon^{-1}(\mathbf{G} - \mathbf{G}')$. Detailed discussion of this numerical procedure is given in sec. \ref{secho}.

\begin{figure}
\begin{center}
 \subfloat[]{\includegraphics[width = 0.33\textwidth]{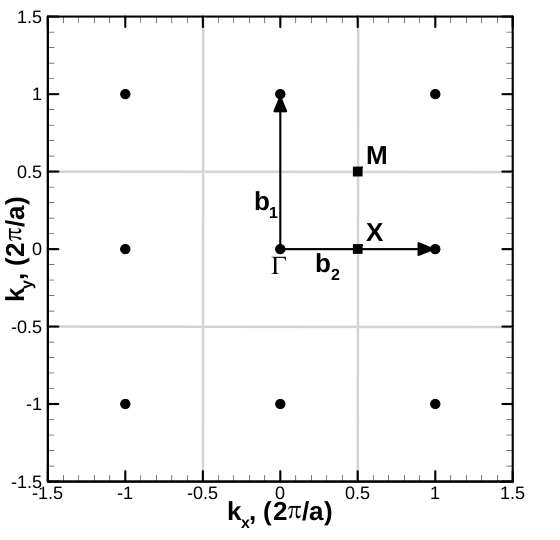}}
 \subfloat[]{\includegraphics[width = 0.66\textwidth]{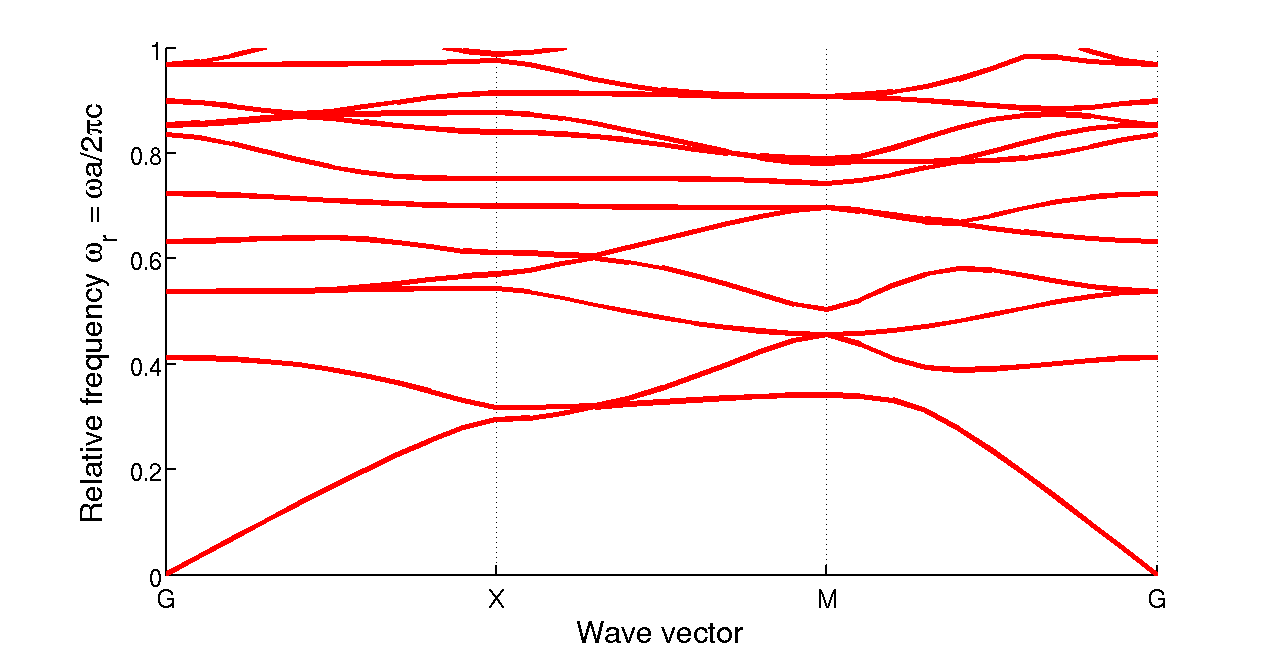}}
 \caption{(a): reciprocal lattice, basis vectors, and high symmetry points of a square lattice of lattice spacing $a$, image from \cite{zabelin}. (b): photonic band structure of such a lattice with circular holes of radius $0.38 a$ in a slab of infinite thickness and dielectric constant $\varepsilon_2 = 9$, computed with PWE.}
 \label{band2d}
 \end{center}
\end{figure}

The plane-wave expansion method has been widely used in different applications, but it is not the purpose of this thesis to go into further detail; so far we outlined the basic numerical procedure of simulating a 2D PHC structure using the method. As a concluding illustration, we show in figure \ref{band2d} (b) the band diagram of a PHC of a square lattice of circular holes in a dielectric medium, computed with the PWE method. Fig. \ref{band2d} (a) illustrates the reciprocal lattice of such a structure, with the reciprocal basis vectors and the high symmetry points given for reference.

\subsection{Guided-mode expansion}

The guided mode expansion (GME), first applied to PHCs by Andreani and Gerace \cite{andreani_2006}, is, similarly to the plane-wave expansion method, in essence an expansion of the modes of the crystal over a given basis set. So far in the PWE, we only considered 2D waveguides with full translational symmetry along the third axis. It is only in this case that the two independent polarizations, TE and TM, can always be defined. In reality, however, 2D PHC-s are usually fabricated in one or more layers of different materials - one of the most straightforward ways to do that is by etching holes in an e.g. silicon slab suspended in air. Without any holes present, such a slab can be viewed as a waveguide due to the difference in refractive indices at the interfaces, i.e. such a slab supports guided modes. Thus, it seems convenient to use as an expansion basis the guided modes of an effective homogeneous slab of the same thickness as the PHC, but with no holes and with an averaged, constant dielectric permittivity

\begin{equation}
\bar{\varepsilon}_2 = \frac{1}{A} \int_{A} \varepsilon_2(\bm{\rho}) \mathrm{d} \bm{\rho},
\end{equation}. 

where A is the elementary cell area, determined based on the PHC structure in consideration. Below, we consider the slab to be sandwiched between two semi-infinite layers of dielectric constant $\bar{\varepsilon}_1 < \bar{\varepsilon}_2$ (in all the computations presented this is just air). The guided modes in such a slab have dispersion given by an implicit trigonometric equation, and the details for the computation of their energies as well as the electric and magnetic field profiles can be found in \cite{andreani_2006}. Those modes are labeled by their in-plane momentum $\mathbf{g}$, their mode number, $\alpha$, and their polarization, TE or TM, i.e. we will denote the magnetic field profile of a specific guided mode as  $\mathbf{H}_{\mathbf{g}, \alpha}(\mathbf{r})$ and specify the polarization separately (for brevity we also label $\mu = (\mathbf{g}, \alpha))$. We can then write a similar expansion as in eq. (\ref{bloch}) for the Bloch modes of the PHC, but expand over guided modes with momenta $\mathbf{G} + \mathbf{k}$ instead of plane waves:

\begin{equation}
\mathbf{H}_{\mathbf{k}n}(\mathbf{r}) = \sum_{\mathbf{G}, \alpha} c_n(\mathbf{k} + \mathbf{G}, \alpha) \mathbf{H}_{\mathbf{k}+\mathbf{G}, \alpha}(\mathbf{r}).
\label{guided}
\end{equation}

This summation is generally a linear superposition of both TE and TM modes, and the resulting Bloch mode has no well-defined polarization, i.e. all of the field components can generally be non-vanishing. However, in a system as the one we will consider, namely consisting of the same upper and lower claddings, those can have a quasi-TE or a quasi-TM nature, especially for low energies, which means that the TE (or correspondingly TM) guided modes give the predominant contribution to the summation in eq. (\ref{guided}). Moreover, the quasi-TE modes are even w.r.t. reflection by the $z = 0$ plane and are the lowest-energy ones, while the quasi-TM modes are odd w.r.t. the same operation. Below we present the general numerical procedure, but in all computations only the first, $\alpha = 1$ (lowest energy) modes were used. In a symmetric structure those are TE modes, and thus the resulting Bloch-modes are quasi-TE.

\subsubsection{Numerical procedure}

For the computations, just as we had an eigenvalue problem \ref{master} in the PWE method, so do we obtain now, once we use eq. (\ref{max}), the expansion in eq. (\ref{guided}), and the normalization condition for guided modes,

\begin{equation}
 \int \mathrm{d}\mathbf{r} \mathbf{H}^*_{\nu}(\mathbf{r}) \mathbf{H}_{\mu}(\mathbf{r}) = \delta_{\mu \nu}.
\label{guidenorm}
\end{equation}

For the complete description with all the relevant equations of the algorithm, the reader is referred to \cite{andreani_2006}, but the bottom line is that one can obtain the expansion coefficients $c_n(\mu)$, and thus the Bloch-modes of the PHC structure via eq. (\ref{guided}), by diagonalizing the Hermitian matrix

\begin{equation}
 \mathcal{H}_{\mu \nu} = \int \frac{1}{\varepsilon(\mathbf{r})}\left(\bm{\nabla} \times \mathbf{H}^*_{\mu}(\mathbf{r})\right)\cdot\left(\bm{\nabla} \times \mathbf{H}_{\nu}(\mathbf{r})\right) \mathrm{d}\mathbf{r}.
\label{guidedeig}
\end{equation}
 
Once the magnetic field of a mode is known, the electric field can also be computed through the relationship

\begin{equation}
 \mathbf{E}(\mathbf{r}) = \frac{ic}{\omega\varepsilon(\mathbf{r})} \mathbf{\bm{\nabla}} \times \mathbf{H}(\mathbf{r}).
\end{equation}

The Fourier expansion $\varepsilon^{-1}(\mathbf{G})$, as given in eq. (\ref{chi}), again enters the matrix for diagonalization, $\mathcal{H}_{\mu\nu}$, and in the GME just as in the PWE, computing this quantity with the inverse-matrix method gives much better results. 

Extrinsic losses of the Bloch modes can also be treated by the GME approach, in a way reminiscent of the Fermi golden rule of quantum mechanics. Namely, the radiative losses can be represented by the imaginary part of their frequency, and this can be estimated through $\Im(\omega) \approx \Im(\omega^2)/(2\omega)$ and

\begin{equation}
-\Im\left(\frac{\omega_{k}^2}{c^2}\right) = \pi \sum_{\mathbf{G'}} \sum_{\lambda, j} | \mathcal{H}_{\mathbf{k}}^{\mathrm{rad}} |^2 \rho_j \left( \mathbf{k} + \mathbf{G}'; \frac{\omega_{k}^2}{c^2} \right), 
\label{imom}
\end{equation}

where $\lambda$ labels polarization of the radiative guided modes (TE or TM), $j = 1, 3$ labels whether those modes are leaky in the upper or the lower cladding, and $\rho_j\left( \mathbf{k} + \mathbf{G}'; \frac{\omega_{k}^2}{c^2} \right)$ is their 1D photonic density of states at fixed wave-vector, and $\mathcal{H}_{\mathbf{k}}^{\mathrm{rad}}$ is the overlap between a guided and a leaky slab mode, 

\begin{equation}
 \mathcal{H}_{\mathbf{k}}^{\mathrm{rad}} = \int \frac{1}{\varepsilon(\mathbf{r})}\left(\bm{\nabla} \times \mathbf{H}^*_{\mathbf{k}}(\mathbf{r})\right)\cdot\left(\bm{\nabla} \times \mathbf{H}_{\mathbf{k}+\mathbf{G}', \lambda, j}^{\mathrm{rad}}(\mathbf{r})\right) \mathrm{d}\mathbf{r}.
\end{equation}

The quality factor of a resonant mode can then also be found through $Q = \omega/(2\Im(\omega))$

\subsubsection{Application to waveguides and cavities}

This method can be easily used for treating defects such as waveguides or cavities in a PHC. As discussed before, one way to form a waveguide is to ``erase'' a whole row of holes, i.e. to introduce a 1-dimensional defect in the crystal. This introduces localized modes within the photonic band gap of the unperturbed crystal, in which the electric field is strongly localized to the defect region. In a similar way one can construct an optical cavity, by just erasing a finite number of holes - one of the most studied such cavities is the L3 cavity, which is constructed by not making three holes in a row of a triangular-lattice-of-circular-holes PHC slab (fig. \ref{cells}). In both cases, one can use the guided-mode expansion method to compute the PHC properties, with the difference that one has to consider ``super-cells'' instead of the normal primitive cell of the crystal in computing the expansion coefficients. Super-cells correspond to a collection of primitive cells, large enough to capture the properties of the introduced defect, and, especially for determining correctly the properties of cavities, large enough so that boundary effects have little to no effect. 

\subsubsection{Results of computations}

\begin{figure}[h!]
\begin{center}
 \subfloat[]{\includegraphics[width = 0.33\textwidth]{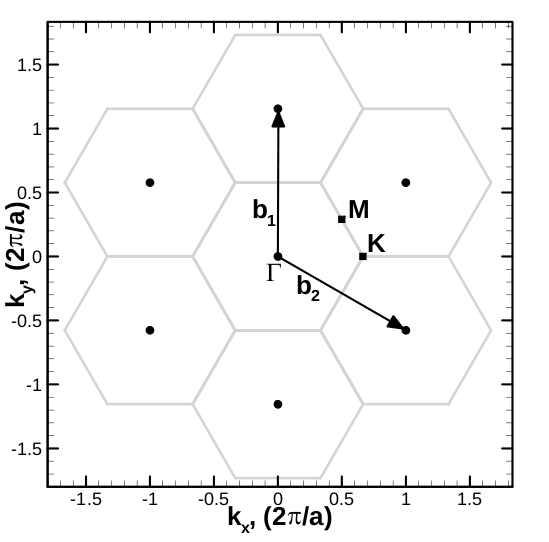}}
 \subfloat[]{\includegraphics[width = 0.66\textwidth]{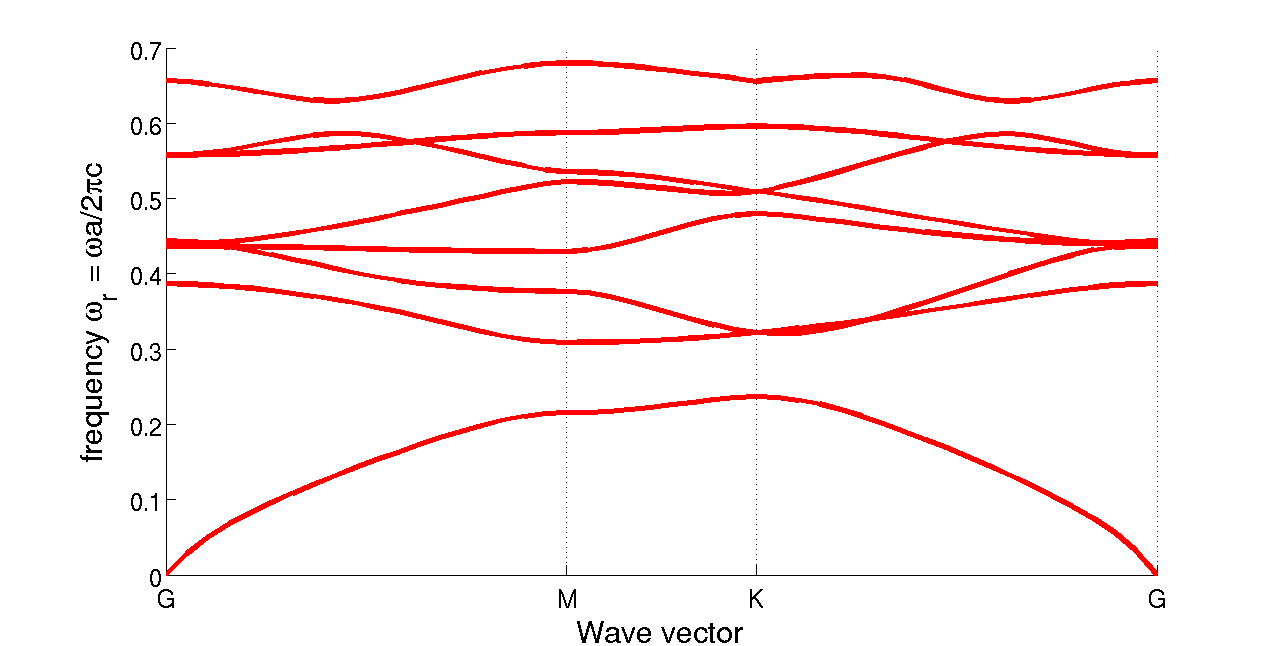}}
 \caption{(a): reciprocal lattice, basis vectors, and high symmetry points of a triangular lattice of lattice spacing $a$, image from \cite{zabelin}. (b): photonic band structure of such a lattice with circular holes of radius $0.3 a$ in a slab of thickness $a$ and dielectric constant $\varepsilon_2 = 12$, computed with GME.}
 \label{band2d1}
 \end{center}
\end{figure}

We illustrate the GME method with three results of computations. First, the photonic energy bands of a PHC with a triangular lattice of circular holes with spacing $a$ between neighboring holes was computed and can be seen in figure \ref{band2d1}. The parameters used were as follows: radius of the holes $r=0.3a$, slab thickness $a$, and dielectric constant of the slab material $\varepsilon_2 = 12$. Furthermore, we limited the $G$-vectors in the computation up to a magnitude $G_{max} = 3$ in units of $\frac{2 \pi}{a}$. 

\begin{figure}[h!]
 \subfloat[]{\includegraphics[type=pdf,ext=.pdf,read=.pdf, width = 0.035\textwidth, trim = 0in 1in 0in 0in, clip = true]{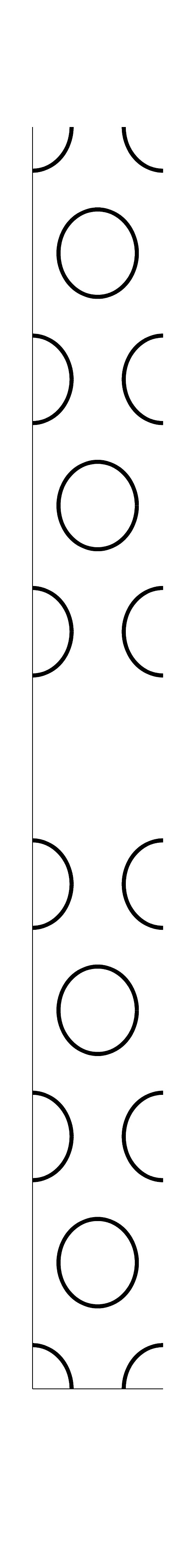}}
 \subfloat[]{\includegraphics[width = 0.52\textwidth, trim = 0in 0in 1in 1in, clip = true]{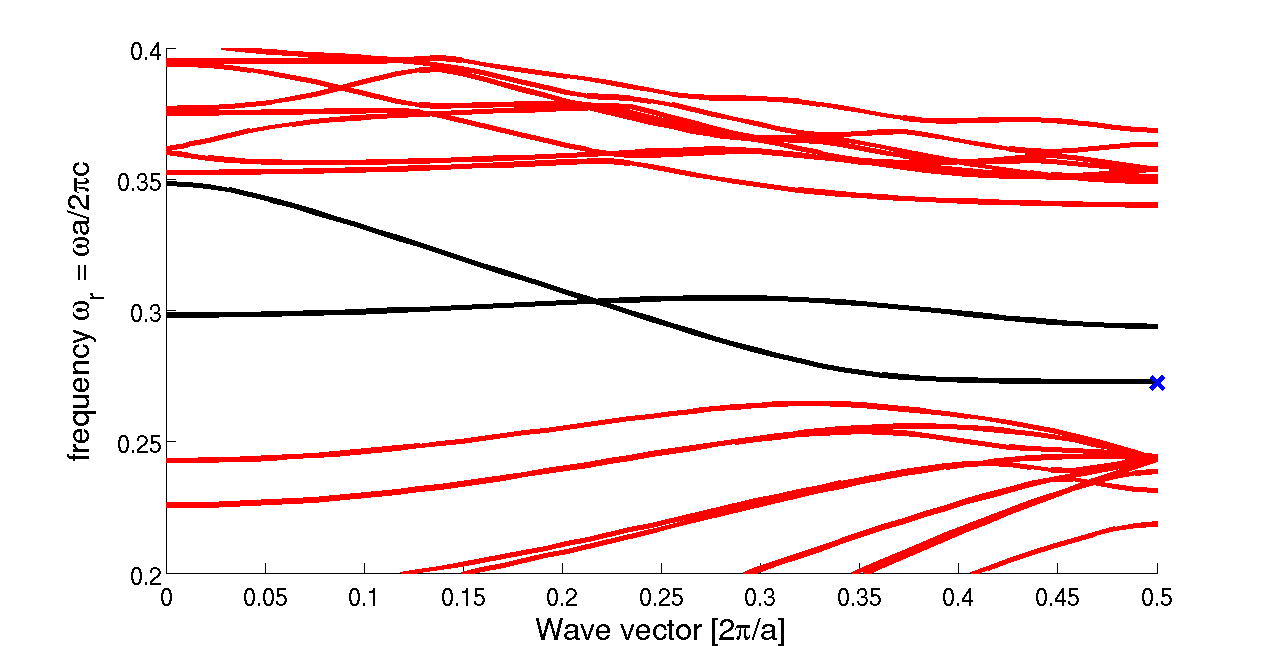}}
 \subfloat[]{\includegraphics[type=pdf,ext=.pdf,read=.pdf, width = 0.42\textwidth, height = 0.24\textwidth, trim = 1in 1.3in 1in 1in, clip = true]{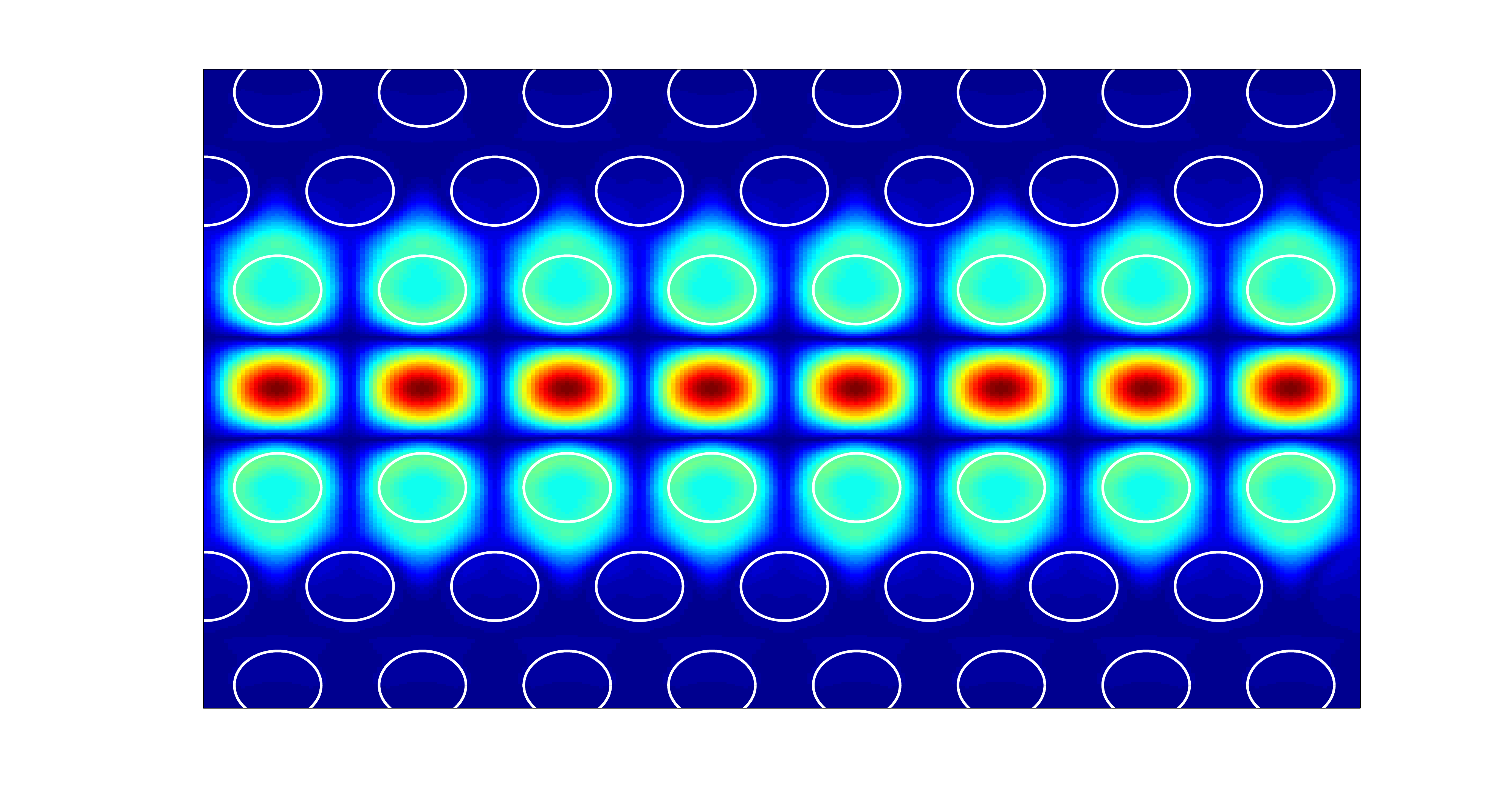}}
 \caption{(a): supercell for the PHC waveguide computation (b): band structure of the waveguide; the two black bands are the guided ones. (c): real-space magnetic field profile of the spatially-even guided mode indicated by a blue cross on the band diagram; the positions of the holes are also plotted}
 \label{guidedband}
\end{figure}

The method was then extended to the computation of a waveguide band diagram, the waveguide consisting of one row of missing holes. The super-cell considered was a square one of dimension $\left[a, 10\frac{\sqrt{3}}{2}a\right]$, and the result was plotted in figure \ref{guidedband}. The parameters used in that computation were, $r = 0.3a$, $d = 0.5a$, and $\varepsilon_2 = 12$. In the figure, one can see the band structure of the PHC (only wavevectors parallel to the direction of the waveguide are considered) and notice the two guided bands which lie in the band-gap of the unperturbed crystal. They correspond to a spatially-even and a spatially-odd band (with respect to the $\sigma_{\vec{k}z}$ mirror plane). The electric field of the spatially-even mode with momentum vector lying on the Brillouin zone edge, i.e. $k = \frac{\pi}{a}$, is also plotted, to demonstrate the localization.

\begin{figure}[h!]
 \subfloat[]{\includegraphics[trim = 1in 0in 1in 0in, clip = true, type=pdf,ext=.pdf,read=.pdf, width = 0.49\textwidth]{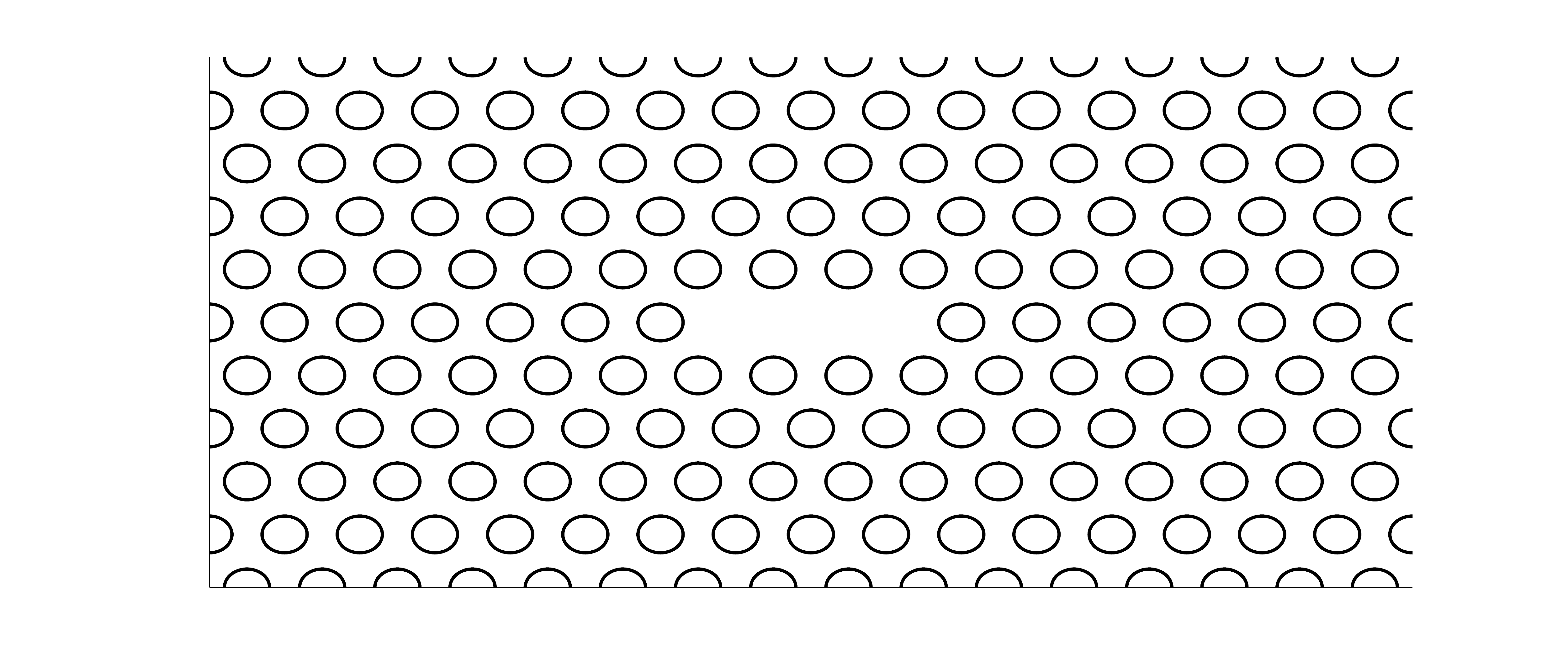}}
 \subfloat[]{\includegraphics[trim = 1in 1in 1in 1in, clip = true, type=pdf,ext=.pdf,read=.pdf, width = 0.49\textwidth]{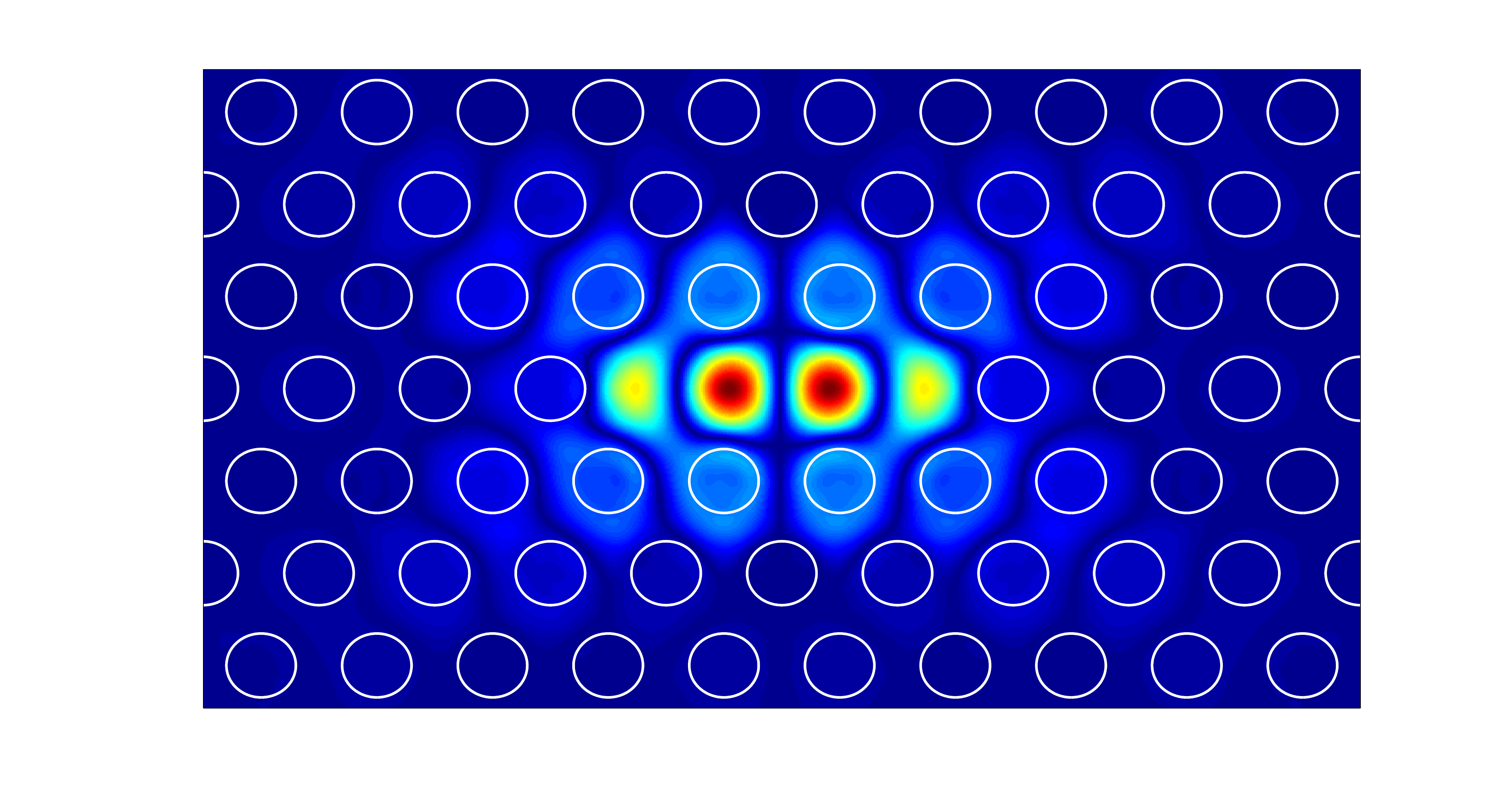}}
 \caption{(a): supercell used in the L3-cavity computation (b): real-space profile of the magnetic field of the main resonant mode inside the cavity (only the part of the supercell where the field is non-vanishing is shown)}
 \label{cells}
\end{figure}

Finally, we also illustrate an L3-cavity computation done with the GME method. For this to be correct, the supercell has to be big enough so that the field is vanishing on its edges, since otherwise the computation will yield the collective excitations of coupled cavities. In other words, what we compute is, in essence, a structure with periodically repeated supercells as the one in fig. \ref{cells}, each containing an L3 cavity. When they are sufficiently big, however, the results can also be viewed as a single-cavity computation. For fig. \ref{cells}, the supercell has size $\left[16a, 10\frac{\sqrt{3}}{2}a\right]$, with the other PHC slab parameters being the same as for the waveguide computation. Using eq. (\ref{imom}), a quality factor of the cavity of $Q = 4659$ was computed. This value is quite low, but as is discussed in detail e.g. in \cite{andreani_L3}, it can be largely improved by slightly displacing or changing the size of the hole immediately on the left and the one immediately on the right of the cavity. With such changes, the quality factor can go well above $10^5$. 

\subsubsection{Discussion of PWE and GME}

The plane-wave expansion method (i.e. the form of the expansion given in eq. (\ref{bloch})) is useful and very fast when computing properties of a simple structure, e.g. a regular crystal with infinite thickness. It can also be used to examine defects such as point defects or waveguides, but to this end the required computational time can significantly increase to obtain satisfactory accuracy. Furthermore, it is not a trivial matter to use it to model a slab of finite thickness, and even less - to treat radiation losses. This, however, comes quite natural in the GME method, as the guided modes used there are already concentrated around the core of the slab waveguide and decay in the claddings. This method can sometimes be more computationally demanding and a little bit more mathematically involved, but it has many advantages. Not only is it straightforward and more precise when treating a slab of finite thickness, it also performs very well when defects such as waveguides or cavities are introduced, and allows the computation of both mode profiles (and from there e.g. cavity volumes) and radiative rates (and so cavity quality factors). It is thus a very versatile method that can be applied in many cases and provide reliable results, while still keeping the numerical complexity to an acceptable level, as is expected from a basis-expansion method. 

Despite all its advantages, the GME is not the best tool for the purposes of this thesis - analyzing disorder in PHC waveguides or cavities, as, first of all, it does not allow for computations of big supercells, since then the diagonalization matrix in \ref{guidedeig} becomes too large, and second of all, computing multiple realizations of disorder becomes cumbersome as each computation has to be done separately and cannot take advantage of the fact that it is only a small perturbation to a pre-computed structure. In sec. \ref{secbme}, however, we outline the Bloch-mode expansion method, which remedies those problems by allowing us to treat disorder as a perturbation over a known structure, which could be first computed with GME. 

\subsection{A numerical subtlety: ``Ho's approach''}

\label{secho}

The ``inverse matrix'' method for computing $\varepsilon^{-1}(\mathbf{G} - \mathbf{G}')$ was first applied both to the plane-wave expansion \cite{ho} and to the guided-mode expansion \cite{andreani_2006} not because of rigorous mathematical formulation of its advantages, but rather because it seems to always give better convergence. To our knowledge, in the case of 2D PHCs there is, to this day, no precise mathematical theory of how the problem should be tackled. However, papers by Li \cite{li} and Shen and He \cite{shen}, which concentrate on the 1D problem, shed some light on the matter, which could better our understanding also in the 2D case.

\subsubsection{Laurent's rule vs. inverse rule for computing Fourier transforms}

The problem of computing a Fourier expansion of a periodic function is that the resulting series is convergent only w.r.t. the $L_2$ integral norm over periodic functions. In other words, the Fourier series of a discontinuous function is not necessarily point-wise convergent around the discontinuities (though the set of points where the series does not converge has to be of zero measure). The quantity $\frac{1}{\varepsilon(\mathbf{r})}$ which enters eq. (\ref{max}) is discontinuous at the hole edges. The electric and magnetic fields $\mathbf{E}(\mathbf{r})$ and $\mathbf{H}(\mathbf{r})$ themselves can also be either continuous or discontinuous at the boundaries depending on the structure and the polarization, but in the most general case they have both a continuous and a discontinuous component. Thus, a Fourier expansion of a product between the two is not necessarily point-wise convergent and thus can produce numerical artifacts. The main result of Li's paper \cite{li} is how to avoid this in a 1D case: faced with a function $h(x) = f(x)g(x)$, where $f(x)$ and $g(x)$ are in general both piece-wise smooth, periodic, and bounded, one has three possibilities: 1, that they have no concurrent discontinuities, 2, that they have only concurrent discontinuities which in the product complement each other to a continuous $h(x)$, and 3, that they also have concurrent discontinuities which do not complement. To have best convergence in the numerics, it turns out that one should split the problem (starting from eq. (\ref{max}) and taking into account the structure and polarizations in consideration) into products of type 1 and 2 only. Then, for type-1 products one should apply the ``Laurent rule'' for computing the Fourier coefficients of $h(x)$,

\begin{equation}
 h_n^{(M)} = \sum_{m = -M}^{M} f_{n-m}g_{m},
\end{equation}

i.e. in our case one can directly compute $\varepsilon^{-1}(\mathbf{G})$ using eq. (\ref{epsdirect}). However, for products of type 2, one should compute this through the ``inverse matrix'' rule, briefly mentioned in sec. \ref{secpwe}:

\begin{equation}
 h_n^{(M)} = \sum_{m = -M}^{M} \left(\sembrack{\frac{1}{f}}^{(M)}\right)^{-1} g_m,
\end{equation}

where $\sembrack{f}$ denotes the Toeplitz matrix of Fourier coefficients of $f$, arranged such that the $(n, m)$-th entry of the matrix is $f_{n-m}$.

\begin{figure}
\begin{center}
 \includegraphics[type=pdf,ext=.pdf,read=.pdf, width = 0.7\textwidth]{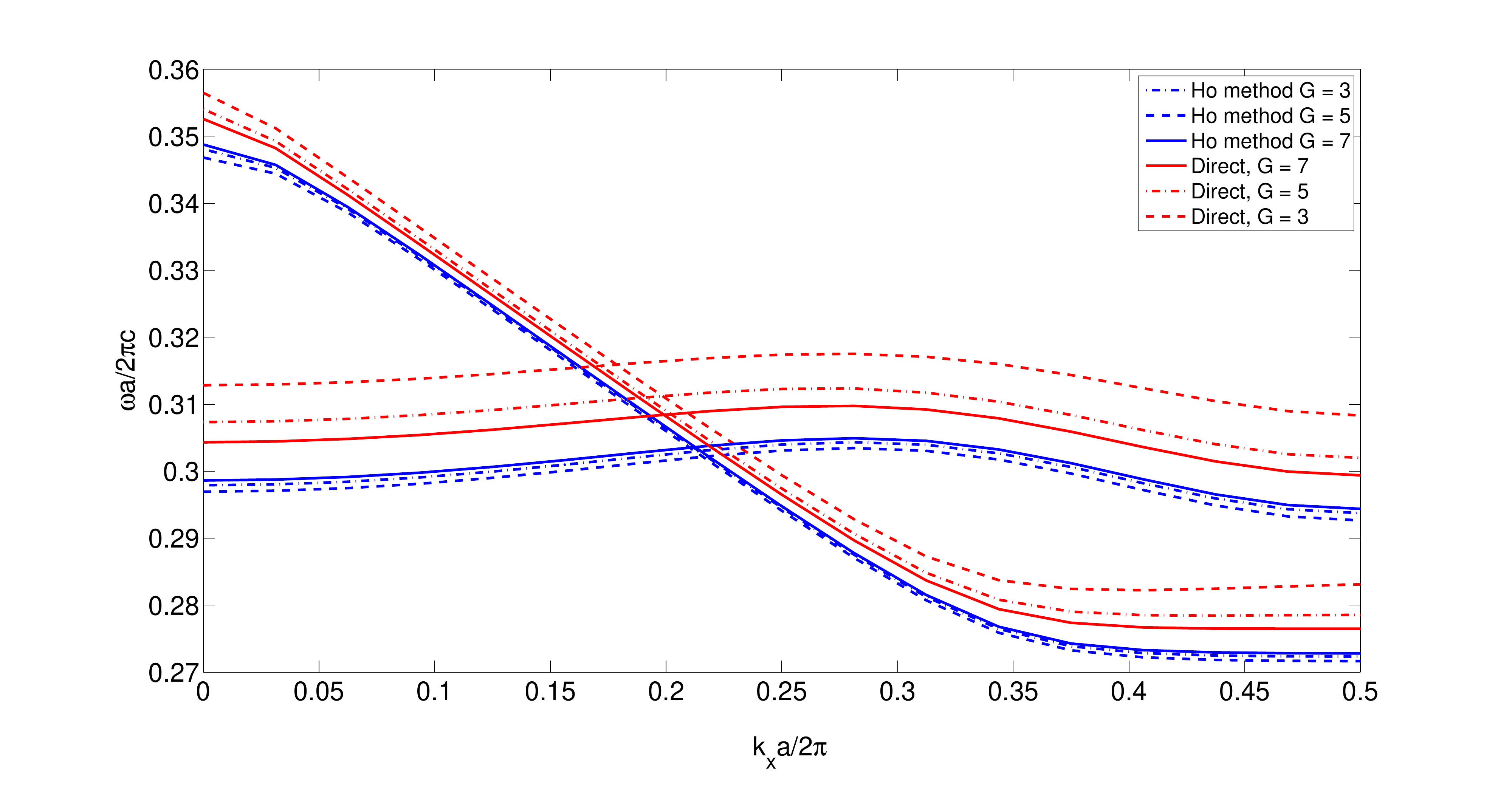} 
 \caption{The guided bands of the waveguide shown in fig. \ref{guidedband} computed using GME where either a direct computation, i.e. using eq. (\ref{epsdirect}) (red) or the Ho method (blue) was used, for $G_{max} = 3$ (dashed), $5$ (dashed-dotted) and $7$ (solid).}
 \label{ho_noho}
\end{center}
\end{figure}

In a 1D-problem, it is often possible to do such a splitting of the problem, and thus to treat the products ``properly''. In a typical 2D case, however, for example in the case of the triangular-lattice of circular holes structures studied here, one would need a very complicated coordinate system with tangential and normal vectors to the hole edges, to be able to split the fields into fully continuous and fully discontinuous at the edges. This, while not necessarily impossible, would still be very arduous mathematically. It turns out, however, that applying only the inverse rule, sometimes called the ``Ho method''  provides very good results in many systems (and in any case, much better results than using only the Laurent's rule, or a ``direct computation''). As an illustration of this point, in fig. \ref{ho_noho} we show the guided bands of a PHC waveguide computed with the two methods, for $G_{max} = 3, 5, 7$ in units of $\frac{2 \pi}{a}$. As can be seen, the convergence of the Ho-method computation is much better, and the same was seen for virtually all computations presented subsequently in sec. \ref{secbme} - both convergence and results reliability appeared to be much better when the inverse-matrix method was applied. Thus, on solely empirical grounds, everywhere in this thesis only the results using this method are presented.

\subsubsection{Inverting a Toeplitz-block Toeplitz matrix}

\label{invtoep}

In the 1D case, the matrix for inversion defined in eq. (\ref{inveps}) is Toeplitz, i.e. the $(n, m)$-th entry is only a function of $n-m$. The numerical complexity of inverting such a matrix is greatly aided by an algorithm proposed by Trench \cite{trench}, bringing it from $N^3$ to $N^2$. In the 2D case, however, the matrix is no longer Toeplitz because the vectors $\mathbf{G}$ have two components. One can, however, construct a Toeplitz-block Toeplitz matrix if one considers $\mathbf{G}$-vectors defined on a square lattice of side-length $2 G_{max}$. The vectors defined this way will have $N_x$ different $G_x$ values and $N_y$ different $G_y$ values (s.t. $N = N_x N_y$), and the differences $\mathbf{G} - \mathbf{G}'$ can be arranged in the following way:

\begin{equation}
 \begin{pmatrix}
  \hat{G}_{11} & \hat{G}_{12} & \cdots & \hat{G}_{1 N_{x}} \\
  \hat{G}_{21} & \hat{G}_{22} & \cdots & \hat{G}_{2 N_{x}} \\
  \vdots  & \vdots  & \ddots & \vdots  \\
  \hat{G}_{N_{x}1} & \hat{G}_{N_{x}2} & \cdots & \hat{G}_{N_{x}N_{x}} \\
 \end{pmatrix}
\end{equation}

where $\hat{G}_{nm}$ are $N_y \times N_y$-dimensional Toeplitz matrices defined as 

\begin{equation}
\hat{G}_{nm} =
  \begin{pmatrix}
   (G_x^n - G_x^m, G_y^1 - G_y^1) & (G_x^n - G_x^m, G_y^1 - G_y^2) & \cdots & (G_x^n - G_y^m, G_y^1 - G_y^{N_y}) \\
   \vdots  & \vdots  & \ddots & \vdots  \\
  (G_x^n - G_y^m, G_y^{N_y} - G_y^1) & (G_x^n - G_y^m, G_y^{N_y} - G_y^2) & \cdots & (G_x^n - G_x^m, G_y^{N_y} - G_y^{N_y})
  \end{pmatrix}
\end{equation}

We then need to compute the matrix $\varepsilon(\mathbf{G} - \mathbf{G}')$ defined over the above arrangement of $\mathbf{G} - \mathbf{G}'$, and invert it, in order to apply the Ho method. Thus, we present the algorithm for inverting a Hermitian Toeplitz-block Toeplitz matrix, following the general algorithm by Wax and Kailath, \cite{wax}, which is itself a generalization of the algorithm for inverting Toeplitz matrices first proved mathematically by Trench \cite{trench} and given a convenient computer implementation form by Zohar, \cite{zohar}. For the sake of generality, we present the inversion of a Toeplitz-block Toeplitz matrix composed of $n+1$ different $p \times p $ dimensional matrices, and their Hermitian conjugates, i.e.

\begin{equation}
 R_{p, n+1} =
 \begin{pmatrix}
  \rho_0 & \rho_1 & \cdots & \rho_n \\
  \rho_1^{\dagger} & \rho_0 & \cdots & \rho_{n-1} \\
  \vdots  & \vdots  & \ddots & \vdots  \\
  \rho_n^{\dagger} & \rho_{n-1}^{\dagger} & \cdots & \rho_0
 \end{pmatrix}
\end{equation}

where each of the block matrices $\rho_0, \rho_1 \dots \rho_n$ are Toeplitz. We start with a note on the notation used: we will use the standard notations, $A^*$ and $A^T$, for complex conjugation and matrix transpose, respectively. We will denote vectors with bold lowercase letters, and $p \times p$ dimensional matrices will be denoted by lowercase Greek letters. Arrays of such matrices, like the matrix $R_{p, n+1}$ given above, will be denoted by capital letters. Finally, $\pi_p$ denotes the square $p \times p$ matrix with units along the cross-diagonal and zeros everywhere else. This matrix plays an important role here because of the persymmetric structure of a Toeplitz matrix. Note that multiplying a matrix on the left by $\pi_p$ is equivalent to exchanging elements which are symmetric along a bisecting horizontal line, while multiplying on the right exchanges the elements symmetric to a bisecting vertical line (and thus a matrix $\rho$ is persymmetric if $\pi_p \rho \pi_p = \rho^T$).

In general, a Toeplitz matrix is fully defined by its first row and first column. If $\mathbf{r}$ and $\mathbf{c}$ are two arbitrary vectors of equal dimension and with an equal first element, we will write as $T = \sembrack{\mathbf{r}, \mathbf{c}}$ the Toeplitz matrix which has $\mathbf{r}$ as its first row and $\mathbf{c}$ as its first column. Because all the matrices $\rho_0 \dots \rho_n$ are Toeplitz, this means that we can store all the information about the matrix $R_{p, n+1}$ in two vectors of $(n+1)p$ elements, $\mathbf{s}$ and $\mathbf{t}$, such that $\mathbf{s}$ contains all the firs rows of the matrices $\rho_0 \dots \rho_n$, while $\mathbf{t}$ contains all of the corresponding first columns, in the following way:

\begin{align}
 \mathbf{s} = \{\overbrace{s_1 \dots s_p}^{(\rho_{0})_{1j}}, \overbrace{s_{p+1} \dots s_{2p}}^{(\rho_{1})_{1j}}, \dots \overbrace{s_{np + 1} \dots s_{(n+1)p}}^{(\rho_{n})_{1j}}\} \\ \nonumber
 \mathbf{t} = \{\underbrace{t_1 \dots t_p}_{(\rho_{0})_{i1}}, \underbrace{t_{p+1} \dots t_{2p}}_{(\rho_{1})_{i1}}, \dots \underbrace{t_{np + 1} \dots t_{(n+1)p}}_{(\rho_{n})_{i1}}\} .
\end{align}

This is simply an efficient way to work with the matrices, from a computer memory point of view. In the description of the algorithm below we will still use the $\rho_m$ matrices, which are readily obtained: 

\begin{equation}
\rho_m = \sembrack{(s_{mp+1} \dots s_{(m+1)p}), (t_{mp+1} \dots t_{(m+1)p})}, \quad m = 0 \dots n
\end{equation}

The algorithm to obtain $R_{p, n+1}^{-1}$ is composed of two main parts: the first is computing the $p\times p$ dimensional matrix $\alpha_n$ and the array of matrices $\hat{W}_n = \{\omega_1, \dots \omega_n\}$ (corresponding to $\hat{\underbar{W}}_n^T$ in \cite{wax}), and the second is reconstructing recursively the full inverse matrix. We begin with the outline of the first part. The algorithm itself is a recursive one, where at each step $\alpha_{m+1}$ and $\hat{W}_{m+1}$ are calculated on the basis of $\alpha_{m}$ and $\hat{W}_{m}$, for $m = 1 \dots n - 1$. The starting values for the iteration are

\begin{align}
 \alpha_1 &= \rho_0 - \rho_1^{\dagger} \rho_0^{-1} \rho_1 \\ \nonumber
 \hat{W}_1 &= \{\omega_1\} = -\pi_p \rho_0^{-1} \rho_1 
\end{align}

Then, for $1 \leq m \leq n-1$ we compute, with the help of the intermediate matrix $\beta_m$:

\begin{align}
 \beta_m &= \pi_p \rho_{m+1} + \sum_{j = 1}^{m}\pi_p \omega_{m-j+1} \pi_p \rho_j \\ \nonumber
 \omega_{j} &= \omega_j - \pi_p \omega_{m-j+1}^* (\alpha_m^{-1})^T \beta_m, \quad j = 1 \dots m\\ \nonumber
 \hat{W}_{m+1} &= \{\omega_1 \dots \omega_m, -(\alpha_m^{-1})^T \beta_m \} \\ \nonumber
 \alpha_{m+1} &= \alpha_m - \beta_m^{\dagger}(\alpha_m^{-1})^T \beta_m
\end{align}
 
The full inverse matrix $R_{p, n+1}^{-1}$ can be fully constructed once $\alpha_n$ and $\hat{W}_n$ are obtained. In the case of PWE and GME computations, one needs this matrix in order to compute the matrix for diagonalization. However, in the Bloch-mode expansion (BME) method outlined in sec. \ref{secbme}, the complete matrix is not needed at once, and thus as long as using computer memory efficiently is concerned, we would like to not store it unless we have to. Hence, we give here the procedure to compute it row by row, where every row is computed on the basis of the previous one. The matrix $R_{p, n+1}^{-1}$ will generally not be Toeplitz, however it will still be Hermitian and persymmetric; hence we only need to compute one quarter of its blocks, e.g. the ones lying above both the main and the secondary diagonal; the rest can be expressed accordingly:

\begin{equation}
 R_{p, n+1}^{-1} =
 \begin{pmatrix}
  \tau_{1, 1} & \tau_{1, 2} & \cdots & \tau_{1, n} & \tau_{1, n + 1} \\
  \tau_{1, 2}^{\dagger} & \tau_{2, 2} & \cdots & \tau_{2,n} & \tau_{1, n} \\
  \tau_{1, 3}^{\dagger} & \tau_{2, 3}^{\dagger} & \cdots & \tau_{2, n-1} & \tau_{1, n-1} \\
  \vdots  & \vdots  & \ddots & \vdots & \vdots \\
 \end{pmatrix}
\end{equation}

We can collect all the unique $\tau$-s in row vectors $T_{m} = \{\tau_{m, m}, \tau_{m, m+1} \dots \tau_{m, n + 2 - m} \}$, where $m$ goes to $\mathrm{max}_m = (n+1)/2$ if $n$ is even and to $\mathrm{max}_m = (n+2)/2$ otherwise. We can compute those recursively, but notice that to compute $T_{m+1}$ only $T_{m}$, $\alpha_n$ and $\hat{W}_n$ are needed; thus, if the operations with the matrix $R_{p, n+1}^{-1}$ which need to be carried out can be split into separate operations over the $T_m$-s, one can optimize the usage of computational memory by only keeping two relatively small arrays instead of the full matrix. The rows themselves are computed by
 
\begin{align}
 T_{1} &= \{\pi_p (\alpha_n^{-1})^T \pi_p, \, \pi_p (\alpha_n^{-1})^T \hat{W}_n \}, \\ \nonumber
 (T_{m+1})_q &\equiv \tau_{m+1, m + q} = (T_{m})_{q} + \omega_{m}^* \pi_p (\alpha_n^{-1})^T \omega_{m + q - 1} - \pi_p \omega_{n -m + 1} \alpha_n^{-1} \omega_{n -m  - q + 2}^*, \\ \nonumber
 &\quad 1 \leq q \leq n + 1 - 2m, \quad 1 \leq m \leq \mathrm{max}_m
\end{align}

For the PWE and GME computations, the main complexity still lies in diagonalizing an $N \times N$ matrix; in the Bloch-mode expansion method, however, using the above algorithm can improve the efficiency greatly, and allows for much bigger structures to be treated within reasonable computational time.

\newpage

\section{Simulating disorder with Bloch-mode expansion}
\label{secbme}

The Bloch-mode expansion (BME) \cite{savona} outlined in this section is an efficient way to treat disorder in PHCs, especially in structures which are too big for a supercell GME computation. Basically, the procedure involves computing the Bloch modes of a structure by any means (including numerical solution of the Maxwell equations, but here we always start from the GME-computed Bloch modes as they are conveniently obtained), and expanding the modes of a slightly perturbed system on the basis of those Bloch modes. In such a way, for example disorder in a waveguide can be treated very efficiently, since computing the Bloch modes of a waveguide as the one in fig. \ref{guidedband} is relatively easy and requires only a small supercell, but then those modes can be used for a BME-computation of a waveguide containing many repetitions of the unit supercell. Furthermore, when all the Bloch modes are included, the computation becomes formally equivalent to a GME computation of the full structure; however, there is the added advantage that in many cases only some of the Bloch modes (e.g. the ones close in energy to the mode of interest) can be considered, and that could greatly diminish the computational time. Lastly, considering which of the Bloch modes overlap mostly with the mode of the disordered structure can provide insight into the physics of the system, and how it can be tuned for certain properties to be controlled. 

\subsection{Theoretical basis of the BME}

\label{bmeth}

To do a Bloch-mode expansion, we start from the magnetic field Bloch modes $\mathbf{H}_{\mathbf{k} n}(\mathbf{r})$ of the regular structure, which  are solutions to the equation

\begin{equation}
 \bm{\nabla} \times \left[ \frac{1}{\varepsilon(\mathbf{r})} \bm{\nabla} \times \mathbf{H}_{\mathbf{k} n}(\mathbf{r})  \right] = \frac{\omega_{\mathbf{k} n}^2}{c^2} \mathbf{H}_{\mathbf{k} n}(\mathbf{r}), 
\label{maxwellb}
\end{equation}

where $\varepsilon(\mathbf{r})$ is the dielectric constant within the regular structure. The modes are normalized according to

\begin{equation}
 \int \mathrm{d}\mathbf{r} \mathbf{H}^*_{\mathbf{k'} n'}(\mathbf{r}) \mathbf{H}_{\mathbf{k} n}(\mathbf{r}) = \delta_{\mathbf{k}'\mathbf{k}} \delta_{n' n}.
\label{modenorm}
\end{equation}

In presence of disorder, the dielectric profile $\varepsilon'(\mathbf{r})$ no longer has the lattice symmetries, thus the momentum $\mathbf{k}$ will no longer be conserved, so we label the eigenmodes of the system by a global index $\beta$. Those eigenmodes are solutions to the equation (coming from Maxwell eqs.)

\begin{equation}
 \bm{\nabla} \times \left[ \frac{1}{\varepsilon'(\mathbf{r})} \bm{\nabla} \times \mathbf{H}_{\beta}(\mathbf{r})  \right] - \frac{\omega_{\beta}^2}{c^2} \mathbf{H}_{\beta}(\mathbf{r}) = 0,
 \label{maxwell}
\end{equation}

and can be expanded on the basis of the Bloch modes of the regular structure:

\begin{equation}
 \mathbf{H}_{\beta}(\mathbf{r}) = \sum_{\mathbf{k} n} U_{\beta}(\mathbf{k}, n)\mathbf{H}_{\mathbf{k} n}(\mathbf{r}).
\label{bme}
\end{equation}

If we insert eq. (\ref{bme}) into eq. (\ref{maxwell}),  we obtain

\begin{equation}
 \sum_{\mathbf{k} n} U_{\beta}(\mathbf{k}, n)\left[\bm{\nabla} \times \frac{1}{\varepsilon'(\mathbf{r})} \bm{\nabla} \times \mathbf{H}_{\mathbf{k} n}(\mathbf{r}) -  \frac{\omega_{\beta}^2}{c^2} \mathbf{H}_{\mathbf{k} n}(\mathbf{r}) \right] = 0.
\end{equation}

Using eq. (\ref{maxwellb}) we can rewrite this as 

\begin{equation}
 \sum_{\mathbf{k} n} U_{\beta}(\mathbf{k}, n) \left[\frac{\omega^2_{\mathbf{k}, n} - \omega^2_{\beta}}{c^2} + \bm{\nabla} \times \eta(\mathbf{r}) \bm{\nabla} \times \right] \mathbf{H}_{\mathbf{k} n}(\mathbf{r}) = 0,
\label{geneig1}
\end{equation}

where we defined $\eta(\mathbf{r}) = \frac{1}{\varepsilon'(\mathbf{r})} - \frac{1}{\varepsilon(\mathbf{r})}$, i.e. the difference of the inverse of the regular and irregular dielectric profiles. If we now multiply eq. (\ref{geneig1}) by $\mathbf{H}^*_{\mathbf{k'} n'}(\mathbf{r})$, integrate over $\mathbf{r}$, and use the normalization in eq. (\ref{modenorm}), we obtain the eigenvalue problem

\begin{equation}
 \sum_{\mathbf{k}' n'} \left[ \frac{\omega^2_{\mathbf{k}n} - \omega^2_{\beta}}{c^2}\delta_{\mathbf{k}\mathbf{k}'} \delta_{n n'} + V_{\mathbf{k} n \mathbf{k}' n'}\right] U_{\beta}(\mathbf{k}', n') = 0, 
\label{eigprob}
\end{equation}

where the matrix $V_{\mathbf{k} n \mathbf{k}' n'}$ carries the information about disorder and is defined as

\begin{equation}
 V_{\mathbf{k} n \mathbf{k}' n'} = \int \mathrm{d} \mathbf{r} \mathbf{H}^*_{\mathbf{k'} n'}(\mathbf{r})\bm{\nabla} \times \left(\eta(\mathbf{r}) \bm{\nabla} \times \mathbf{H}_{\mathbf{k} n}(\mathbf{r}) \right)
\end{equation}

The last integral is over the volume of the considered system. We can modify it using some vector calculus identities and considering periodic boundary conditions, such that surface integral terms vanish:

\begin{align}
\nonumber
 V_{\mathbf{k} n \mathbf{k}' n'} &= \int \mathrm{d} \mathbf{r} \left[ (\eta(\mathbf{r}) \bm{\nabla} \times \mathbf{H}_{\mathbf{k} n}(\mathbf{r})) ( \bm{\nabla} \times \mathbf{H}_{\mathbf{k'} n'}^* (\mathbf{r})) - \bm{\nabla} \cdot (\mathbf{H}_{\mathbf{k'} n'}^* (\mathbf{r})\times (\eta(\mathbf{r}) \bm{\nabla} \times \mathbf{H}_{\mathbf{k} n}(\mathbf{r}))) \right] \\ \nonumber
 &= \int \mathrm{d} \mathbf{r} \eta(\mathbf{r}) (\bm{\nabla} \times \mathbf{H}_{\mathbf{k} n}(\mathbf{r})) (\bm{\nabla} \times \mathbf{H}_{\mathbf{k'} n'}^*(\mathbf{r})) - \iint_{\partial V} \mathrm{d} \mathbf{s} \eta(\mathbf{r}) \mathbf{H}_{\mathbf{k}n}^*(\mathbf{r}) \times ( \bm{\nabla} \times \mathbf{H}_{\mathbf{k}n}(\mathbf{r})) \\
&= \int \mathrm{d} \mathbf{r} \eta(\mathbf{r}) (\bm{\nabla} \times \mathbf{H}_{\mathbf{k} n}(\mathbf{r})) (\bm{\nabla} \times \mathbf{H}_{\mathbf{k'} n'}^*(\mathbf{r})) .
\end{align}

When $\eta(\mathbf{r})$ is non-zero only within the slab (which is always the case if the claddings are just some uniform material), the above integral only needs to be evaluated there. As mentioned earlier, we will use GME to compute the Bloch modes, expanding the latter as in eq. (\ref{guided}). Thus the matrix elements, written using the guided modes $\mathbf{H}_{\mu}(\mathbf{r})$, read

\begin{align}
 V_{\mathbf{k} n \mathbf{k}' n'} &= \sum_{\mu, \mu'} c_n(\mu) c_{n'}^*(\mu') \int \mathrm{d} \mathbf{r} \eta(\mathbf{r}) (\bm{\nabla} \times \mathbf{H}_{\mu}(\mathbf{r})) (\bm{\nabla} \times \mathbf{H}_{\mu'}^*(\mathbf{r})) \\ \nonumber
 &= \sum_{\mu, \mu'} c_n(\mu) c_{n'}^*(\mu') \int \mathrm{d} \mathbf{r} \eta(\mathbf{r}) \bar{\varepsilon}_2^2 \frac{\omega_{\mu} \omega_{\mu'}}{c^2} \mathbf{E}_{\mu}(\mathbf{r})  \mathbf{E}_{\mu'}^*(\mathbf{r}), 
\end{align}

where in the last equality we used the relation $\mathbf{E}(\mathbf{r}) = \frac{ic}{\omega \varepsilon(\mathbf{r})} \bm{\nabla} \times \mathbf{H}(\mathbf{r})$ and the fact that for the guided modes $\varepsilon(\mathbf{r})$ is simply the average slab permittivity $\bar{\varepsilon}_2$. In this case summation over $\mu$ stands for summation over all the included reciprocal lattice vectors $\mathbf{G}$ and guided modes $\alpha$, for fixed $\mathbf{k}$ and $\mathbf{k}'$. With the expression for the electric field of a guided mode given in appendix A of \cite{andreani_2006}, and with the definitions of the coefficients $A_{2\mu}$ and $I_{2\pm}$ given therein, we finally write (for a structure with complete symmetry w.r.t. reflection through the $z=0$ plane, s.t. $A_{2\mu} = A_{2\mu}^* = B_{2\mu}$)

\begin{align}
 V_{\mathbf{k} n \mathbf{k}' n'} &= \sum_{\mu, \mu'} c_n(\mu) c_{n'}^*(\mu') \eta(\mathbf{g} - \mathbf{g}') M_{\mu \mu'} \\ \nonumber
 M_{\mu \mu'} &= \bar{\varepsilon}_2^2 \frac{\omega_{\mu}^2}{c^2} \frac{\omega_{\mu'}^2}{c^2} \hat{\epsilon_{\mathbf{g}}} \cdot \hat{\epsilon_{\mathbf{g}'}} 2 A_{2\mu'}^* A_{2\mu} (I_{2+} + I_{2-}).
\label{bmemain}
\end{align}

The Fourier transform $\eta(\mathbf{g} - \mathbf{g}') = \varepsilon'^{-1}(\mathbf{g} - \mathbf{g}') - \varepsilon^{-1}(\mathbf{g} - \mathbf{g}')$ yields much better results when each of the two components is computed with the Ho method, and it is here that the algorithm for inverting a Toeplitz-block Toeplitz matrix defined in sec. \ref{invtoep} comes in very handy. For large structures, like the ones we would like to treat with BME, the matrices $\varepsilon'(\mathbf{g} - \mathbf{g}')$ and $\varepsilon(\mathbf{g} - \mathbf{g}')$ which need be inverted become very big, causing problems both from computational time and from computational memory perspective. The algorithm, however, solves both problems, first by reducing the complexity and second by removing the need to store the whole matrices at once.

Once the eigenvalue problem in eq. (\ref{eigprob}) is solved so that the expansion coefficients $U_{\beta}(\mathbf{k}', n')$ are known, the radiation losses of the disordered modes can also be computed in a perturbative approach analogous to the Fermi golden rule in Quantum Mechanics, namely by computing the overlap integrals between the modes $\mathbf{H}_{\beta}(\mathbf{r}) $ and the leaky modes of the effective slab, 

\begin{align}
 \mathcal{H}_{\beta, \mathbf{g}}^{\mathrm{rad}} &= \int \mathrm{d} \mathbf{r}  \frac{1}{\varepsilon'(\mathbf{r})} (\bm{\nabla} \times \mathbf{H}_{\beta}^*(\mathbf{r})) \cdot (\bm{\nabla} \times \mathbf{H}_{\mathbf{g}\lambda j}^{\mathrm{rad}}(\mathbf{r})) \\ \nonumber
 &= \sum_{\mathbf{k}' n'} \sum_{\mathbf{G}' \alpha} U_{\beta}^*(\mathbf{k}', n') c_{n'}^*(\mu') \int \mathrm{d} \mathbf{r}  \frac{1}{\varepsilon'(\mathbf{r})} \left(\bm{\nabla} \times  \mathbf{H}_{\mu'}^*(\mathbf{r})\right) \cdot (\bm{\nabla} \times \mathbf{H}_{\mathbf{g}\lambda j}^{\mathrm{rad}}(\mathbf{r})), 
\end{align}

where the leaky modes $\mathbf{H}_{\mathbf{g}\lambda j}^{\mathrm{rad}}(\mathbf{r})$ are given in appendix D of \cite{andreani_2006} and are labeled by their total in-plane momentum $\mathbf{g}$ and polarization $\lambda \in \{\mathrm{TE}, \mathrm{TM}\}$, and $j \in \{1, 3\}$ distinguishes between modes leaking in the two cladding layers. The amplitudes

\begin{equation}
 \mathcal{H}_{\mathbf{\mu}', \mathbf{g}}^{\mathrm{rad}} = \int \mathrm{d} \mathbf{r}  \frac{1}{\varepsilon'(\mathbf{r})} \left(\bm{\nabla} \times  \mathbf{H}_{\mu'}^*(\mathbf{r})\right) \cdot (\bm{\nabla} \times \mathbf{H}_{\mathbf{g}\lambda j}^{\mathrm{rad}}(\mathbf{r}))
\end{equation}

are similar to the ones computed in Appendix E of \cite{andreani_2006}, but in this case the disorder is manifest in the integration measure $\frac{1}{\varepsilon'(\mathbf{r})}$, which would in general break the translational invariance of the system within the slab - thus, the integrals are proportional to $\delta_{\mathbf{k} \mathbf{k}'}$ only in the claddings. Therefore, for example for a structure with identical upper and lower claddings, and for TE polarization of both the guided and the radiative modes, and with the notation of ref. \cite{andreani_2006}, the amplitude is given by two terms: the claddings contribution


\begin{equation}
 \mathcal{H}_{\mathbf{\mu}', \mathbf{g}}^{\mathrm{rad}, (1)} =  \delta_{\mathbf{k} \mathbf{k}'} \frac{\omega_{\mu'}^2}{c^2} \frac{\omega_r}{c}  \hat{\epsilon_{\mathbf{g}}} \cdot \hat{\epsilon_{\mathbf{g}'}} \bar{\varepsilon}_1^2  \varepsilon^{-1}_1 (\mathbf{G}-\mathbf{G}')  B_{1\mu'}^* \left(W_{1r}I_{1+} + X_{1r}I_{1-}+W_{3r}I_{1-}+X_{3r}I_{1+} \right) ,
\end{equation}

and the slab contribution,

\begin{equation}
 \mathcal{H}_{\mathbf{\mu}', \mathbf{g}}^{\mathrm{rad}, (2)} = \frac{\omega_{\mu'}^2}{c^2} \frac{\omega_r}{c}  \hat{\epsilon_{\mathbf{g}}} \cdot \hat{\epsilon_{\mathbf{g}'}} \bar{\varepsilon}_2^2  \varepsilon'^{-1}_2 (\mathbf{g}-\mathbf{g}')  A_{2\mu'}\left[\left(W_{2r} + X_{2r}\right)I_{2-} + \left(X_{2r} +  W_{2r}\right)I_{2+}\right],
\end{equation}

where now $\varepsilon^{-1}_1 (\mathbf{G}-\mathbf{G}')$ is the Fourier transform of the inverse of the dielectric profile within the cladding and $ \varepsilon'^{-1}_2 (\mathbf{g}-\mathbf{g}')$ - within the slab. For a more general structure and polarization pattern, the amplitudes can be obtained analogously, following ref. \cite{andreani_2006}. The loss rate $\gamma_{\beta}$ of a mode is then characterized by an imaginary part of the eigenenergy $\omega_{\beta}$. This can be found by computing the imaginary part of $\frac{\omega_{\beta}^2}{c^2}$, given by 

\begin{equation}
 \Gamma_{\beta} = -\Im\left(\frac{\omega_{\beta}^2}{c^2}\right) = \pi \sum_{\mathbf{k}, \mathbf{G}} \sum_{\lambda, j} | \mathcal{H}_{\beta, \mathbf{g}}^{\mathrm{rad}} |^2 \rho_j \left( \mathbf{g}; \frac{\omega_{\beta}^2}{c^2} \right), 
\end{equation}

and approximating $\gamma_{\beta} \approx c^2 \Gamma_{\beta}/(2 \omega_{\beta})$. Just as in eq. (\ref{imom}), $\rho_j \left( \mathbf{g}; \frac{\omega_{\beta}^2}{c^2} \right)$ is the 1D photonic density of states of radiative modes, at fixed wave-vector $\mathbf{g}$.

\subsection{Application to gentle confinement cavities}

\label{seckur}

\begin{figure}[h]
\begin{center}
\begin{minipage}{0.4\textwidth}
\begin{flushleft}
 \subfloat[]{\includegraphics[width = \textwidth]{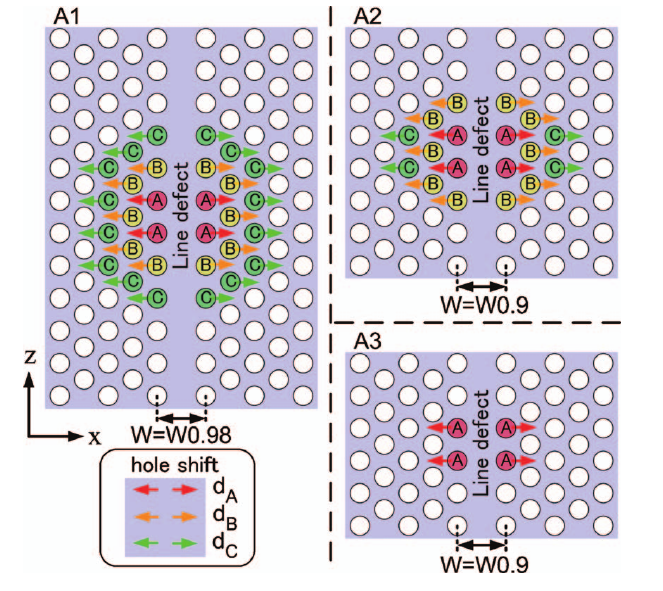}}
\end{flushleft}
\end{minipage}
\begin{minipage}{0.59\textwidth}
\begin{flushright}
 \subfloat[]{\includegraphics[trim = 1in 0in 1in 0in, clip = true, type=pdf,ext=.pdf,read=.pdf, width = \textwidth]{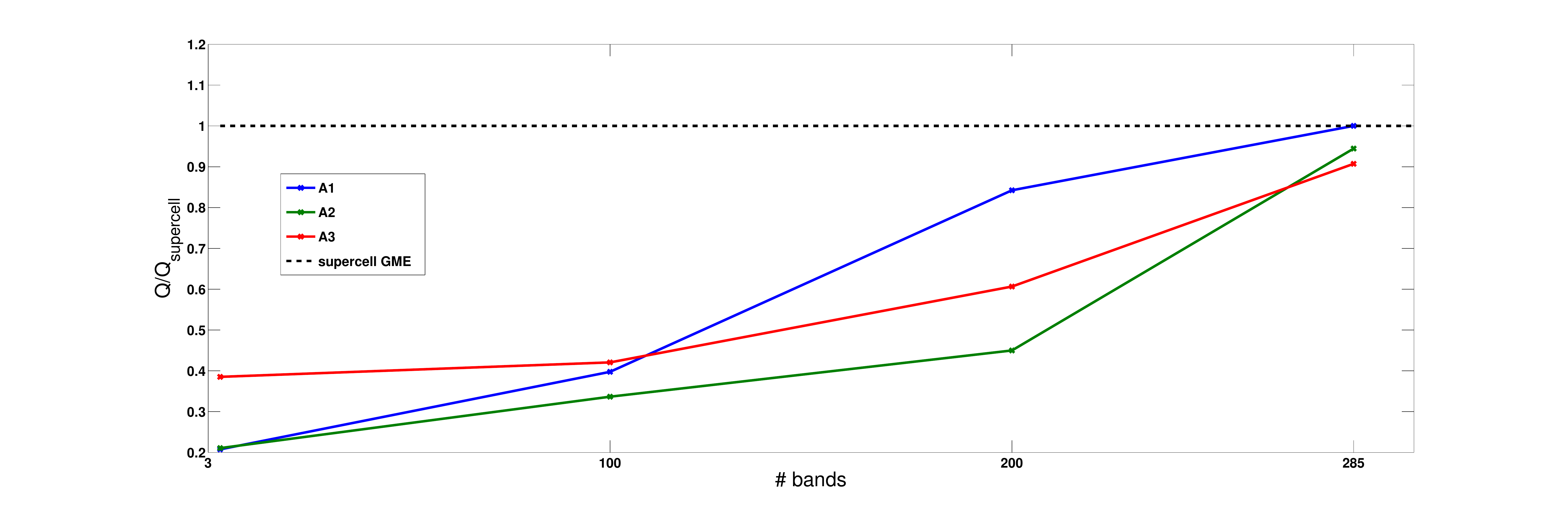}} \\
 \subfloat[]{\includegraphics[trim = 1in 0in 1in 0.5in, clip = true,type=pdf,ext=.pdf,read=.pdf, width = \textwidth]{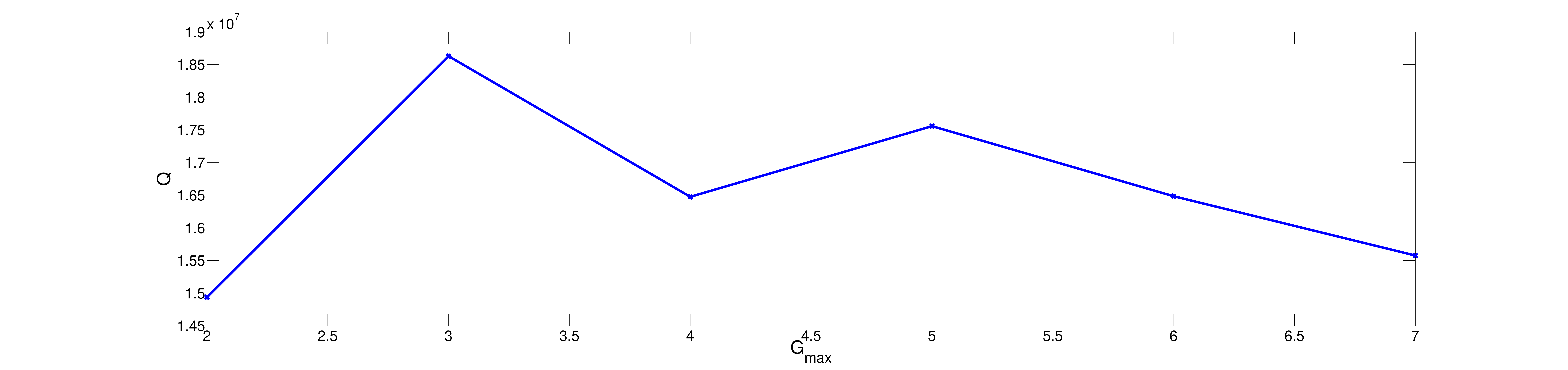}}
\end{flushright}
\end{minipage}
 \caption{(a): the A1, A2 and A3 cavities as defined in \cite{kuramochi}. (b): convergence w.r.t. number of bands of the BME-computed quality factor, normalized to the supercell GME-computed one. (c): dependence on $G_{max}$ of the BME-computed quality factor of the A1 cavity.}
\label{kurcav}
\end{center}
\end{figure}

As a first test of the BME method, we apply it to computing the resonant modes of three cavities called ``A1'', ``A2'' and ``A3'' by Kuramochi et. al. \cite{kuramochi}. The cavities are based on a waveguide structure, in which around a certain position, the regular-structure holes are slightly displaced. The resulting disorder serves to localize some of the modes of the waveguide around the disorder position, creating cavities with very high theoretical quality factors. The cavities are illustrated in fig. \ref{kurcav} (a), where besides the standard waveguide parameters we have considered so far, there are some additional ones. The width of the line defect in units of $\sqrt{3}a$ is labeled as $W0.98$ for the A1 cavity and $W0.9$ for the A2 and A3 cavities (so essentially, the guide in fig. \ref{guidedband} is a W1 waveguide), while the displacement of the holes is labeled by three parameters, $d_{A}$, $d_{B}$ and $d_{C}$ which are however related to a single parameter $x$ s.t. $d_{A} = x$, $d_{B} = 2x/3$, $d_{C} = x/3$.

\begin{figure}[h]
\begin{center}
 \subfloat[]{\includegraphics[trim = 1in 0in 1in 0in, clip = true, type=pdf,ext=.pdf,read=.pdf, width = 0.49\textwidth]{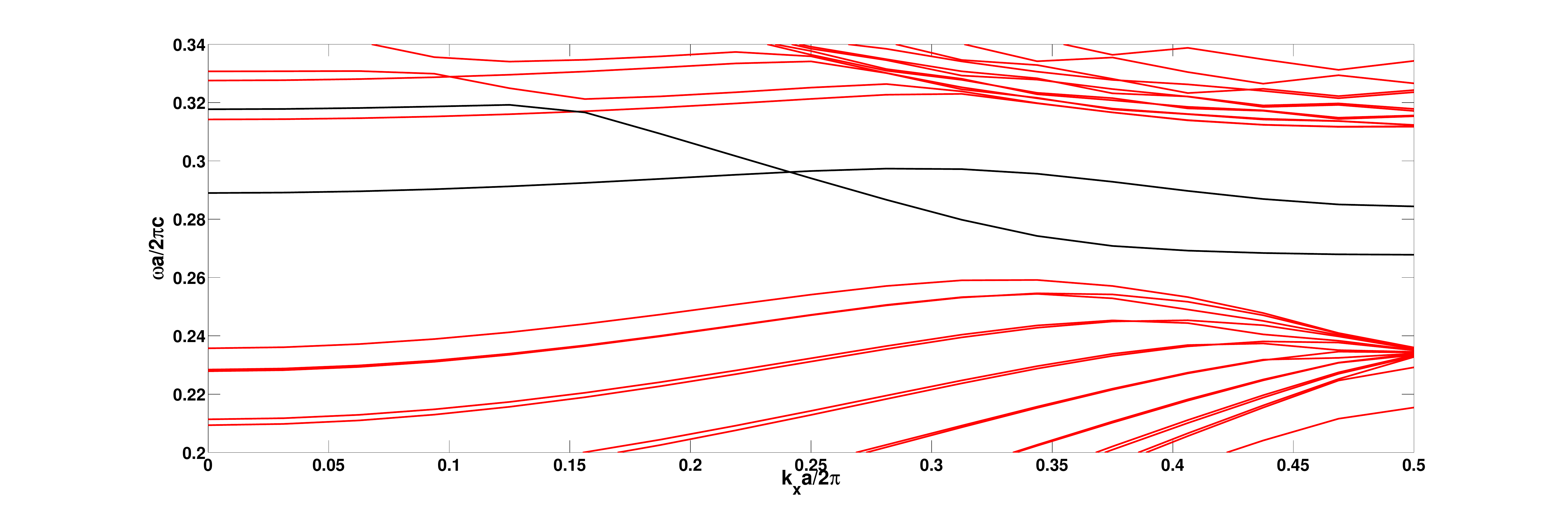}} 
 \subfloat[]{\includegraphics[trim = 1in 0in 1in 0in, clip = true, type=pdf,ext=.pdf,read=.pdf, width = 0.49\textwidth]{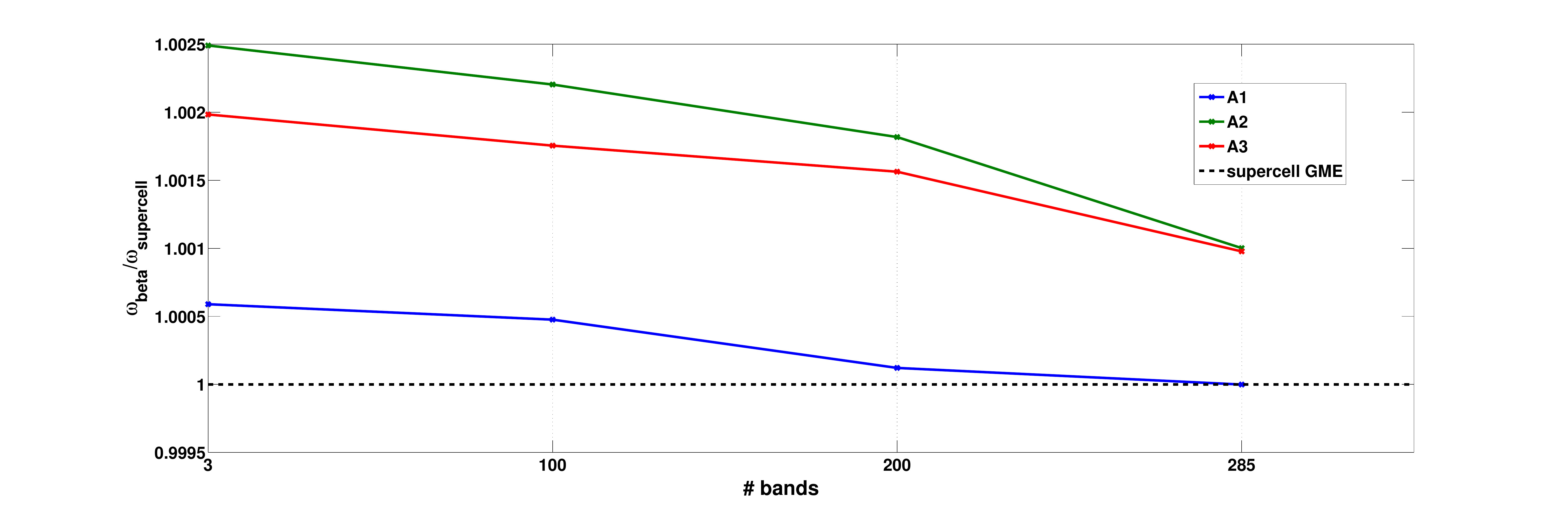}} 
 \caption{(a): band structure of the waveguide on which the A1 cavity is based, the two guided bands are colored in black. (b): BME-computed energies of the main resonant modes of the three cavities normalized to the GME values for different number of bands included in the computation.}
 \label{bands}
\end{center}
\end{figure}

The way we treated those structures with the BME is as follows: first, we computed the band structure of a regular waveguide using the GME method and a supercell of dimension $\left[a, 15.96\frac{\sqrt{3}}{2}a\right]$ for the A1 case and $\left[a, 15.8\frac{\sqrt{3}}{2}a\right]$ for the A2 and A3 ones. Then, using the so-computed Bloch modes, the procedure outlined in sec. \ref{bmeth} was applied to a waveguide composed of 16 of the above-mentioned unit cells, with the corresponding cavity disorder introduced in their middle. The parameters chosen were as in \cite{kuramochi}, namely: for A1, $x = 0.0214a$ (corresponding to the maximum Q found in \cite{kuramochi}), $r = 0.257a$, $d = 0.486a$, and $\varepsilon_2 = 11.972$, and for A2 and A3, $x = 0.0278a$, $r = 0.25a$, $d = 0.472a$, and $\varepsilon_2 = 11.972$. For such parameters, the band structure is quite similar to the one for the guide shown in fig. \ref{guidedband}, with the exception that the spatially-even guided band is not fully lying in the band gap (fig. \ref{bands}, (a)).

Apart from that, we restrict most of the computations to $G_{max} = 2 (\frac{2\pi}{a})$, so that a GME computation with the full supercell of size $\left[16a, 15.96\frac{\sqrt{3}}{2}a\right]$ is easily carried out, for the purposes on comparison. In fact, when all the Bloch modes of the regular structure are included in the BME, the approach is formally equivalent to using GME; however, the application of the inverse rule separately to $\varepsilon'(\mathbf{g} - \mathbf{g}')$ and $\varepsilon(\mathbf{g} - \mathbf{g}')$ to compute $\eta(\mathbf{g} - \mathbf{g}')$ in eq. (\ref{bmemain}) breaks the formal equivalence in the finite-dimensional matrix situation (though, results are still expected to be close to each other).

A natural thing to do to make a BME computation efficient is to consider only Bloch modes with energies close to the energy of the mode in investigation. In other words, as the waveguide has two guided bands in a band-gap, and the cavity mode itself will be in the band-gap (the main resonant mode is the spatially-even one with lowest energy), we should include the Bloch modes of those two guided bands, and perhaps also add the modes of the top valence and the bottom conduction bands. In the case of the cavities considered here, the disorder is symmetric w.r.t. the plane orthogonal to the slab passing through the center of the guide. Thus, the modes will still have the spatially-even and spatially-odd symmetry, and if we are interested in the main resonance mode, which is spatially-even, we do not need to include the spatially-odd guided band (or any other spatially-odd bands) as it will be fully decoupled. In the most basic computation, then, three bands were included - the spatially-even guided band, the top conduction band, and the lowest spatially-even valence band (which turns out to not be the lowest valence band). The quality factors of the cavities obtained with this computation, estimated through the loss rates $\gamma_{\beta}$, were not very satisfactory when compared to a full-supercell GME computation or the FDTD computation done in \cite{kuramochi}. The respective values can be seen in columns 2, 4 and 5 of table \ref{qcomparison}. 

The fact that the GME values correspond closely to the FDTD results shows that indeed the two methods, based on a very different foundation, agree nicely in their predictions. The fact that the BME computation was so far away from the GME one could only imply that more bands need be taken into account. Indeed, when all the 285 bands that are computed for $G_{max} = 2 (\frac{2\pi}{a})$ were taken into account, the results (shown in column 3 of table \ref{qcomparison}) came close to the GME ones, as expected from the mathematical equivalence of the diagonalization problems in eq. (\ref{guidedeig}) and eq. (\ref{bmemain}) in this case. 

\begin{table}[h]
\begin{center}
{
\renewcommand{\arraystretch}{1.2}
\begin{tabular*}{\textwidth}{@{\extracolsep{\fill}}|c|c c c c|}
\hline
  & 3-bands BME & 285-bands BME & supercell GME & FDTD from \cite{kuramochi} \\
\hline 
A1 & $1.5 \times 10^7$ & $7.2 \times 10^7$ & $7.2 \times 10^7$ & $7 \times 10^7$ \\ 
A2 & $1.0 \times 10^6$ & $4.5 \times 10^6$ & $4.8 \times 10^6$ & $5 \times 10^6$ \\ 
A3 & $0.65\times 10^6$ & $1.5 \times 10^6$ & $1.7 \times10^6$ & $1\times 10^6$ \\ \hline
\end{tabular*}
}
 \end{center}
\caption{Quality factors of the A1, A2 and A3 cavities with different computations}
\label{qcomparison}
\end{table}

Thus, it turns out that the high quality factor of the resonant mode is not mainly due to the Bloch modes lying closest in energy, but is instead due to a complex interference of all the Bloch modes of the system, even the ones lying very far away in the band diagram. To substantiate this claim, in fig. \ref{kurcav} (b) we present the quality factors of the three cavities for 3, 100, 200, and 285 bands included in the computation. It can be seen that their increase takes place gradually, and is not due to any single band, but is rather a complex property of the full system. However, as can be seen from fig. \ref{bands} (b), the energies of the resonant modes are much better converged w.r.t. number of bands than the quality factors, which means that only the loss rates require many number of bands for an exact result. As long as energies are concerned, a 3-band computation is already adequate: the 3-band computed energies are correspondingly $99.94\%$, $99.85\%$ and $99.90\%$ of the 285-band computed ones.

Another parameter determining the convergence of the BME and the GME alike is the number of $\mathbf{G}$-vectors included, determined by $G_{max}$. Therefore, in fig. \ref{kurcav} (c) we plot the dependence of the 3-band BME quality factor on $G_{max}$ (notice that we can relatively easily use BME to go to values of $G_{max}$ for which it would be almost impossible and certainly not practical to use the GME). As can be seen, the dependence on $G_{max}$ is much weaker than the dependence on number of bands, and even the $G_{max} = 2 (\frac{2\pi}{a}$ result is already realiable, as illustrated also by the fact that in the all-bands case it agrees with the FDTD and GME computations.

\begin{figure}[h]
\begin{center}
 \subfloat[]{\includegraphics[trim = 1in 1in 1in 0in, clip = true, type=pdf,ext=.pdf,read=.pdf, width = 0.33\textwidth]{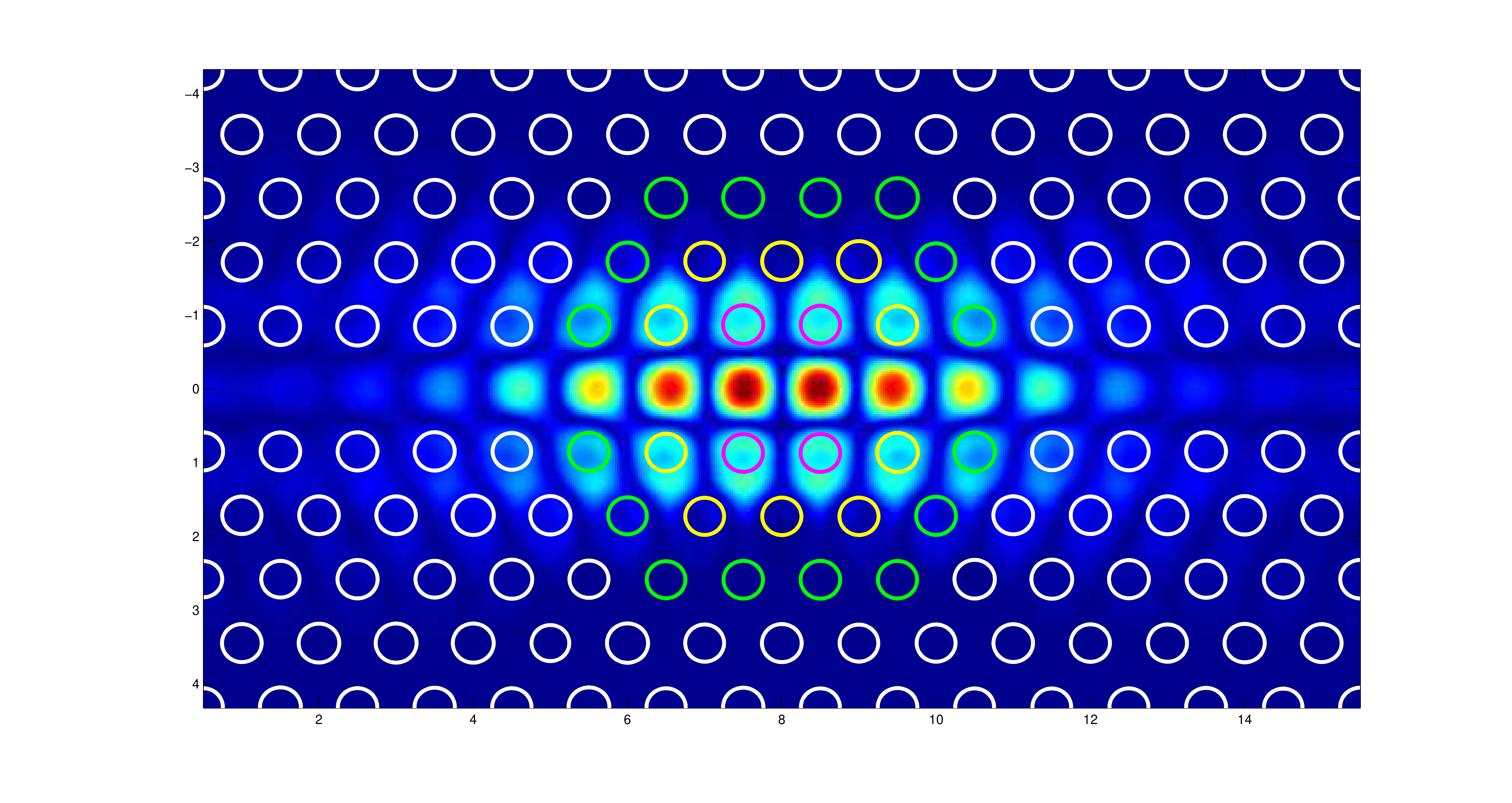}} 
 \subfloat[]{\includegraphics[trim = 1in 1in 1in 0in, clip = true, type=pdf,ext=.pdf,read=.pdf, width = 0.33\textwidth]{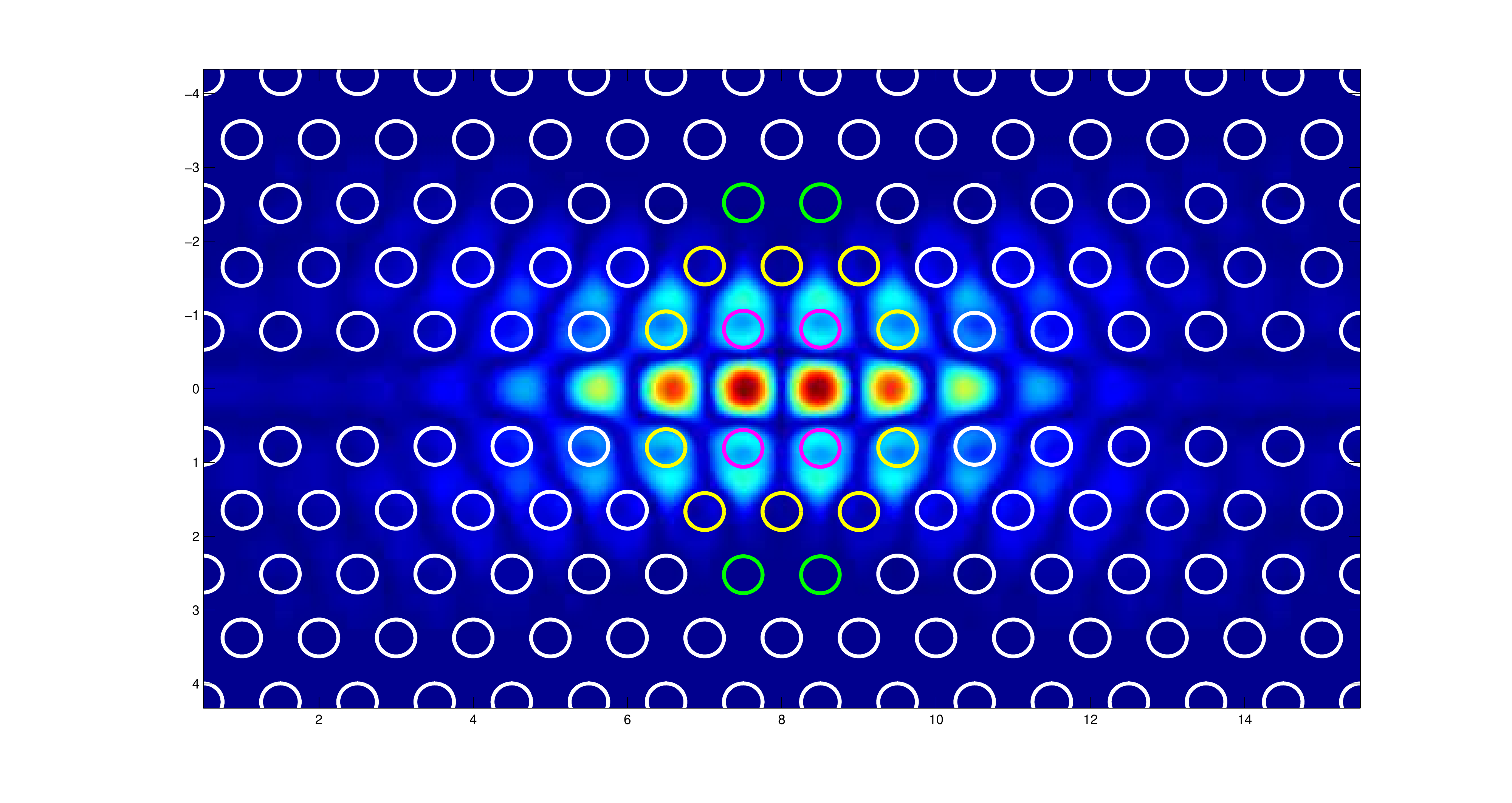}} 
 \subfloat[]{\includegraphics[trim = 1in 1in 1in 0in, clip = true, type=pdf,ext=.pdf,read=.pdf, width = 0.33\textwidth]{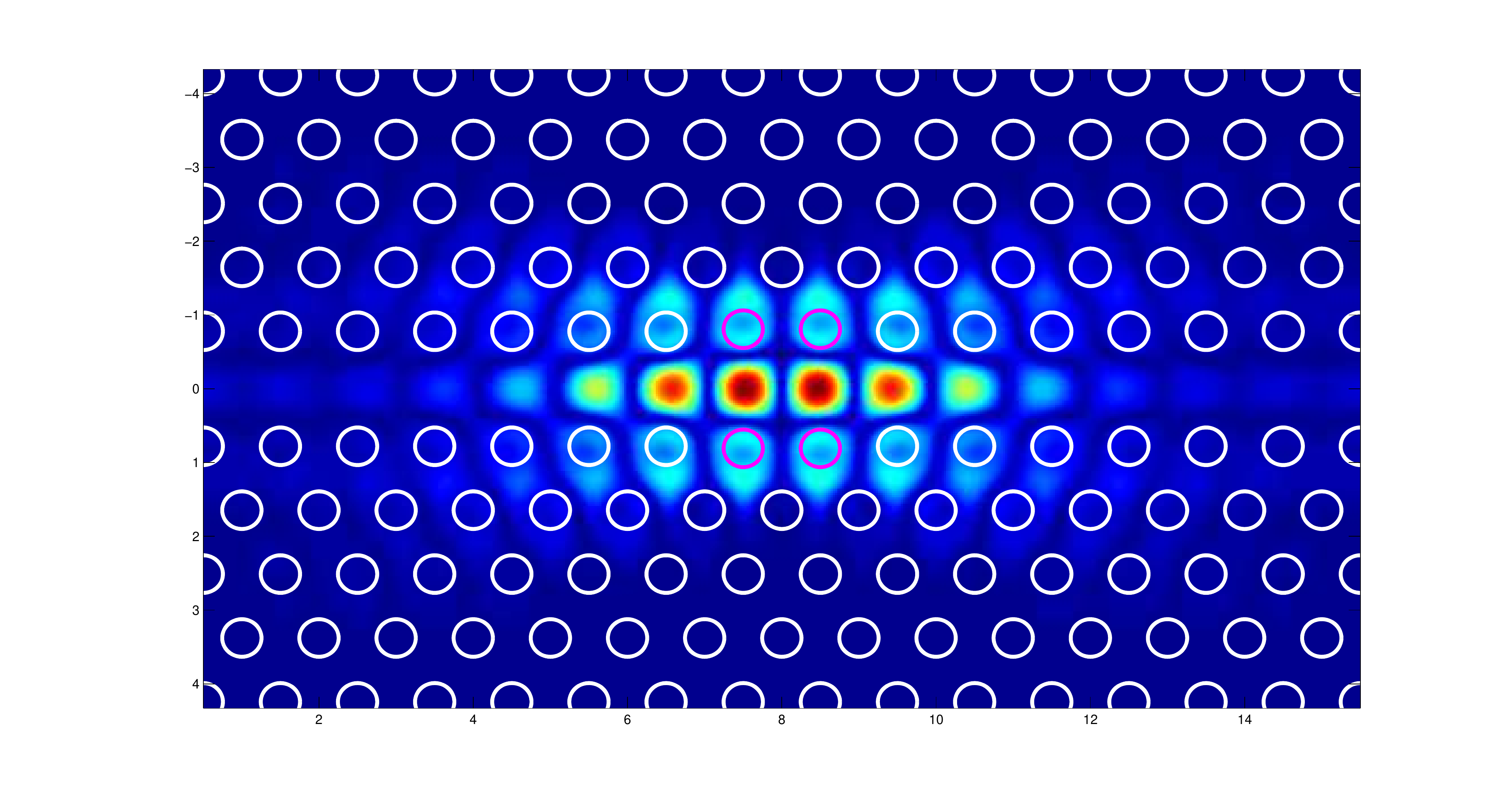}} 
 \caption{Magnetic field profiles of the main resonant mode in (a) A1 cavity, (b) A2 cavity, and (c) A3 cavity. The displaced holes are colored in the fashion of fig. \ref{kurcav} (a).}
 \label{kurcavprof}
\end{center}
\end{figure}

The magnetic field profiles of the main resonant mode in the three cavities are illustrated in fig. \ref{kurcavprof}, for 285 bands and $G_{max} = 2 (\frac{2\pi}{a})$. The qualitative appearance of the profiles, however, changes very little with the number of bands and with $G_{max}$, in the same way that, as we already mentioned, the energies change very little. One reason why so many bands are needed to get the correct quality factors in the case of such strong localization as is present in the considered cavities is that the Bloch modes are essentially extended functions. Thus, for application to cavities it might be wiser to expand over some more localized basis, such as the photonic Wannier functions \cite{busch}. However, the theoretical Q-s cited in table \ref{qcomparison} are never achieved in practice, due to the presence of random disorder. In fact, the experimentally measured quality factors cited in \cite{kuramochi} are 800 000 for A1, 390 000 for A2, and 240 000 for A3. Conveniently enough, in the presence of disorder the BME seems to work well even with 3 bands, and is furthermore a convenient tool to quantify the effects of said disorder.

To illustrate this last point, disorder was included in the A1 cavity computation, in the form of random radius fluctuations of all the wholes with mean $r_0 = 0.257a$ and standard deviation $\sigma = 0.0014a$. The details of computing $\eta(\mathbf{g} - \mathbf{g}')$ in this case can be inferred from sec. \ref{fourirr}. To capture better the effects of the small hole fluctuations, $G_{max} = 3 (\frac{2\pi}{a})$ was used, and the waveguide length was extended to 32 elementary cells. The delicate interference which produces the very high theoretical Q-value of the cavity is broken by disorder, and the quality factor is brought down to $1.07 \times 10^6$ in a computation with the 2 guided bands only (the second guided band was included since disorder breaks the symmetry and allows mixing). This is now quite weakly affected by adding more bands, with the value increasing very slightly to $1.09 \times 10^6$ in a computation with 196 bands (all of which are spatially even in the absence of disorder, i.e. expected to couple more to the resonance mode than the spatially-odd ones). Thus, it is essentially only necessary to choose a proper disorder model to obtain numerically a certain quality factor, including the experimentally measured value of 800 000. And since in the presence of disorder including very few bands is sufficient (a claim which is also examined in the subsequent section), different disorder models can be handled very efficiently with a BME computation.

\subsection{Application to hole-edge disorder in waveguides}

\label{secirr}

In this section, we apply the BME to the problem of quantifying the effects of hole disorder in a W1 waveguide. While there is some existing literature on the matter, e.g. \cite{andreani_2004}, \cite{andreani_2005} and \cite{hughes}, the approach presented here has several advantages. First, we model disorder as a general hole roughness given by an arbitrary deviation from a constant radius, $R(\phi) = R_0 + \delta R(\phi)$, which, to our knowledge, has so far never been done, while an SEM analysis of PHC structures given in \cite{begin} suggests that fine features of the holes are non-negligible. Second, with the help of BME we are able to compute very big structures, e.g. going to a waveguide of length 256 elementary cells, and to that we are able to average over many realizations in order to compute density of states. Finally, the method also allows us to compute not only the energies and radiative rates, but also the electromagnetic modes of the disordered guides, which had not been done prior to \cite{savona}. 

\subsubsection{Fourier transform of irregular holes}

\label{fourirr}

As mentioned before, to compute the matrix elements $V_{\mathbf{k}n,\mathbf{k}'n'}$ given in eq. (\ref{bmemain}), and specifically the quantity $\eta(\mathbf{g} - \mathbf{g}') = \varepsilon'^{-1}(\mathbf{g} - \mathbf{g}') - \varepsilon^{-1}(\mathbf{g} - \mathbf{g}')$ entering there, we need to compute $\varepsilon'(\mathbf{g} - \mathbf{g}')$ and $\varepsilon(\mathbf{g} - \mathbf{g}')$ as matrices and then use matrix inversion. The second of those is easily computed for a regular waveguide with perfectly circular holes, in the fashion of eq. (\ref{epsdirect}). Namely, let $\zeta$ be an index running over all holes included in the supercell, with $\bm{\rho}_{\zeta}$ denoting their positions and $R_{\zeta}$ - their radii (i.e. we allow for different radii of the holes, while still assuming they are perfectly spherical). Then we have

\begin{equation}
 \varepsilon(\mathbf{g}) = \varepsilon_2 \delta_{\mathbf{g}, 0} + \frac{\varepsilon_1 - \varepsilon_2}{A} \sum_{\zeta} \mathrm{e}^{i\mathbf{g}\bm{\rho}_{\zeta}} \frac{2\pi R_{\zeta}}{g}J_1(gR_{\zeta}),
\end{equation}

where again $J_1$ denotes Bessel function of the first kind and $g$ is the magnitude of $\mathbf{g}$.

The quantity $\varepsilon'(\mathbf{g})$, on the other hand, can be very tricky to compute, and essentially depends on the disorder model we wish to simulate. Here, we will first assume that the disordered dielectric profile is constant in the $z$-direction, i.e. $\varepsilon'(\mathbf{r}) = \varepsilon(\bm{\rho}, 0)$, where $\bm{\rho}$ labels the position in the x-y plane. Then, we will consider a slab with holes centered at the same positions $\bm{\rho}_{\zeta}$ ($\zeta = 1, 2, \dots$) as the holes of the regular guide, but of arbitrary shape given by a contour $R_{\zeta}(\phi)$ (in polar coordinates), i.e. we would like to treat disorder due to irregular hole-edge. In such a setup we have

\begin{equation}
 \varepsilon'(\bm{\rho}) = \varepsilon_2 + \sum_{\zeta}(\varepsilon_1 - \varepsilon_2) S_{\zeta}(\bm{\rho} - \bm{\rho}_{\zeta}),
\end{equation}

where $S_{\zeta}(\bm{\rho})$ is a function defined as 

\begin{equation}
S_{\zeta}(\rho, \phi) = \left\{
 \begin{array}{l l}
 1, & \quad \rho \leq R_{\zeta}(\phi) \\
 0, & \quad \rho > R_{\zeta}(\phi) \\
 \end{array} \right.
\end{equation}

To compute the Fourier transform $\varepsilon'(\mathbf{g})$, we start from the Fourier transform of a single function $S(\bm{\rho})$ as defined above, for some hole profile $R(\phi)$:

\begin{equation}
S(\mathbf{G}) = \int_{A} \mathrm{e}^{-i\mathbf{G}\bm{\rho}} \mathrm{d^2}\bm{\rho} = \int_{0}^{2\pi} \mathrm{d}\phi \int_{0}^{R(\phi)} \rho \mathrm{d} \rho \mathrm{e}^{-iG\rho \cos(\theta - \phi)},
\label{S(G)}
\end{equation}

where $\mathbf{G} = (G, \theta)$ in polar coordinates. Assume that the hole profile can be represented as deviating slightly from a circular shape, $R(\phi) = R_0 + \delta R(\phi)$. Then,

\begin{equation}
S(\mathbf{G}) = \int_{0}^{2\pi} \mathrm{d}\phi \left[ \int_{0}^{R_0} + \int_{R_0}^{R_0 + \delta R(\phi)}\right] \rho \mathrm{d} \rho \mathrm{e}^{-iG\rho \cos(\theta - \phi)} = S_0(\mathbf{G}) + \delta S(\mathbf{G}),
\label{S(G)1}
\end{equation}

where we split the result into a circular hole expression, $S_0(\mathbf{G})$ - which is well-known,  and a correction term, $\delta S(\mathbf{G})$. For the latter we have,

\begin{equation}
\delta S(\mathbf{G}) = \int_{0}^{2\pi} \mathrm{d}\phi \int _{R_0}^{R_0 + \delta R(\phi)} \rho \mathrm{d} \rho \mathrm{e}^{-iG\rho \cos(\theta - \phi)} = \int_{0}^{2\pi} \mathrm{d}\phi \int _{R_0}^{R_0 + \delta R(\phi)} \rho \mathrm{d} \rho \mathrm{e}^{iG\rho \sin\left(\theta - \phi + \frac{\pi}{2}\right)},
\label{dS(G)}
\end{equation}

and we make use of the following identity involving Bessel functions of the first kind:

\begin{equation}
 \mathrm{e}^{i \omega \sin \varphi} = \sum_{m = -\infty}^{\infty} J_m (\omega)\mathrm{e}^{im\varphi},
\end{equation}

to get

\begin{align}
\delta S(\mathbf{G}) &=  \int_{0}^{2\pi} \mathrm{d}\phi \int_{R_0}^{R_0 + \delta R(\phi)} \rho \mathrm{d} \rho \sum_{m = -\infty}^{\infty} J_m (G\rho)\mathrm{e}^{im\left(\theta - \phi + \frac{\pi}{2}\right)} \nonumber \\
&=  \int_{0}^{2\pi} \mathrm{d}\phi \sum_{m = -\infty}^{\infty} i^m \mathrm{e}^{im\left(\theta - \phi \right)} \int_{R_0}^{R_0 + \delta R(\phi)} \rho \mathrm{d} \rho J_m (G\rho)
\label{dS(G)1}.
\end{align}

For $\delta R$ small, we can expand the $\mathrm{d}\rho$ integrals to first order,

\begin{equation}
\int_{R_0}^{R_0 + \delta R(\phi)} \rho \mathrm{d} \rho J_m(G\rho) \approx \delta R(\phi) R_0 J_m(GR_0). 
\label{approx}
\end{equation}

Since $\delta R(\phi)$ is a $2\pi$ - periodic function, we can take its Fourier series expansion, 

\begin{equation}
 \delta R(\phi) = \sum_{n = -\infty}^{\infty} C_{n} e^{in\phi},
 \label{Rfour}
\end{equation}

with reality condition $C_{-n} = C_n^*$. Plugging (\ref{approx}) and (\ref{Rfour}) in (\ref{dS(G)1}) yields

\begin{align}
 \delta S(\mathbf{G}) &= R_0 \sum_{m, n = -\infty}^{\infty} i^m \mathrm{e}^{im\theta} J_m(GR_0) \int_{0}^{2\pi} \mathrm{d}\phi \mathrm{e}^{-im\phi} C_n \mathrm{e}^{in\phi}  \nonumber \\
 &= 2 \pi R_0 \sum_{m = -\infty}^{\infty} i^m \mathrm{e}^{im\theta} C_m J_m(GR_0) 
\end{align}

The Fourier expansion of the dielectric profile in presence of disorder can be directly expressed using this result, since

\begin{equation}
 \varepsilon'(\mathbf{g}) = \varepsilon_2 \delta_{\mathbf{g}, 0} + \frac{\varepsilon_1 - \varepsilon_2}{A} \sum_{\zeta} \mathrm{e}^{i\mathbf{g}\bm{\rho}_{\zeta}} S_{\zeta}(\mathbf{g}),
\end{equation}

where A is the area of the supercell, $\zeta$ runs over all the holes centered at $\bm{\rho}_{\zeta}$ present in the supercell, and $S_{\zeta}(\mathbf{g}) = S_{0, \zeta}(\mathbf{g}) + \delta S_{\zeta}(\mathbf{g})$ is the Fourier transform corresponding to a $R_{\zeta}(\phi) = R_{0, \zeta} + \delta R_{\zeta}(\phi)$ profile. In the computations done here, we further simplify this result by taking $R_{0, \zeta} = R_0$ - constant for all holes (both in the regular and in the disordered case), i.e. only the hole-edge disorder $\delta R(\phi)$ is different for every hole in the supercell. If we characterize this by a set of Fourier-expansion parameters $C_{m, \zeta}$ for hole number $\zeta$, the expression is finally

\begin{equation}
 \varepsilon'(\mathbf{g}) = \varepsilon_2 \delta_{\mathbf{g}, 0} + \frac{2 \pi R_0}{A}(\varepsilon_1 - \varepsilon_2) \sum_{\zeta}\mathrm{e}^{i\mathbf{g}\bm{\rho}_{\zeta}}\left[\frac{J_1(gR_0)}{g} + \sum_{m = -\infty}^{\infty} i^m \mathrm{e}^{im\theta} C_{m, \zeta} J_m(gR_0) \right].
\label{epsprime}
\end{equation}

\subsubsection{Disorder models}

To simulate the disorder, for each hole within the supercell, a set of coefficients $\{C_{m, \zeta}\}$ was pseudo-randomly generated, with several requirements determining the underlying distribution. First, we label the standard deviation in the radius fluctuation as $\sigma$, i.e. $\langle \delta R^2(\phi) \rangle = \sigma^2$, which, together with $\langle C_{m_1}^* C_{m_2} \rangle \propto \delta_{m_1m_2}$ imposes $\langle \sum_m |C_m|^2 \rangle = \sigma^2$. Next, an obvious choice is that the fluctuations increase or decrease the radius of the hole with equal probability, i.e. $\langle \delta R \rangle = 0$. This results in $\langle C_0 \rangle = 0$. For $m \neq 0$, $\langle C_m \rangle = 0$ also holds, since $C_m$ is a complex number with random phase. Finally, we characterize how finely detailed the disorder is by a correlation angle $\delta$. If we define exponential correlations, such that $\langle \delta R(\phi) \delta R(\phi')\rangle \propto e^{-\frac{|\phi - \phi'|}{\delta}}$,  the $C_m$ coefficients have to follow

\begin{equation}
 \sum_{m = -\infty}^{\infty}\langle |C_m|^2 \rangle e^{im\phi} = \sigma^2 e^{-\frac{|\phi|}{\delta}}.
\end{equation}

In other words, the averages $\langle |C_m|^2 \rangle$ are distributed like the Fourier series coefficients of $e^{-\frac{|\phi|}{\delta}}$, resulting in

\begin{equation}
 \langle |C_m|^2 \rangle = \sigma^2 \int_{-\pi}^{\pi} e^{-\frac{|\phi |}{\delta}} e^{-im\phi} \mathrm{d}\phi,
\end{equation}

which tells us that the coefficients ${C_m}$ have to be generated according to

\begin{align}
 \quad  \langle C_m \rangle = 0, \quad \langle |C_m|^2 \rangle = \frac{\sigma^2}{\pi} \frac{\delta}{1 + \delta^2 m^2}\left(1 - \mathrm{e}^{-\frac{\pi}{\delta}}(-1)^m\right).
\label{expcor}
\end{align}

Alternatively, we could require Gaussian correlations, such that $\langle \delta R(\phi) \delta R(\phi')\rangle \propto e^{-\frac{(\phi - \phi')^2}{2 \delta^2}}$, and follow the same procedure to find a different requirement for the average of the squares of the coefficients:

\begin{equation}
 \langle |C_m|^2 \rangle = \sigma^2 \int_{-\pi}^{\pi} e^{-\frac{\phi^2}{2 \delta^2}} e^{-im\phi} \mathrm{d}\phi,
\label{gausint}
\end{equation}

The result of this computation involves Dawson integrals, defined as $F(z) = e^{-z^2}\int_0^z e^{t^2} \mathrm{d}t$, and is:

\begin{align}
\langle |C_m|^2 \rangle = \sigma^2 (-1)^{m} \frac{\delta}{2 \pi} e^{- \frac{\pi^2}{2\delta^2}} \Im\left(F\left( \frac{m}{\sqrt{2}} \delta + i \frac{\pi}{\sqrt{2} \delta}\right) - F\left( \frac{m}{\sqrt{2}} \delta - i \frac{\pi}{\sqrt{2} \delta}\right)\right).
\label{gausscor}
\end{align}

For very small values of $\delta$, the Dawson function becomes hard to handle numerically, but the integrand in (\ref{gausint}) becomes extremely well localized around 0, i.e. if we compute the integral with boundaries $-\infty$ to $\infty$ instead of $-\pi$ to $\pi$, we would add only an exponentially small correction. The former boundary conditions are much easier to compute, and thus we can approximate

\begin{equation}
 \langle |C_m|^2 \rangle \approx \sigma^2 \frac{\delta}{\sqrt{2 \pi}} e^{-\frac{\delta^2 m^2}{2}}, \quad  \delta \mathrm{-small} 
\end{equation}

This approximation is already very good for values of $\delta \approx \frac{2 \pi}{10}$, but was used only for values $\delta < \frac{2 \pi}{50}$, where it is virtually an exact result.

\begin{figure}
\begin{center}
 \includegraphics[trim = 1in 1in 1in 1in, type=pdf,ext=.pdf,read=.pdf,width = 0.23\textwidth]{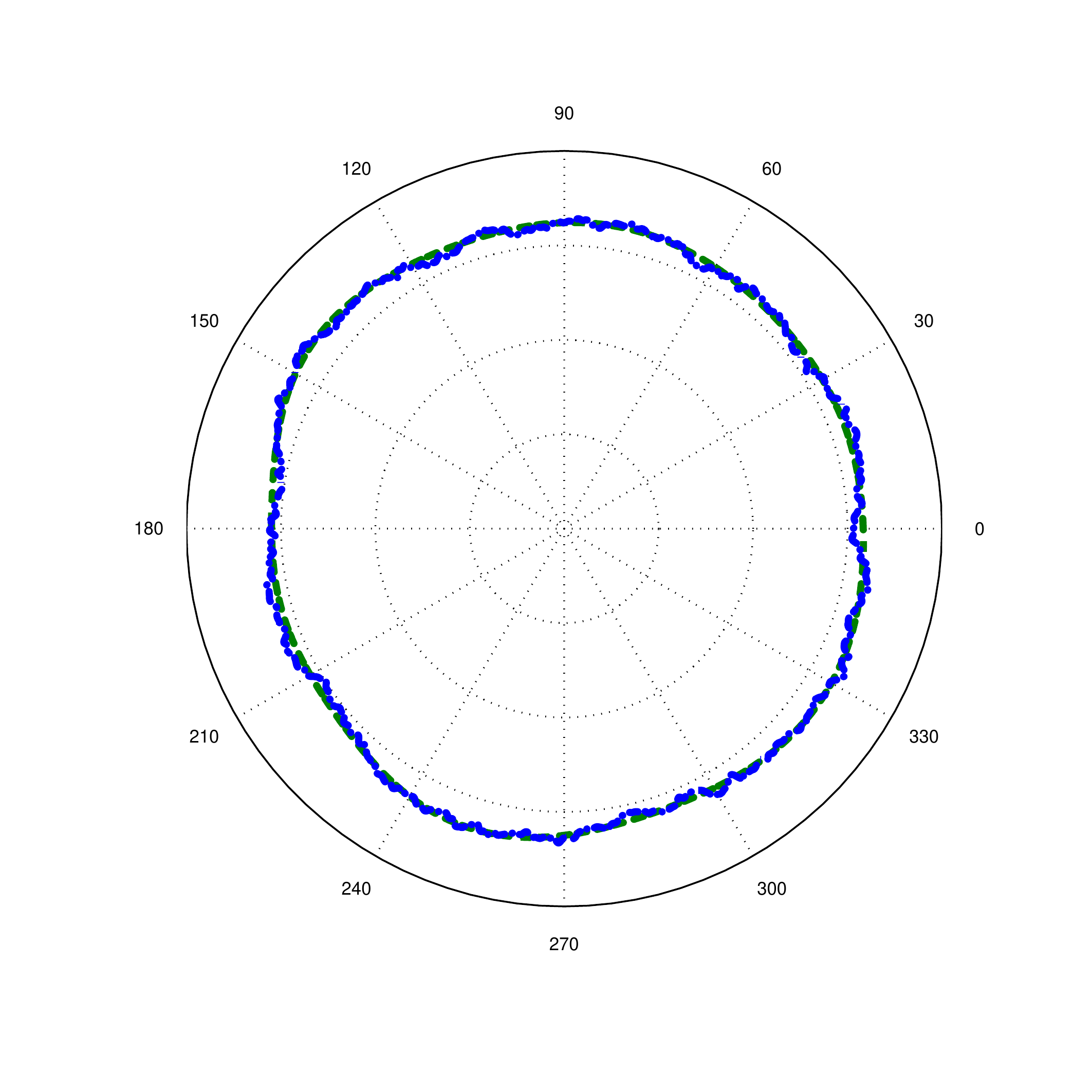}
 \includegraphics[trim = 2in 1in 2in 3in, type=pdf,ext=.pdf,read=.pdf,width = 0.76\textwidth]{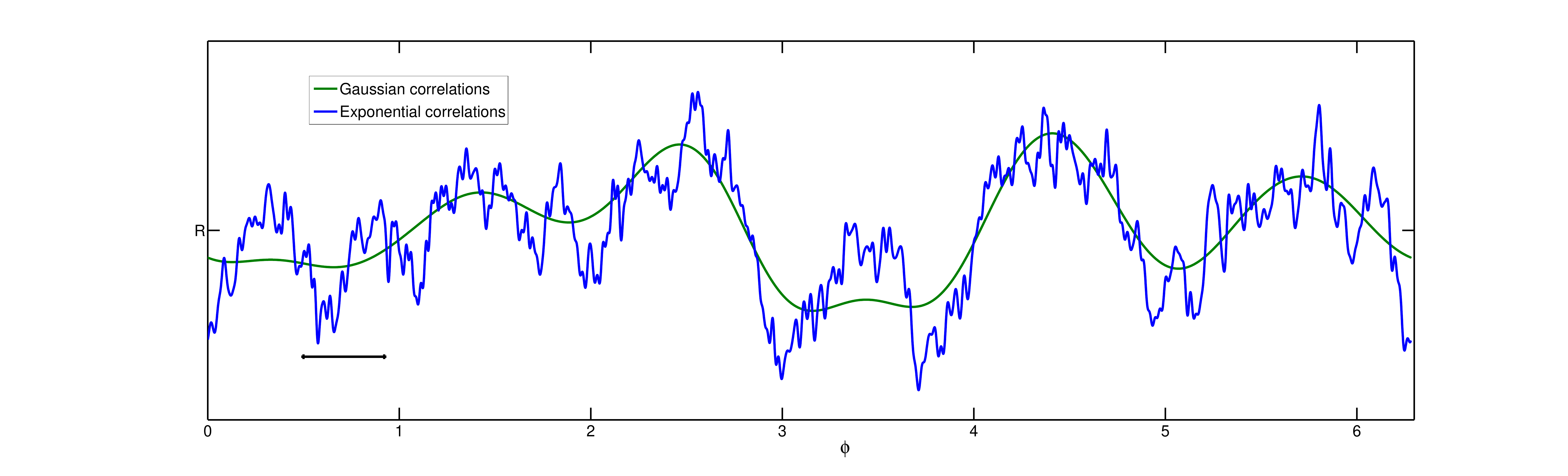}
 \caption{Hole profile in polar coordinates (left) and as a plot vs. $\phi$ (right) for the same random seed of $C_m$ coefficients, but when either Gaussian-correlation scaling (green) or exponential-correlation scaling (blue) was used. The black line shows the corresponding correlation length.}
 \label{exp_gauss}
 \end{center}
\end{figure}

To show the difference between the two disorder models - exponential vs. Gaussian correlations - in fig. \ref{exp_gauss} we plot one realization of the hole edge when either of the two correlations was used, in polar coordinates in (a), and the radius R as a function of the angle $\phi$ in (b). The black line in fig. \ref{exp_gauss} (b) shows the correlation length corresponding to the correlation angle $\delta = 0.06 (2\pi)$ which was used. The same random seed was used in both cases, which means that the same starting random numbers $C_m$ with zero mean were normalized either as in eq. (\ref{expcor}) for exponential correlations, or as in eq. (\ref{gausscor}) for Gaussian ones. As can be seen, the fluctuations of the radius with Gaussian correlations correspond much more closely to the correlation length given, while the exponential correlations contain fluctuations on every length scale. Furthermore, intuitively, Gaussian correlations seem more likely to arise in the lab because of the Central Limit Theorem. Thus, unless explicitly stated otherwise, everywhere below Gaussian correlations were used..

\subsubsection{Numerical results}

\label{holeirr}

\begin{figure}[h]
\begin{center}
 \includegraphics[trim = 1in 1in 1in 1in, type=pdf,ext=.pdf,read=.pdf,width = 0.25\textwidth]{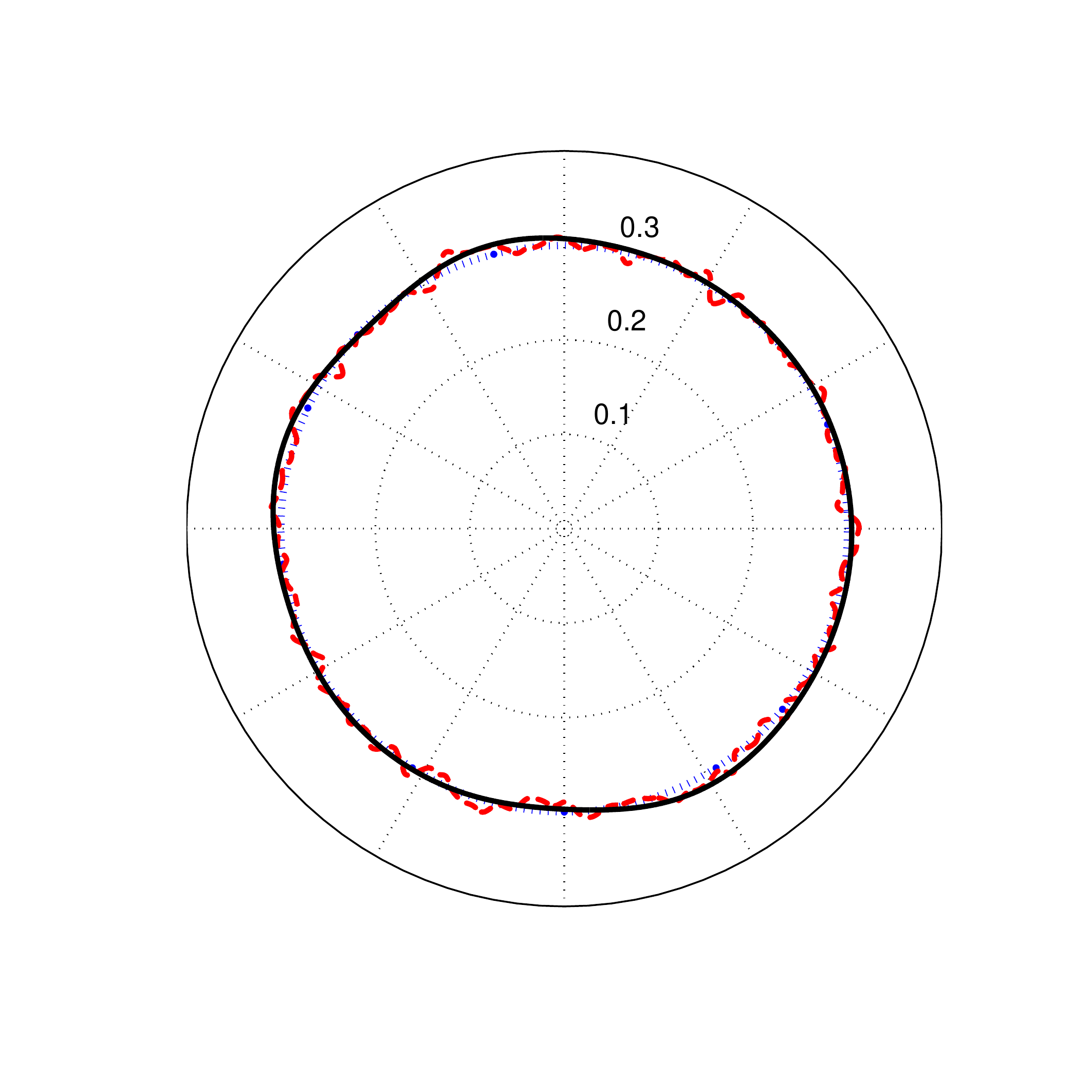}
 \includegraphics[trim = 2in 1in 2in 3in, type=pdf,ext=.pdf,read=.pdf,width = 0.74\textwidth]{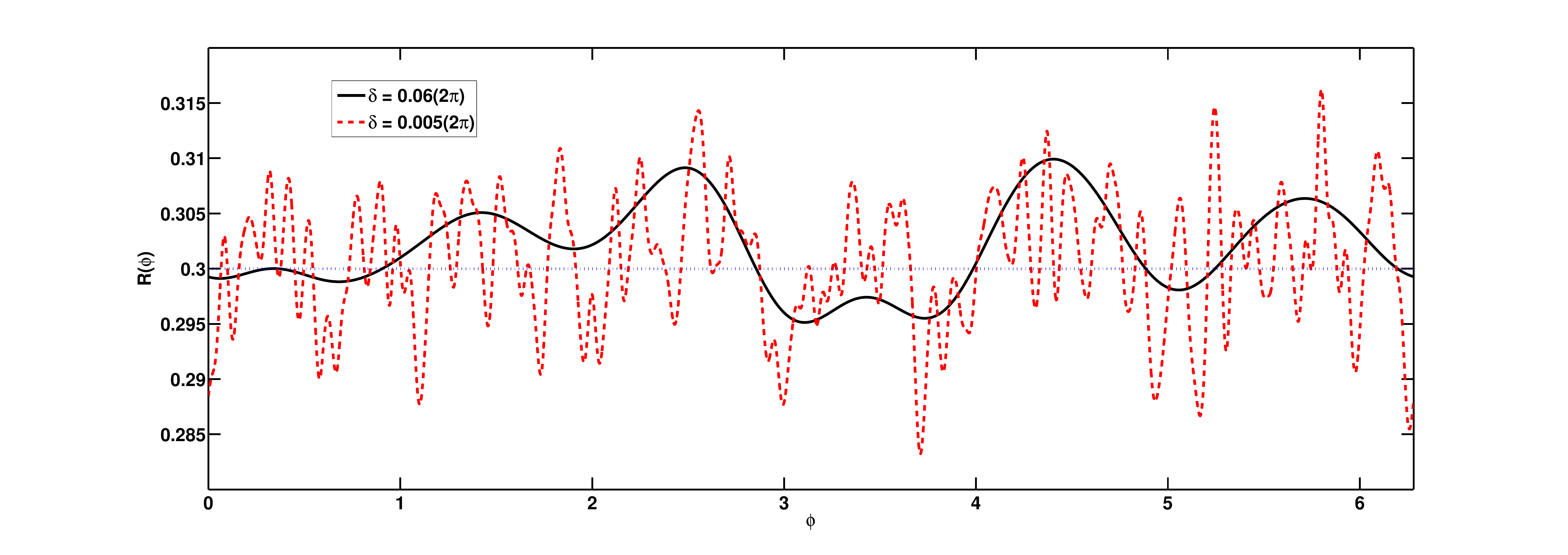}
 \caption{Hole profile in polar coordinates (left) and as a plot vs. $\phi$ (right) for the same random seed of $C_m$ coefficients, with $\delta = 0.06(2\pi)$ (black) and $\delta = 0.005(2\pi)$ (red). The blue dotted line indicates the regular hole.}
 \label{corr_ang}
 \end{center}
\end{figure}

As a starting point for the computations, we choose standard deviation of the radius fluctuations of $\sigma = 0.006a$, which is consistent with (if not a bit higher than) current state-of-the art structures made in the lab. The correlation angle is not so easily measured and so harder to define, so values of $\delta = 0.06(2\pi)$ (corresponding to hole-roughness as plotted in fig. \ref{exp_gauss}) and $\delta = 0.005(2\pi)$ (corresponding to much finer roughness) are chosen. For completeness, computations with the same values, but $\sigma = 0.002a$ are also performed, so that the effect of the disorder ``magnitude'' $\sigma$ and the correlation angle $\delta$ can be studied separately. To visualize the difference in correlation angle, in fig. \ref{corr_ang} we show a realization of a hole profile with the two different values for $\delta$, for $\sigma = 0.006a$.

In this section, we present simulations of a W1 waveguide with 256 elementary supercells, with radius of the (unperturbed) holes $r=0.3a$, slab thickness $0.5a$, and dielectric constant of the slab material $\varepsilon_2 = 12$, i.e. the regular guide is the same as the one shown in fig. \ref{guidedband}. For the disorder BME computation, there are three parameters which control the precision of the computation: the number of bands included, the number of $\mathbf{G}$-vectors defined by $G_{max}$, and the highest number $m_{max}$  of moments included in the summation of eq. (\ref{epsprime}). The effect of all three of those is discussed in more detail in sec. \ref{bmedisc} for a smaller waveguide (of 32 elementary supercells), with the assumption that a similar convergence behavior will apply. For the moment, we set $G_{max} = 3 (\frac{2\pi}{a})$ and include only the two guided bands of the guide. 

\begin{figure}[h]
\begin{center}
 \subfloat[]{\includegraphics[trim = 2in 1in 1in 1in, type=pdf,ext=.pdf,read=.pdf,width = \textwidth]{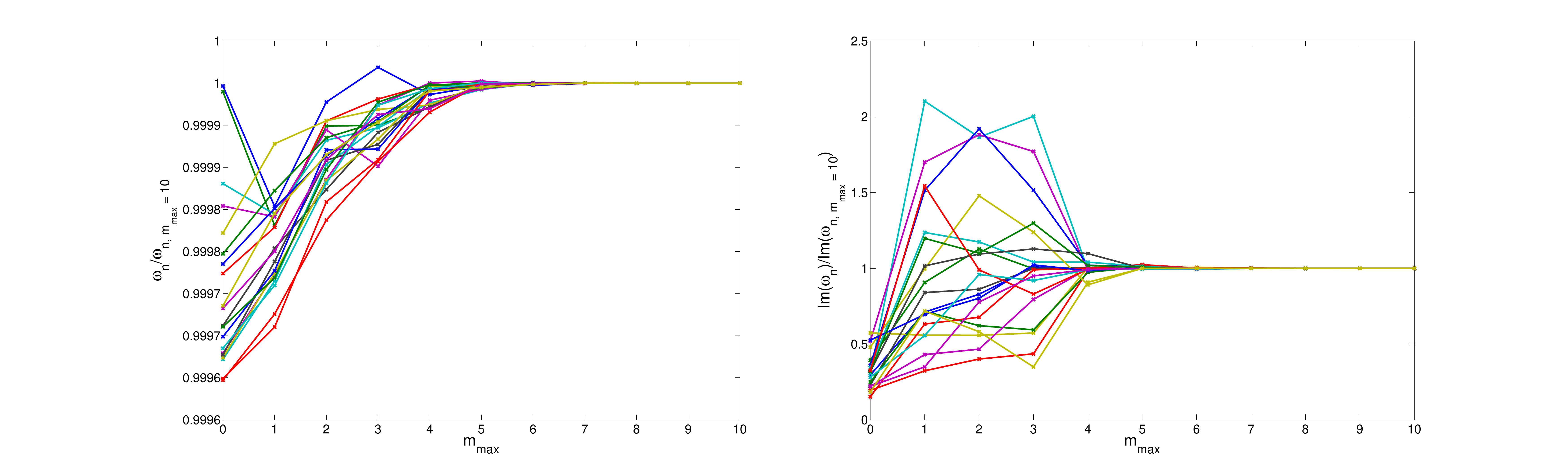}} \\
 \subfloat[]{\includegraphics[trim = 2in 1in 1in 1in, type=pdf,ext=.pdf,read=.pdf,width = \textwidth]{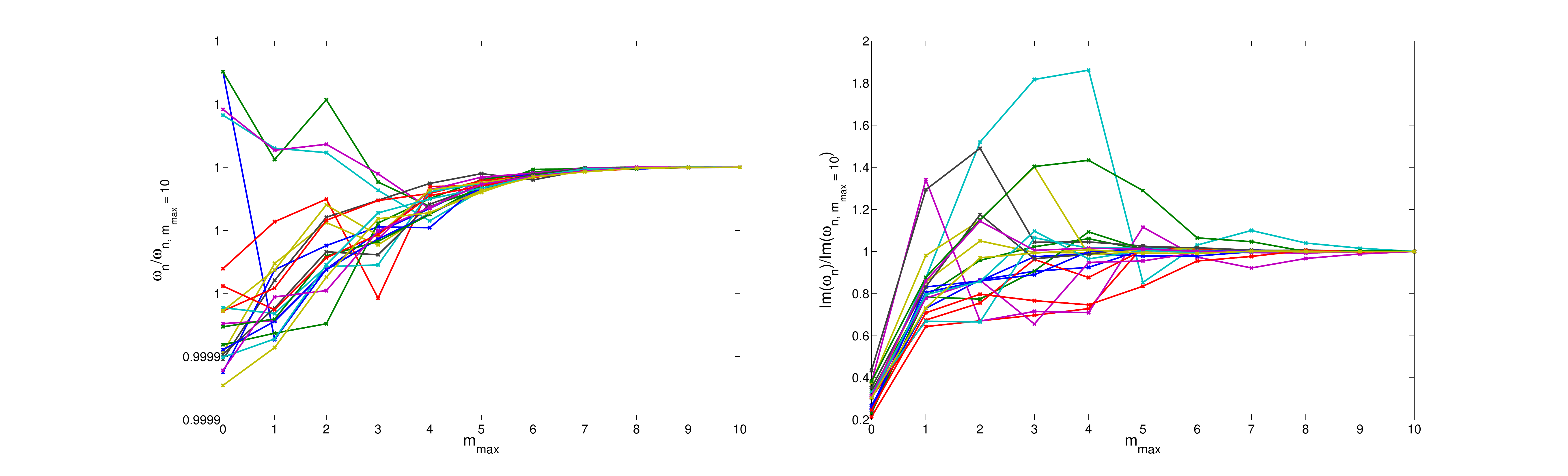}}
 \caption{Convergence with $m_{max}$ of the energies (left) and the radiative rates (right) of the lowest 20 states in the band-gap of a disordered PHC waveguide of 256 supercells. $\sigma = 0.006a$ was used everywhere, while the correlation angle is $\delta = 0.06 (2\pi)$ in (a) and $\delta = 0.005 (2\pi)$ in (b).}
 \label{mconv}
 \end{center}
\end{figure}

To decide on a suitable $m_{max}$, we take a look at fig. \ref{mconv}, where the energies and the radiative rates of the 20 lowest modes are plotted w.r.t. $m_{max}$, for $\sigma = 0.006a$ and the two different correlation angles $\delta$. As can be seen, for $m_{max} = 10$ the computation is fully converged in both cases, and even for $m_{max} = 5$ the results are converged to a satisfactory precision. Indeed, the radiative rates of the $\delta = 0.005(2\pi)$ structure converge the slowest, but even for them $95.9 \%$ of all the 512 of them are within $10\%$ of the corresponding $m_{max} = 10$ value. This precision is quite satisfactory both with respect to the approximations of the algorithm and with respect to the fact that a slightly different disorder realization can have a much bigger effect on the rates. Thus, we choose to stick to $m_{max} = 5$ for the other computations shown in this section. 

This is actually a very interesting result. As can be seen in the figure, the difference between keeping radius fluctuations only, i.e. $m_{max} = 0$, and including higher moments as well, can be quite significant: for example, all the plotted loss rates have $m_{max} = 0$ values which are between 10 and 50 $\%$ of the corresponding $m_{max} = 10$ values. The overall fluctuations, however, are mostly due to the first 5 moments, after which the result appears to converge nicely. This shows that the roughest disorder in the hole edge plays the biggest part in determining the modes of the structure; as the features included become finer and finer, the results are affected less and less. This is not unexpected: due to its wave-like properties, the electromagnetic field should generally not be affected much by features smaller than some characteristic scale, e.g. the wavelength. 

\begin{figure}[h]
\begin{center}
 \includegraphics[trim = 3in 1in 2in 1in, type=pdf,ext=.pdf,read=.pdf,width = \textwidth]{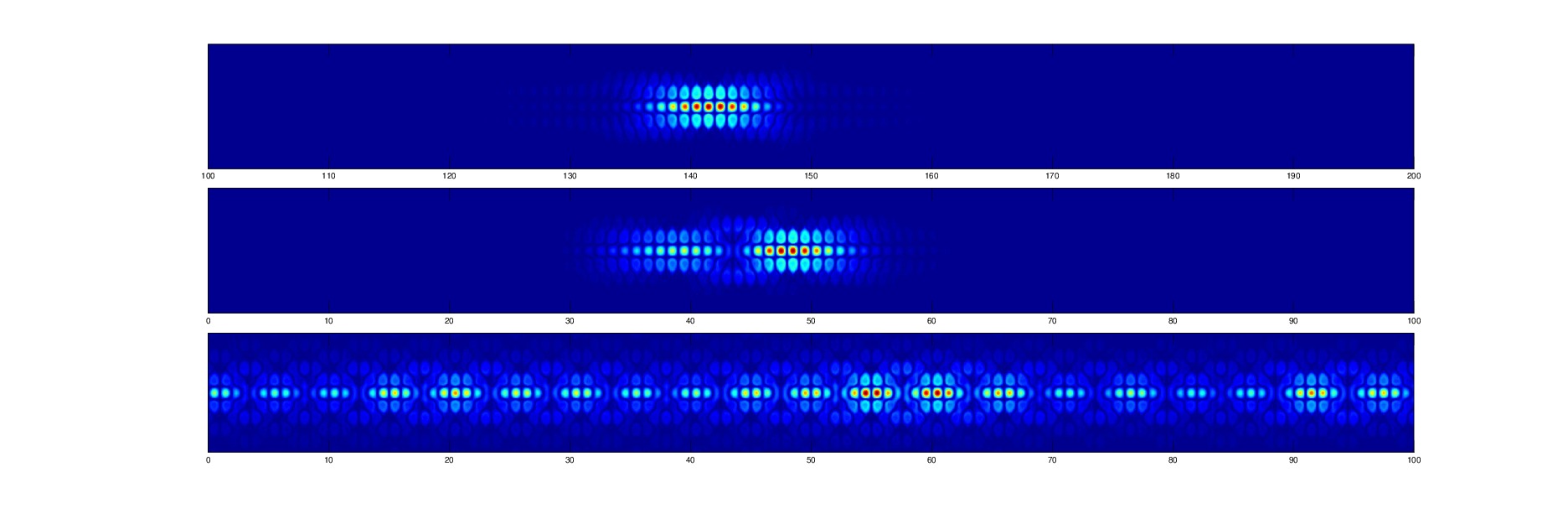}
 \caption{Magnetic field profile over selected portions of a W1 waveguide with 256 elementary cells and disorder parameters $\sigma = 0.006a$ and $\delta = 0.06 (2\pi)$. The modes correspond respectively to frequencies $\frac{\omega a}{2\pi c} = 0.27177$ (lowest frequency in the band gap), $\frac{\omega a}{2\pi c} = 0.27189$ and $\frac{\omega a}{2\pi c} = 0.27270$.}
 \label{guidedprof}
 \end{center}
\end{figure}

As an illustration of the power of the BME algorithm, in fig. \ref{guidedprof} we show the magnetic field profiles of three different modes of a $\sigma = 0.006a$, $\delta = 0.06(2\pi)$ structure. As can be seen, the BME method shows that the presence of disorder can induce Anderson localization of light into the PHC. The top image is the ground mode of the system - it has energy $\frac{\omega a}{2\pi c} = 0.27177$ lying below the band edge, and is fully localized within a region of around 10 elementary cells. The middle image is a mode with $\frac{\omega a}{2\pi c} = 0.27189$, still below the band edge, but close to it. As can be seen, it is still localized, but within a larger area and contains two lobes - such a character is common for most of the states around the band edge: they exhibit two or more main lobes with decaying tails. Finally, for a mode with energy well within the guided band, e.g. the mode of frequency $\frac{\omega a}{2\pi c} = 0.27270$ plotted in the bottom of the figure, the state is extended over the full waveguide, even though fluctuations in intensity are of course visible because of the random disorder. 

\begin{figure}[h]
\begin{center}
 \includegraphics[type=pdf,ext=.pdf,read=.pdf,width = \textwidth]{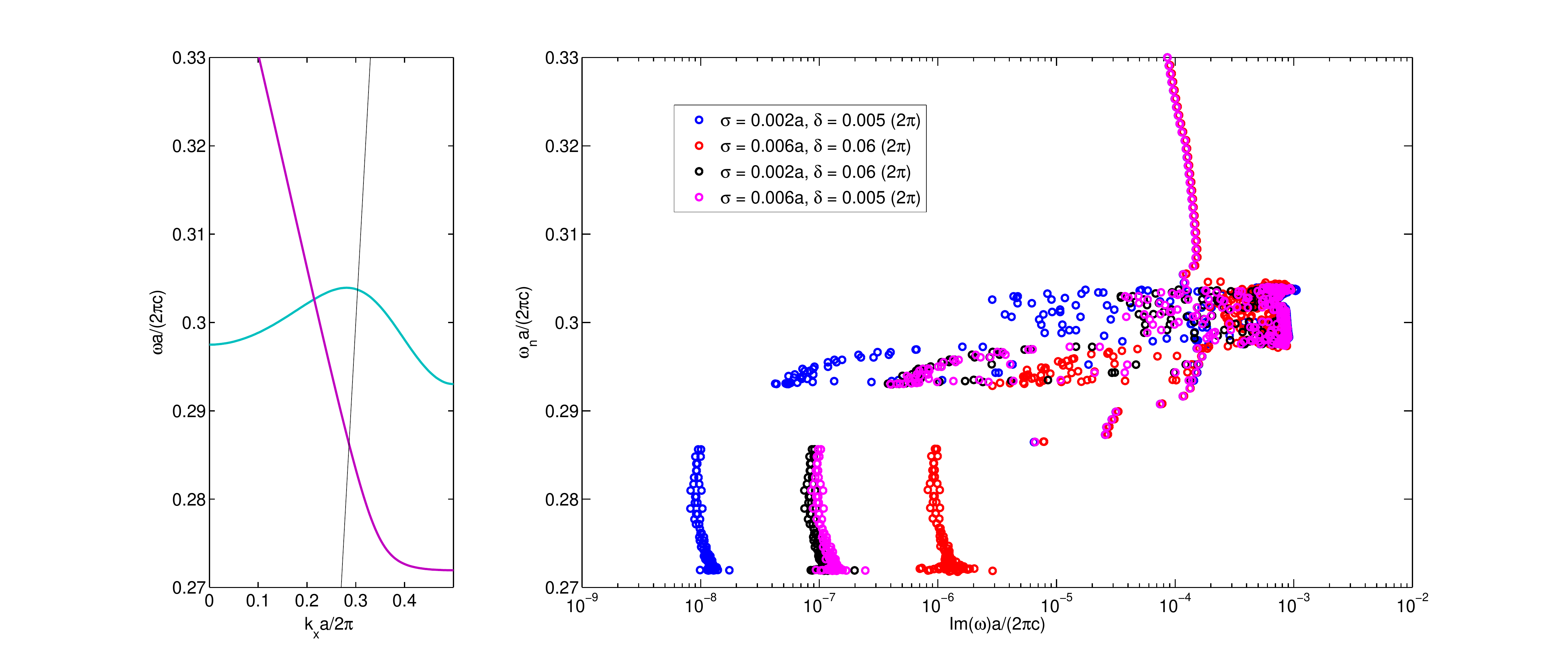}
 \caption{Radiation loss rates of all the states inside the band gap for four different sets of disorder parameters, with Gaussian correlations. On the left, the bands of the regular structure and the light cone (black line) are given for reference.}
 \label{ej_pr}
 \end{center}
\end{figure}

Radiation loss rates of the modes were also computed; they are shown for all four combinations of disorder parameters in fig. \ref{ej_pr}. The behavior of the loss rates for a single realization can be described by considering several energy regions: for lowest energies, due to the strong localization due to random disorder, the rates are scattered around a wide range. For slightly higher energies, in the region between the band edge and the light cone, the modes are no longer localized and show little fluctuation in $\gamma_{\beta}$, up until the point when the states enter the light cone, when the radiative rates increase greatly. This is when the loss rates become of mostly intrinsic nature, which is also testified by the fact that they no longer depend strongly on the different disorder parameters. Perhaps the most striking result visible in the figure is the large effect which the correlation length can have on the loss rates. As can be seen, the difference in the low-modes $\gamma_{\beta}$-s between a $\delta = 0.06(2\pi)$ and a $\delta = 0.005(2\pi)$ structure is approximately an order of magnitude, and is equivalent to the difference between a $\sigma = 0.006a$ and $\sigma = 0.002a$ structure. 

\begin{figure}[h]
\begin{center}
 \includegraphics[type=pdf,ext=.pdf,read=.pdf,width = \textwidth]{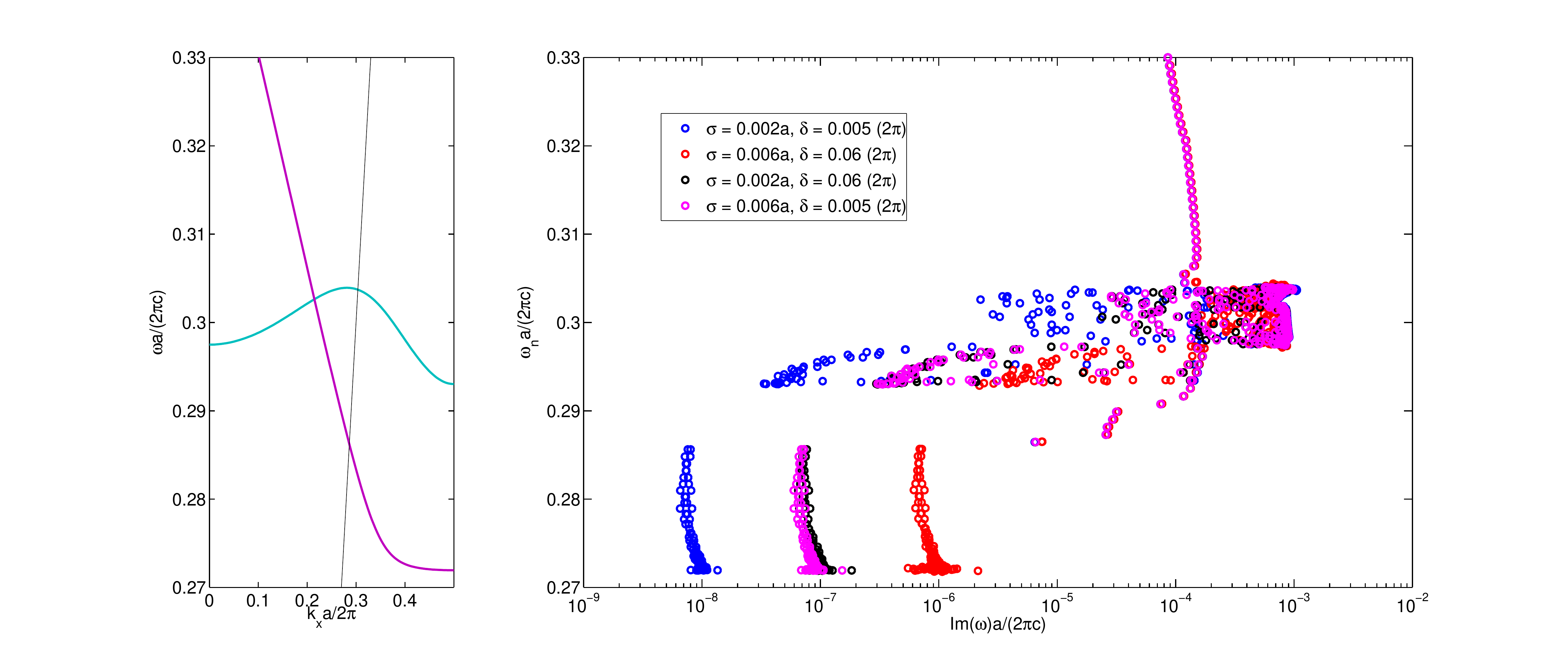}
 \caption{Radiation loss rates of all the states inside the band gap for four different sets of disorder parameters, with exponential correlations. On the left, the bands of the regular structure and the light cone (black line) are given for reference.}
 \label{ej_pr_exp}
 \end{center}
\end{figure}

For the sake of comparison between exponential and Gaussian correlations, in fig. \ref{ej_pr_exp} we present a plot of the radiative rates when the exponentially correlated disorder model of eq. (\ref{expcor}) was used, for the same four sets of the parameters $\sigma$ and $\delta$. Two small differences between the results with the two models can be observed: first, the loss rates in the Gaussian case are slightly higher, which can be expected as for the same correlation angle, in the exponential case there are features much finer than the correlation length which are still included. For the same reason, the effect of changing $\delta$ from $0.06 (2\pi)$ to $0.005 (2\pi)$ is slightly smaller in the exponential case, which can be inferred from the fact that the $\sigma = 0.006a, \delta = 0.005(2\pi)$ values lie slightly closer to the $\sigma = 0.006a, \delta = 0.06(2\pi)$, as is the case for the $\sigma = 0.002a, \delta = 0.005(2\pi)$ and $\sigma = 0.002a, \delta = 0.06(2\pi)$ pair.

\begin{figure}[h]
\begin{center}
 \subfloat[]{\includegraphics[trim = 1in 0.5in 0.5in 1in, type=pdf,ext=.pdf,read=.pdf,width = 0.45\textwidth]{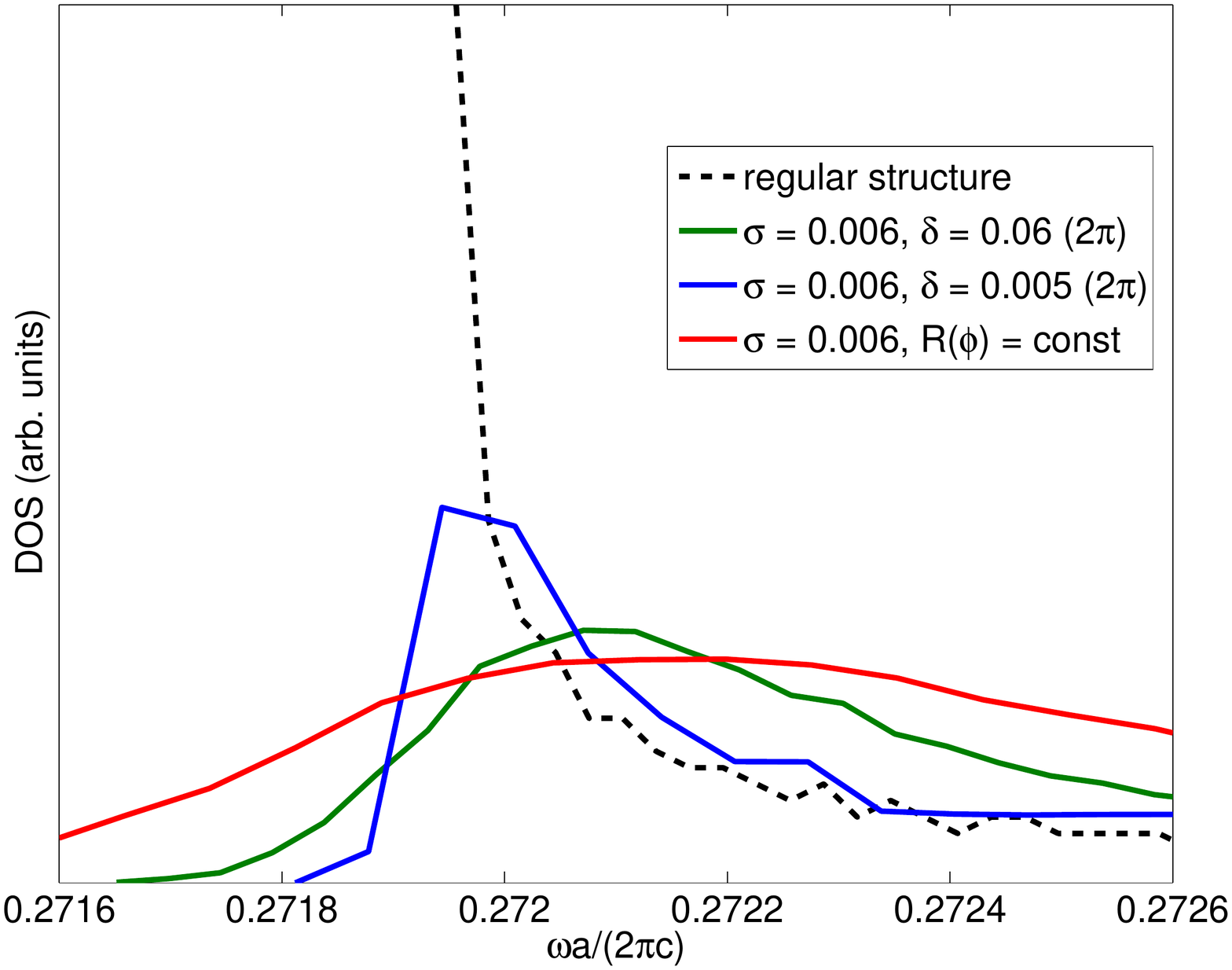}}
 \subfloat[]{\includegraphics[trim = 0.5in 0.5in 1in 1in, type=pdf,ext=.pdf,read=.pdf,width = 0.45\textwidth]{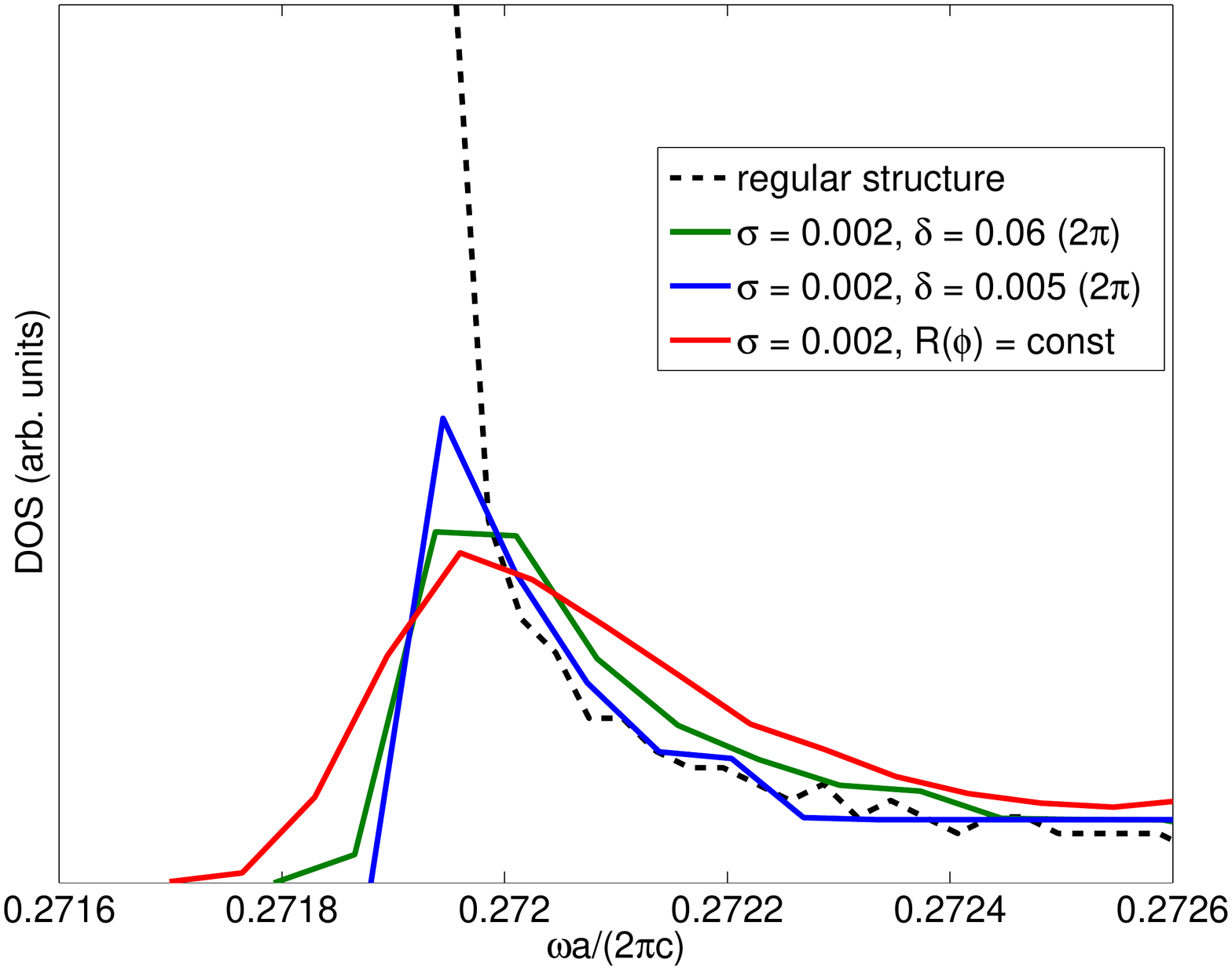}}
 \caption{(a): density of states of the regular structure (black) and of three different realizations of disorder with $\sigma = 0.006a$. (b): same for $\sigma = 0.002a$.}
 \label{DOS}
 \end{center}
\end{figure}

Finally, the density of states (DOS) of a structure could also be computed after averaging over 300 disorder realizations. In panel (a) of fig. \ref{DOS}, we show the DOS of the regular structure (black) - with a Van Hove singularity at the band edge - and of a disordered structure with $\sigma = 0.006a$ and $\delta = 0.06 (2\pi)$ (green) and $\delta = 0.005(2\pi)$ (blue). For comparison, the red line shows the DOS of a structure in which only constant radius fluctuations were considered, with a standard deviation again $\sigma = 0.006a$ (in other words, a case in which $\langle C_{0}^2 \rangle = \sigma^2$, $C_{m \neq 0} = 0$). All this is the same in panel (b), but for $\sigma = 0.002a$. In all the cases in the presence of disorder, the Van Hove singularity is no longer present, in exchange for a Lifshitz tail of the DOS. The interesting result now is that the smearing of the energies around the band edge in the presence of disorder is strongly affected by the correlation length, which is obvious especially in the $\sigma = 0.006a$ picture (for the $\sigma = 0.002a$ picture, all the effects of disorder are scaled down and thus higher precision of the computation is needed for the same result to be clearly visible). All in all, we see that the correlation angle $\delta$ of the hole-edge disorder can have a great effect both on the distribution of energies and of radiative rates. Thus, generally, $\delta$ is a parameter which deserves consideration in the production of PHCs in the lab - for example, if lower loss rates are desired, decreasing $\delta$ is almost as imperative as decreasing the overall disorder magnitude $\sigma$. 

In view of previous related work, the findings presented here are consistent with the results of \cite{hughes}, where it was found, within perturbation theory, that local field effects due to irregular hole shapes blueshift and broaden the band structure. Our results extend this finding in several ways: first, the density of states shows not only the blueshifted states, but also the redshifted ones lying in the Lifshitz tail. Second, the method allowed us to compute the electromagnetic profiles of the disordered modes. And finally, we could also compute the radiative rates and quantify the way the correlation angle $\delta$ affects them. Next we consider the experimental results of Le Thomas et. al. \cite{houdre}, which show that fluctuations in the hole area are the main source of degradation of the guided modes around the band edge (i.e. the spreading of the DOS in fig. \ref{DOS}). The scaling of disorder effects with the hole area deviations is also discussed in \cite{savona} and \cite{krauss}. While the authors of \cite{houdre} do not observe a significant effect coming specifically from the shape of the holes after the area fluctuations are taken into account, their analysis does not specifically relate to different correlation lengths of the disorder, but rather to some general fluctuations in the shape of the holes. In our model, hole area fluctuations are not separated from the shape fluctuations, but we see indications that different correlation lengths could have an effect, even after the area differences are subtracted. This would make sense intuitively, since the dielectric profile is ``smoothened'' by the electromagnetic field equations, and area deviations coming from small features are ``ignored''. Below we employ some statistics to go further into this point, specifically for the loss rates (with the current data, it is hard to quantify the spread of the DOS in all four cases). Within our model, the area of the holes has a slightly higher expectation value than the area of the regular holes, $\pi R_0^2$, given by

\begin{equation}
 \langle A \rangle = \pi \sigma^2 + \pi R_0^2.
\label{areafluct}
\end{equation}

Looking at the radiative rates presented in fig. \ref{ej_pr}, we see that both for $\delta = 0.06(2\pi)$ and $\delta = 0.005(2\pi)$, when $\sigma$ is decreased 3-fold, the radiative rates change by approximately an order of magnitude. This is consistent with the fact that the shift in the mean value of the area in eq. (\ref{areafluct}) is proportional to $\sigma^2$. However, the same ten-fold change is observed when changing the correlation angle $\delta$ for a fixed value of $\sigma$, despite the fact that eq. (\ref{areafluct}) is independent of it. Still, this does not mean that the area fluctuations are overall the same for different correlation angles, since the standard deviation $\sigma_{A} = \sqrt{\langle \langle A^2 \rangle - \langle A \rangle^2 \rangle}$ depends non-trivially on $\delta$, due to:

\begin{equation}
 \langle A^2 \rangle = \pi^2 \langle \left(\sum_{m \neq 0} |C_m|^2 + (C_0 + R_0)^2 \right) \left(\sum_{n \neq 0} |C_n|^2 + (C_0 + R_0)^2 \right) \rangle.
\end{equation}

We can numerically compute this standard deviation for the disorder realizations of fig. \ref{ej_pr}, to find that its change with changing either $\sigma$ or $\delta$ is almost the same. Namely, we compute $\frac{\sigma_A(\sigma = 0.006a, \delta = 0.06(2\pi))}{\sigma_A(\sigma = 0.006a, \delta = 0.005(2\pi))} \approx 3.5 \approx \frac{\sigma_A(\sigma = 0.002a, \delta = 0.06(2\pi))}{\sigma_A(\sigma = 0.002a, \delta = 0.005(2\pi))}$ for the two cases of changing $\delta$, and $\frac{\sigma_A(\sigma = 0.006a, \delta = 0.06(2\pi))}{\sigma_A(\sigma = 0.002a, \delta = 0.06(2\pi))} \approx 3.0 \approx \frac{\sigma_A(\sigma = 0.006a, \delta = 0.005(2\pi))}{\sigma_A(\sigma = 0.002a, \delta = 0.005(2\pi))}$ for the two cases of changing $\sigma$. Thus, a change in the loss rates of a full decade between the two correlation lengths seems too much, if it is only due to the induced area fluctuations. Unfortunately, at this point no stronger statement can be made, but with a different statistical model, one can set the expectation value of eq. (\ref{areafluct}) to 0, and separately compare the effects of $\sigma_A$ and $\delta$, for a conclusive statement of weather the correlation length has an effect on its own. 

\subsection{Discussion of the method}

\label{bmedisc}

The results of the previous section, while convincing and logical, are not at all conclusive because of the several approximations in the BME computation. Here we try to analyze those separately, using a 32-supercell waveguide with the same irregular-holes disorder as a toy model. In all the figures here, we used $\sigma = 0.006a$, $\delta = 0.06 (2\pi)$, but the results with the three other possibilities: $\sigma = 0.006a$, $\delta = 0.005 (2\pi)$; $\sigma = 0.002a$, $\delta = 0.06 (2\pi)$ and $\sigma = 0.002a$, $\delta = 0.005 (2\pi)$, are presented in the Appendix.

\begin{figure}[h]
\begin{center}
 \subfloat[]{\includegraphics[trim = 1in 0.9in 0.3in 1in,type=pdf,ext=.pdf,read=.pdf,width = 0.49\textwidth]{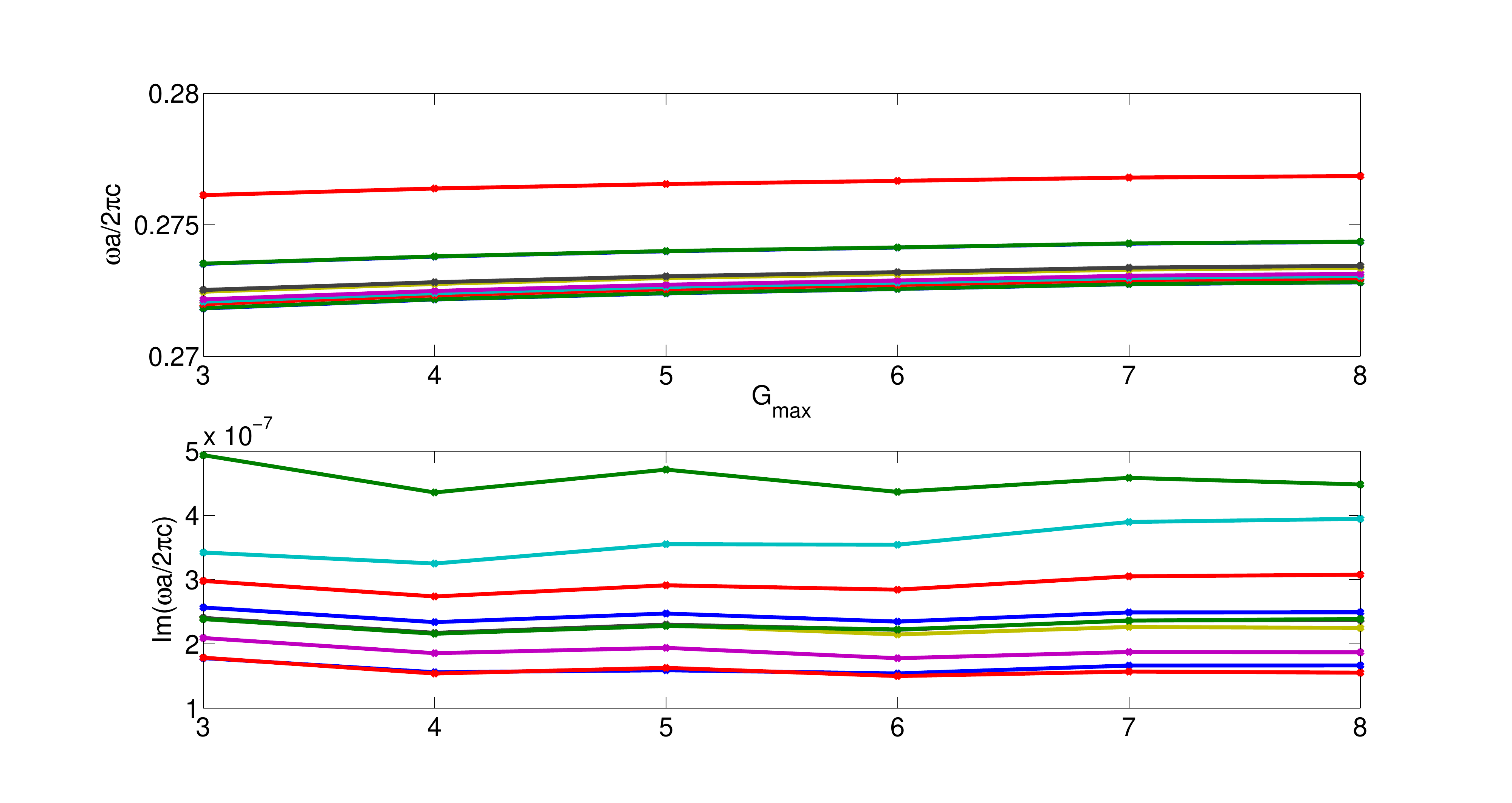}}
 \subfloat[]{\includegraphics[trim = 1in 0.9in 0.3in 1in, type=pdf,ext=.pdf,read=.pdf,width = 0.49\textwidth]{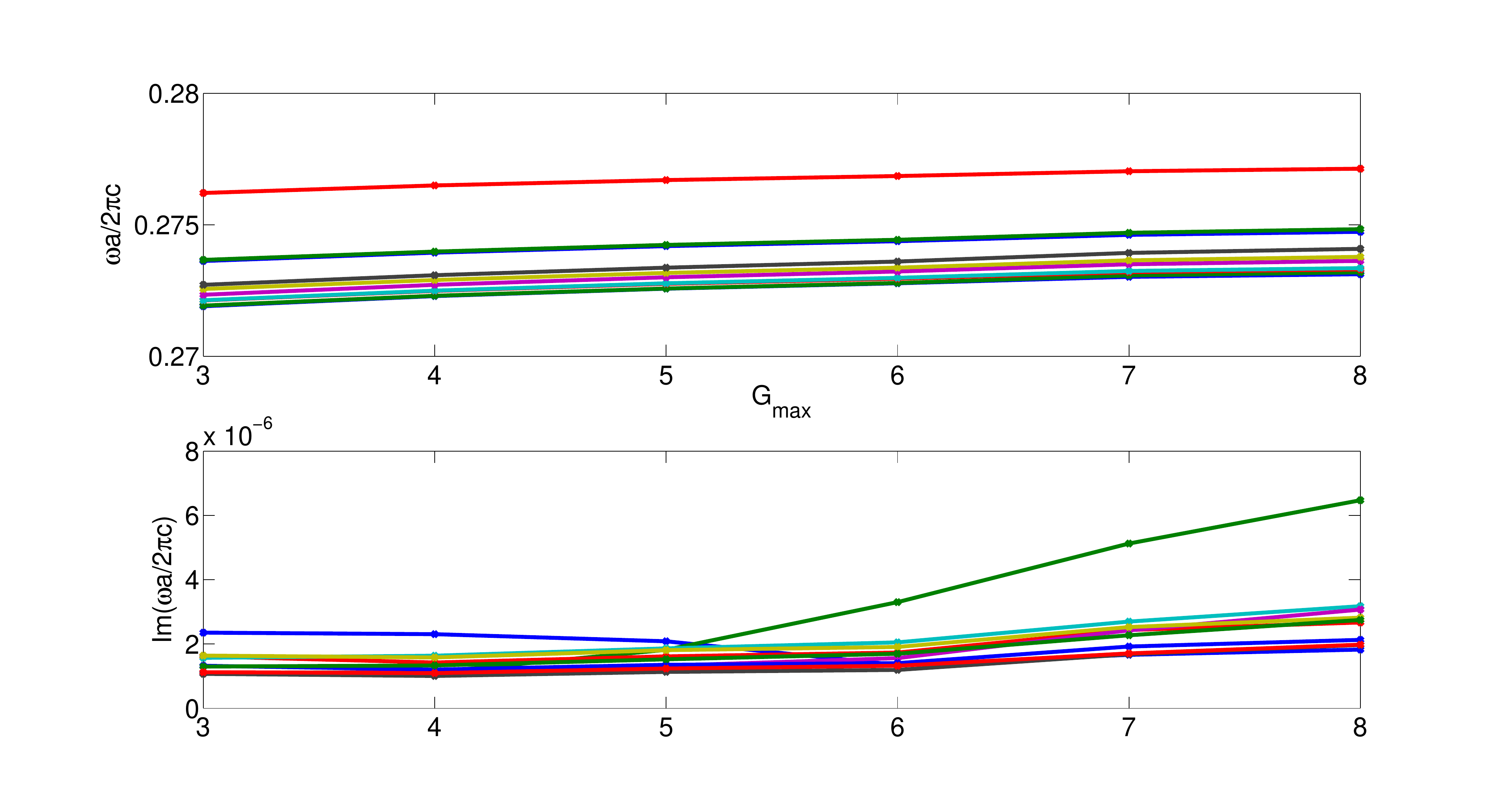}}
 \caption{(a): dependence on $G_{max}$ of the energies (up) and radiative rates (down) of the 10 lowest modes of a 32-supercell W1 guide with $\sigma = 0.006a$, $\delta = 0.06(2\pi)$ and $m_{max}=0$. (b): same for $m_{max} = 10$.}
 \label{m0_conv}
 \end{center}
\end{figure}

We start with the convergence of the modes with $G_{max}$. In fig. \ref{m0_conv} (a) we plot the energies and radiative rates for the lowest 10 modes vs. $G_{max} = 3, 4, 5, 6, 7$ and $8 (\frac{2\pi}{a})$, for $m_{max} = 0$. As can be seen, increasing $G_{max}$ does not appear to change much any of the properties of the system. This is slightly different when we consider the same plots, but for $m_{max} = 10$, shown in fig. \ref{m0_conv} (b), where the energies change in just the same way as in the $m_{max} = 0$ case, but the loss rates change more. Of course, when finer features are included in the computation, one might need a bit higher Fourier modes (correspondingly higher $\mathbf{G}$ vectors) to get convergence. Notice, however, that the radiative rates predominantly change in parallel. This means that one would expect the results with low $G_{max}$ to still be, at least qualitatively, correct. The reason why the values change is simply that including more terms in the expansion of the potential $\varepsilon'(\mathbf{r})$ corresponds in a way to considering a \textit{different} potential, i.e. one with different local and global minima. But this is the same as considering a slightly different disorder realization; thus, especially for the averaging results taken in the computation of the DOS in fig. \ref{DOS}, increasing $G_{max}$ should have little effect. 

It should also be noted that the constant increase of the energies with $G_{max}$ in both cases is predominantly due to the increase in energies of the regular structure, rather than the BME computation of the irregular one. In other words, as can be seen in fig. \ref{ho_noho}, even when using the Ho method, the bands are not fully convergent for $G_{max} = 3$, and shift slightly up in energy when this value is increased. This is actually the main contribution to the shift of the disordered energies.

\begin{figure}[h]
\begin{center}
 \subfloat[]{\includegraphics[trim = 1in 0.8in 0.3in 2.5in, type=pdf,ext=.pdf,read=.pdf,width = 0.49\textwidth]{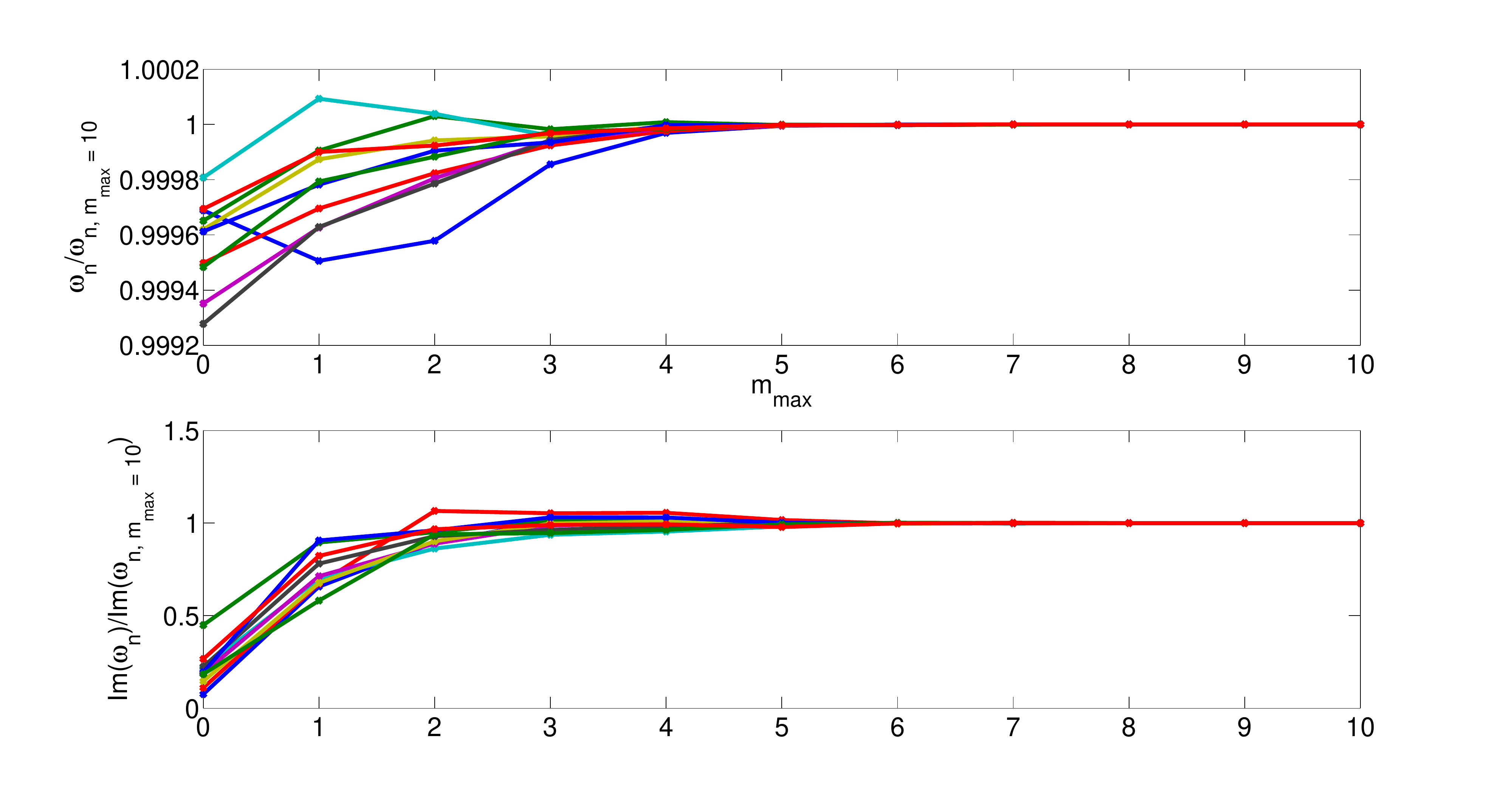}}
 \subfloat[]{\includegraphics[trim = 1in 0.8in 0.3in 2.5in, type=pdf,ext=.pdf,read=.pdf,width = 0.49\textwidth]{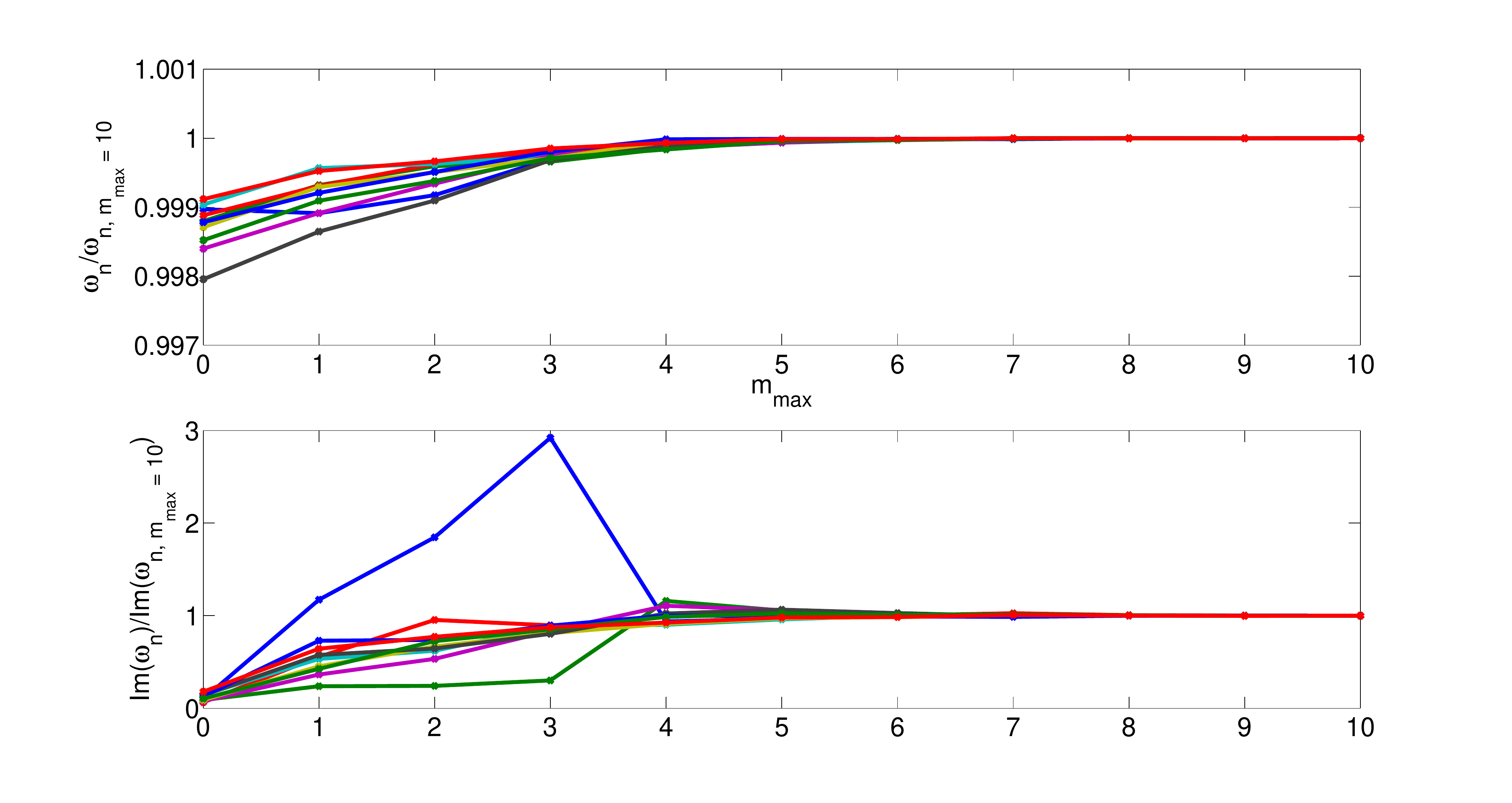}}
 \caption{(a): dependence on $m_{max}$ of the energies (up) and radiative rates (down) of the 10 lowest modes of a 32-supercell W1 guide with $\sigma = 0.006a$, $\delta = 0.06(2\pi)$ and $G_{max}=3(\frac{2\pi}{a})$. (b): same for $G_{max} = 7(\frac{2\pi}{a}$).}
\label{G3_conv}
 \end{center}
\end{figure}

The results of sec. \ref{holeirr} are very much based on the conclusion drawn from fig. \ref{mconv} that moments higher than $m = 5$ have little effect on the computation. Since we saw that increasing $G_{max}$ could produce an overall effect, it is imperative that we test this claim for high $G_{max}$ as well. To that end, in fig. \ref{G3_conv} (a) we present the convergence with $m_{max}$ of energies and rates of the ten lowest modes for $G_{max} = 3 (\frac{2\pi}{a})$, while in panel (b) we show the same, but for $G_{max} = 7 (\frac{2\pi}{a})$. As can be seen everywhere, the same result as in fig. \ref{mconv} is obtained: the energies and loss rates can change significantly with the addition of the lowest moments, but the effect is only up to $\approx m = 5$; beyond that, the computation is converged. Thus, the claim that fine features of the hole-edge disorder do not contribute to the overall disorder effects appears to hold even when very high $\mathbf{G}$-vectors are included.

\begin{figure}[h]
\begin{center}
 \includegraphics[type=pdf,ext=.pdf,read=.pdf,width = \textwidth]{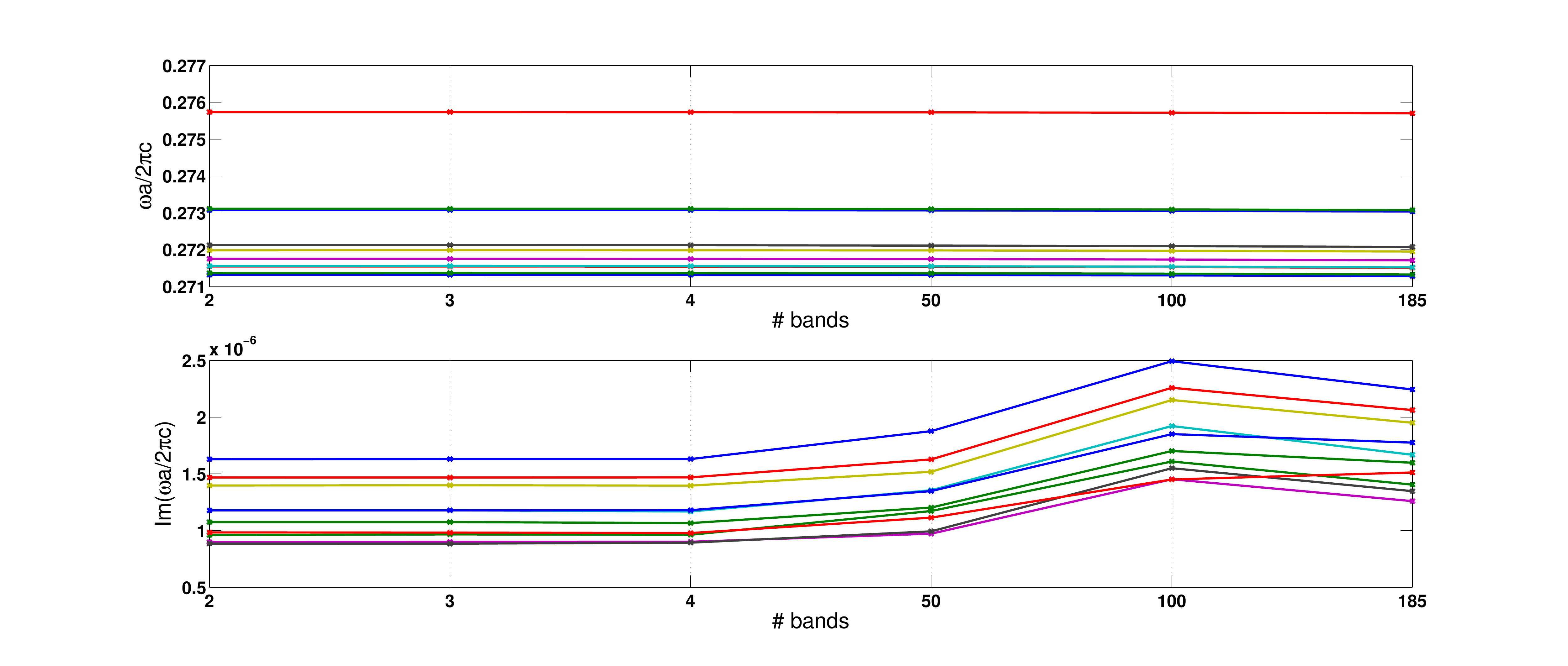}
 \caption{Dependence of the energies (up) and loss rates (down) on the number of bands included in the BME for the 10 lowest modes of a waveguide.}
 \label{bands_conv}
 \end{center}
\end{figure}

The last thing to verify is how the computation is affected by adding more bands. It was hinted in sec. \ref{seckur} that including more bands has a huge effect on the radiative rates only, and only when some state resulting from complex interference is investigated, which is not the case of random disorder. The same result can be seen in fig. \ref{bands_conv}, where we plot the energies and loss rates of a guide when different number of bands were included, for $G_{max} = 2(\frac{2\pi}{a})$ and $m_{max} = 10$. The 2-band computation involves simply the two guided bands; next the highest conduction band was added, and afterwards the lowest valence band. Finally, computations with all bands from 1 to 50, 1 to 100 and 1 to 185 were carried out. As can be seen, the energies stay virtually constant, while the radiative rates increase slightly when many bands are added. Still, the 2-band loss rates are $\approx 70\%$ of the 185-bands values, which shows again that in the case of disorder, a few bands are sufficient to reveal many of the properties of the system, and provide a good estimate (lower bound) of the radiative rates. 

\clearpage

\section{Conclusions and Outlook}

The BME formalism outlined here was illustrated through application to a 2D PHC of a triangular lattice of circular holes in a dielectric slab. An interesting future application of the method was already mentioned in the end of sec. \ref{secirr} - namely, to investigate separately the effects of hole-area fluctuations and hole-shape fluctuations on the modes of a structure. In addition to that, the method can of course equally well be applied to any 2D slab-based PHC structure, and furthermore, with a little extra mathematics, it can be extended to 3D structures. Here we also give two open problems for which it can prove very useful.

\subsection{Simulating surface roughness}

\begin{figure}[h!]
 \begin{minipage}{0.49\textwidth}
\begin{flushleft}
 \subfloat[]{\includegraphics[type=pdf,ext=.pdf,read=.pdf,width = 0.8\textwidth]{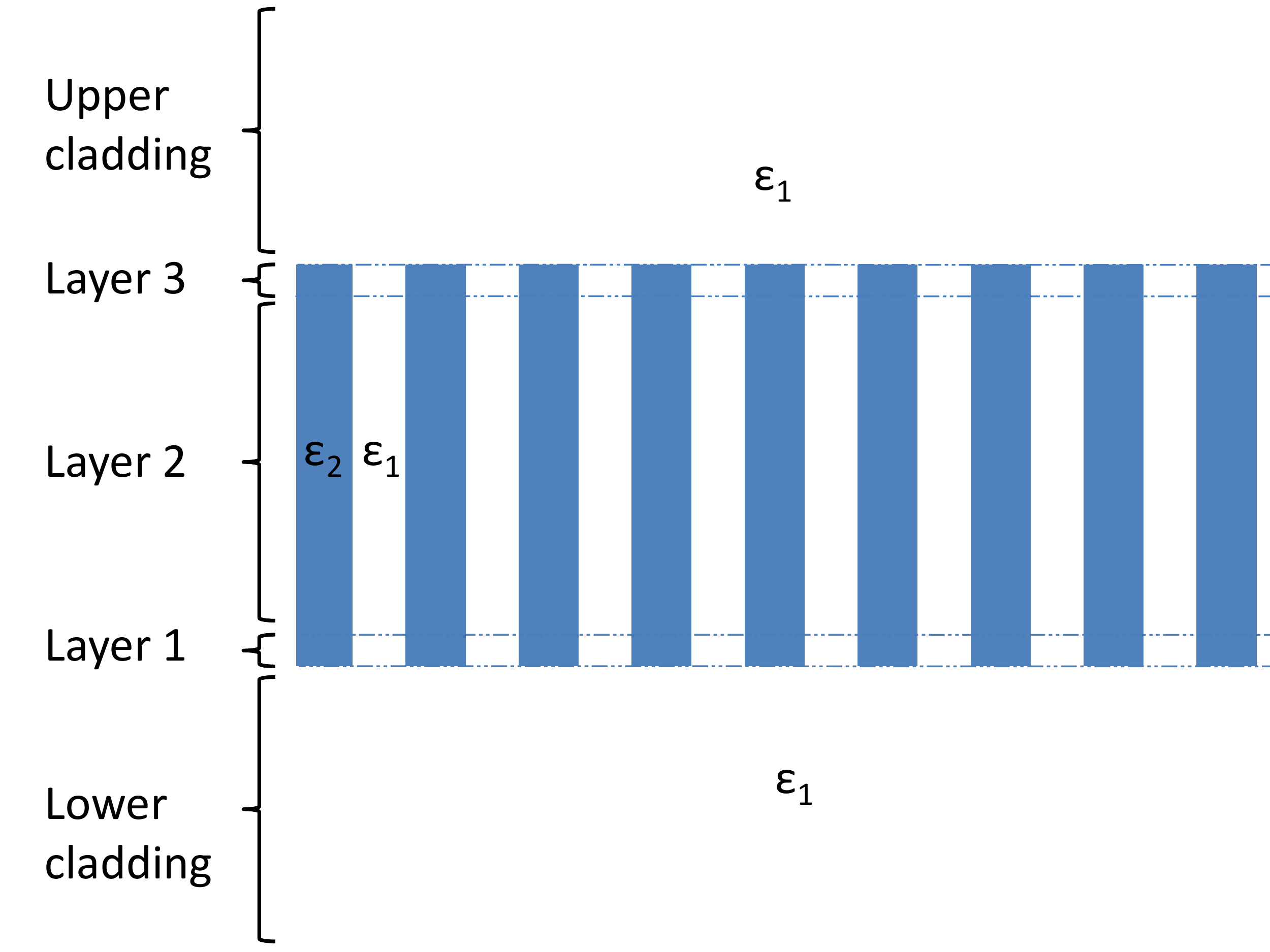}}
\end{flushleft}
\end{minipage}
 \begin{minipage}{0.49\textwidth}
\begin{center}
 \subfloat[]{\includegraphics[type=pdf,ext=.pdf,read=.pdf,width = 0.8\textwidth]{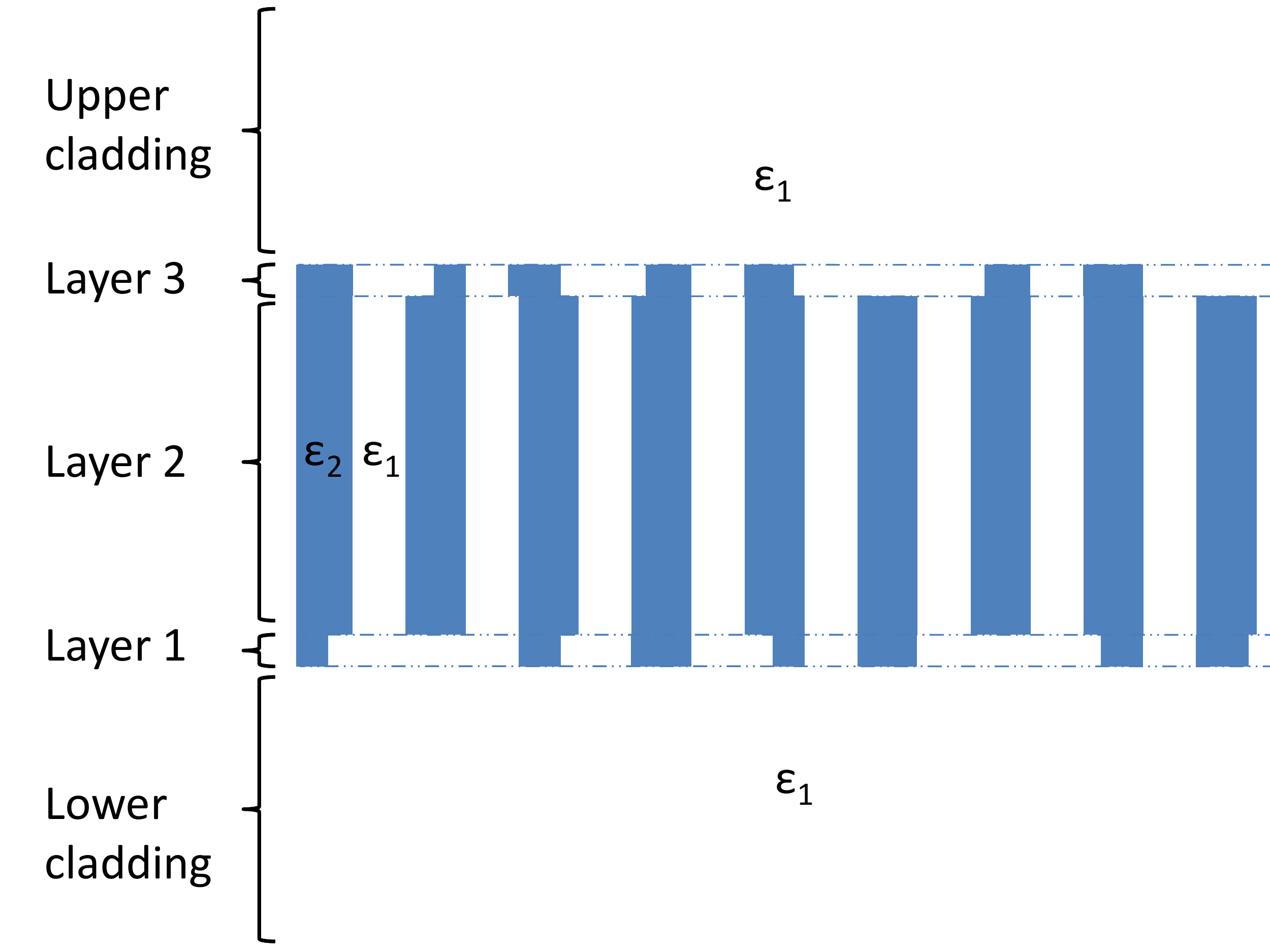}}
\end{center}
\end{minipage}
\caption{(a): cross-section of the PHC without disorder: it can be viewed as composed of 5 layers instead of 3. (b): adding random fluctuations (constant in the vertical direction) to layers 1 and 3 simulates roughness at the interfaces.}
\label{layers}
\end{figure}

In sec. \ref{secirr} we applied the Bloch-mode expansion formalism in order to obtain interesting results about how variations in the 2D-shape of holes in a PHC structure affect its electromagnetic modes. Another interesting question, into which the BME could give some insight, is what role roughness at the interfaces between the slab and the claddings plays. This has currently never been numerically simulated, so it is an essence still an open question. We will briefly outline how the BME method can be expanded to include such disorder. To that end, one can start from a many-layer GME computation to compute the Bloch modes: while the approach was outlined for three layers in \cite{andreani_2006} (i.e. slab and two cladding layers), it is straightforward to generalize to $N$ layers, as in sec. 3 of \cite{zabelin}. In a similar fashion, the BME formulas are straightforward to extend; once this is done, the simplest model one can consider is to view the regular PHC as composed of five layers (fig. \ref{layers}, (a)): the two claddings and the slab, plus two thin layers at the interfaces. The Bloch modes of such a structure can be computed with GME, and in fact, as long as the thin layers are from the same material as the slab, a 3-layer GME calculation is sufficient. We can then simulate surface roughness by leaving the slab (layer 2) and claddings unperturbed, while changing the profile of the thin layers (1 and 3) to some arbitrary shape (fig. \ref{layers}, (b)). As long as this change does not depend on $z$, it can be conveniently treated in a BME formalism, with the matrix $V_{\mathbf{k} n \mathbf{k}' n'}$ determined by the disordered profiles of the two thin layers, $\varepsilon_1(\mathbf{r})$ and $\varepsilon_3(\mathbf{r})$. The proposed model, while very simplified, could provide plenty of insight into the effect of surface disorder on the PHC modes. 

\subsection{Quantum dots embedded in a PHC structure}

Quantum dots embedded in PHC structures promise to open up almost unlimited possibilities for applications starting from simple light emission and/or lasing to implementation of qubits and qubit gates for quantum information processing. With the ever-advancing sophistication of our ability to produce such integrated structures in the lab \cite{benson}, it is imperative that numerical simulations follow suit. The electromagnetic modes of a PHC structure which can be computed in a BME formalism can be the starting point for such simulations. For example, in a Maxwell-Schr\"{o}dinger framework as the one developed and applied in \cite{savona_2004}, one models the presence of the dots by adding a (generally nonlocal) linear susceptibility tensor $\hat{\chi}(\mathbf{r}, \mathbf{r}', \omega)$, which can be computed to a different degree of sophistication based on the quantum-dot wavefunction models one wishes to include. This, together with the electromagnetic mode of the background medium is sufficient to explore the radiative properties both of a single quantum dot (energy-shift and radiative lifetime) and of a collection of dots (radiative coupling and collective modes). The fact that the BME makes it easy to compute small variations to a starting structure, then, would also make it possible to optimize structures for the tuning of the quantum dot properties.

\subsection{Conclusion}

In this thesis, we summarized some basis-expansion methods for numerically simulating 2D Photonic Crystals, namely plane-wave expansion and guided-mode expansion for simulating regular periodic structures, and Bloch-mode expansion for the addition of disorder. The main results of this work are based on the Bloch-mode expansion, a method which allows for very large structures (compared to the limitations of e.g. FDTD or supercell-GME) to be simulated, and yet can be used to accurately compute not only the frequencies and loss rates of the modes, but also their real-space profiles. In sec. \ref{secbme}, we first tested the method against some gentle-confinement cavities for which results already appear in the literature, as well as against guided-mode expansion with a big supercell, and saw that in the case of strongly-confined modes, the BME produces quantitatively reliable quality factors only when many bands are included, even though even a few bands can give good information about the mode profiles and their energies. In the presence of disorder, however, BME appears to be fully quantitatively reliable even for a low number of bands, as illustrated by the small change in quality factors of the cavities between a 2-band and a 195-band computation with disorder.

A similar observation was made for the case of disorder in W1 waveguides, and a 2-band BME was applied to the problem of quantifying the effect of hole-edge disorder on the modes of the waveguide. This is still an open question, and the results obtained are the main innovation of this work. Namely, we found that if we split the hole profile into Fourier components, only the first $\approx 5$ moments can have a strong influence on both the energies and the loss rates of a guided mode, and for $m_{max} = 10$ the computations were always fully converged. The explanation to this lies most probably in the fact that sufficiently fine structures become ``invisible'' for an electromagnetic wave of a certain wavelength, i.e. the underlying Maxwell equations smoothen the potential so that only the big features matter. This resulted in the fact that for the same magnitude of disorder, $\sigma = 0.006a$ or $\sigma = 0.002a$, a difference of correlation angle between $\delta = 0.06(2\pi)$ and $\delta = 0.005(2\pi)$ can produce an order of magnitude of difference in the loss rates of the guided modes. Furthermore, for higher correlation angle, we saw a stronger smearing of the DOS of the waveguide around the guided-band edge, which suggests stronger localization of the modes and so degradation of the dispersive slow-light regime. It will be interesting to extend this work by refining the statistical model of disorder and considering more sets of the disorder parameters. 

\section{Acknowledgement}

M. Minkov would like to thank prof. Vincenzo Savona for the helpful guidance throughout the Master studies, and for many useful discussions of all the results presented here.

\clearpage

\appendix
\section{Results as in sec. \ref{bmedisc}, but for different disorder parameters}

\begin{figure}[h]
\begin{center}
 \subfloat[]{\includegraphics[trim = 1in 0.8in 1in 1in, type=pdf,ext=.pdf,read=.pdf,width = 0.49\textwidth]{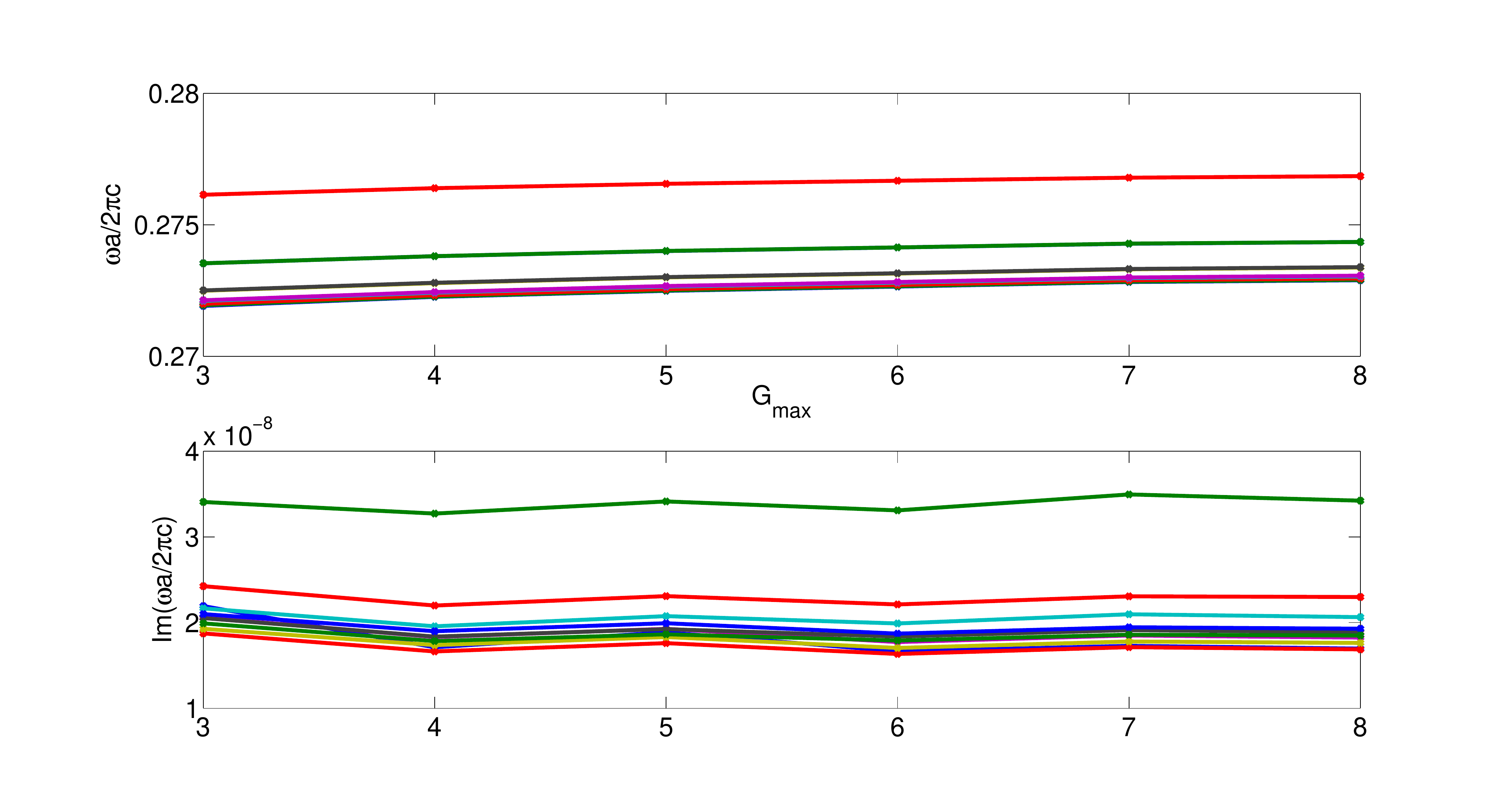}}
 \subfloat[]{\includegraphics[trim = 1in 0.8in 1in 1in, type=pdf,ext=.pdf,read=.pdf,width = 0.49\textwidth]{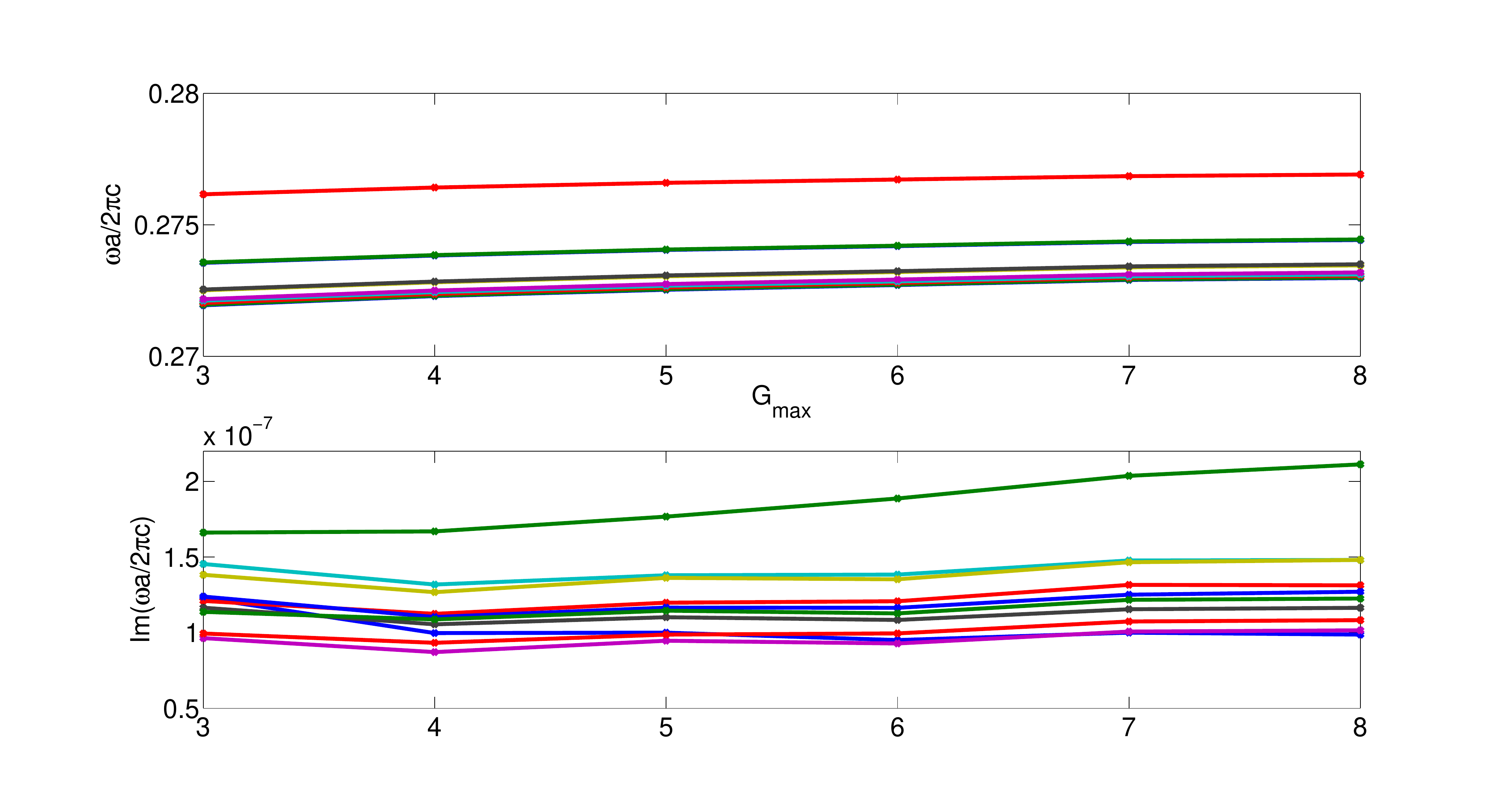}}
 \caption{(a): dependence on $G_{max}$ of the energies (up) and radiative rates (down) of the 10 lowest modes of a 32-supercell W1 guide with $\sigma = 0.006a$, $\delta = 0.005(2\pi)$ and $m_{max}=0$. (b): same for $m_{max} = 10$.}
 \label{app1}
 \end{center}
\end{figure}

\begin{figure}[h]
\begin{center}
 \subfloat[]{\includegraphics[trim = 1in 0.8in 1in 1in, type=pdf,ext=.pdf,read=.pdf,width = 0.49\textwidth]{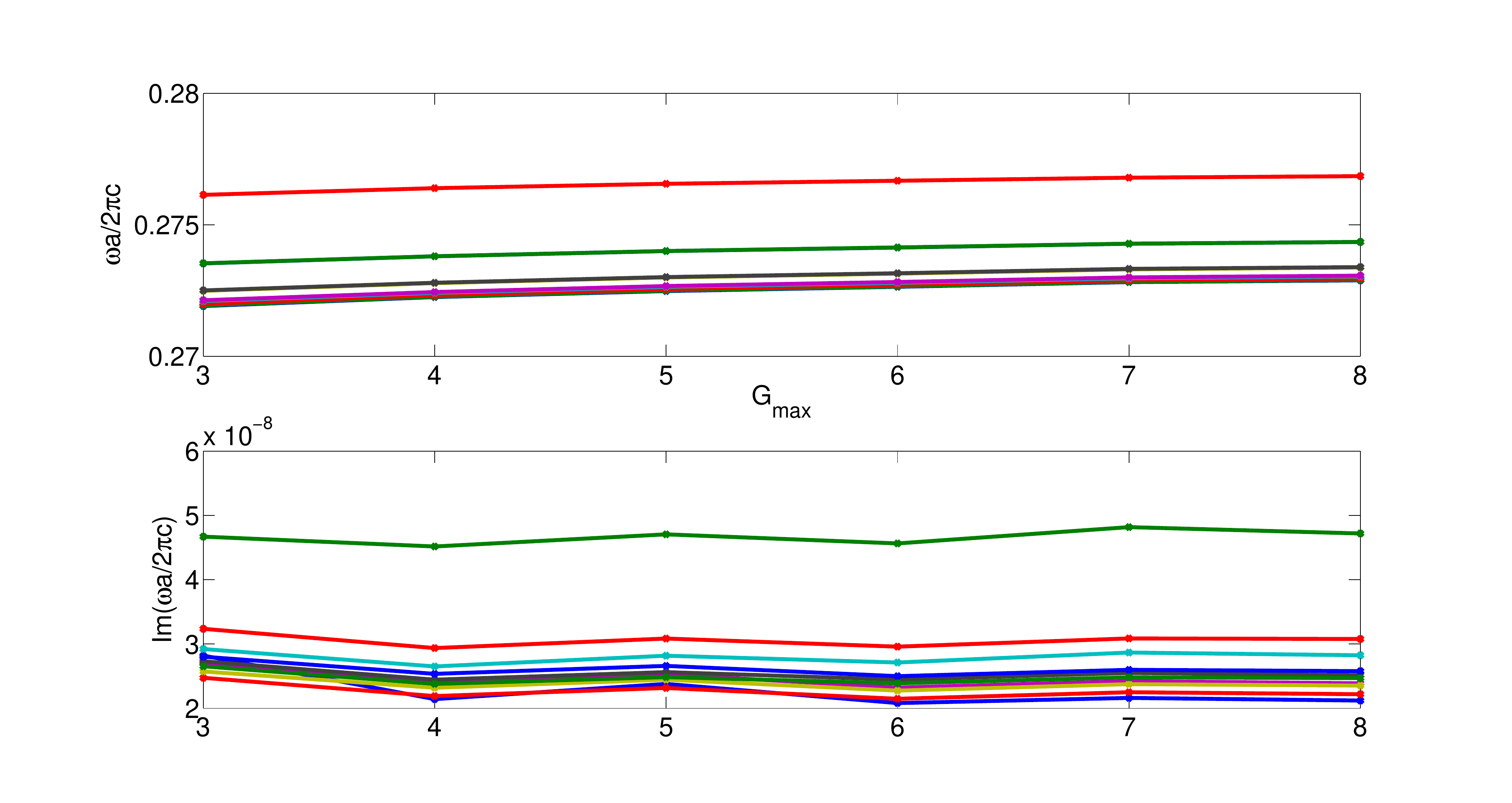}}
 \subfloat[]{\includegraphics[trim = 1in 0.8in 1in 1in, type=pdf,ext=.pdf,read=.pdf,width = 0.49\textwidth]{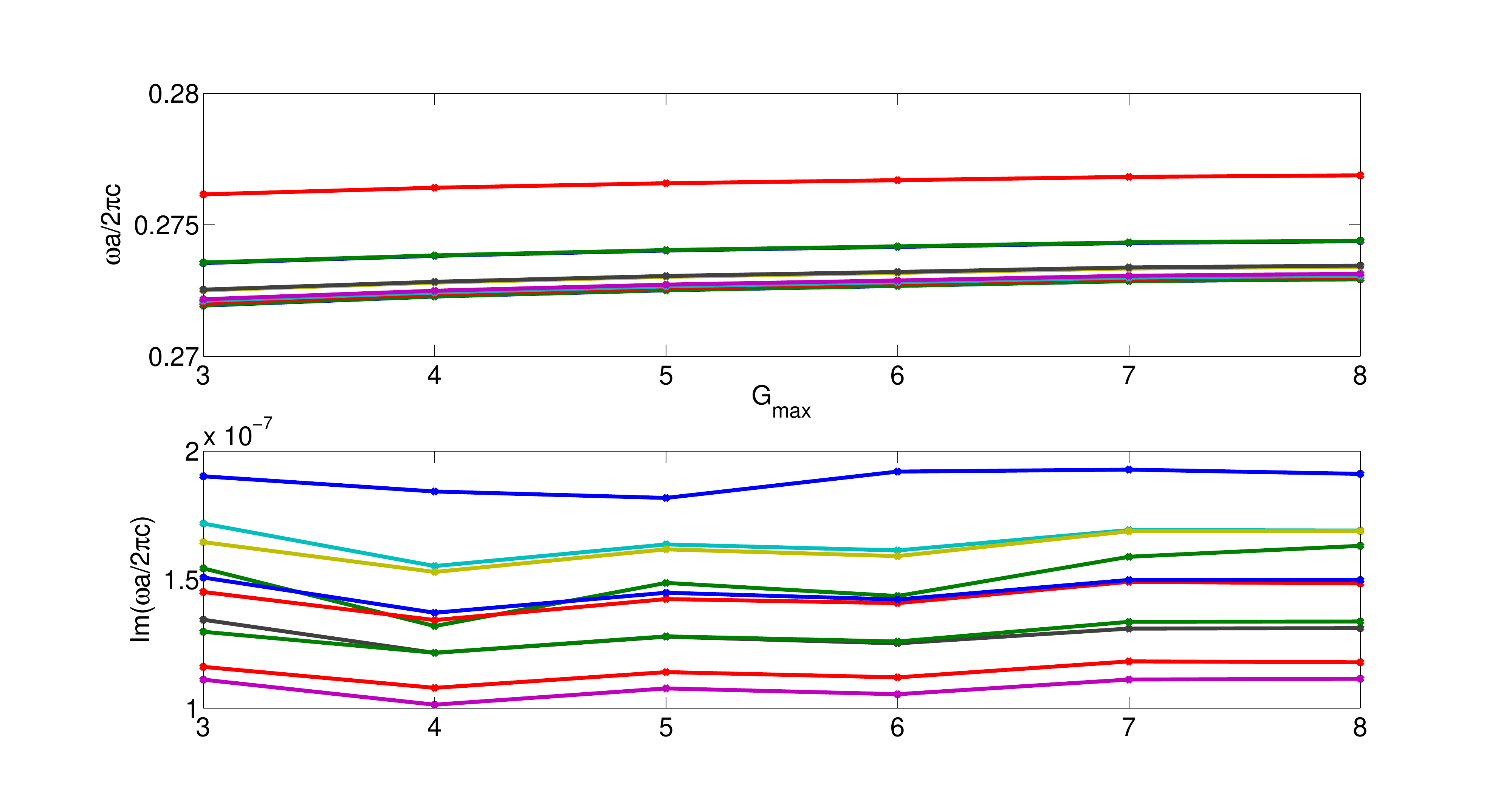}}
 \caption{(a): dependence on $G_{max}$ of the energies (up) and radiative rates (down) of the 10 lowest modes of a 32-supercell W1 guide with $\sigma = 0.002a$, $\delta = 0.06(2\pi)$ and $m_{max}=0$. (b): same for $m_{max} = 10$.}
 \label{app2}
 \end{center}
\end{figure}

\begin{figure}
\begin{center}
 \subfloat[]{\includegraphics[trim = 1in 0.8in 1in 1in, type=pdf,ext=.pdf,read=.pdf,width = 0.49\textwidth]{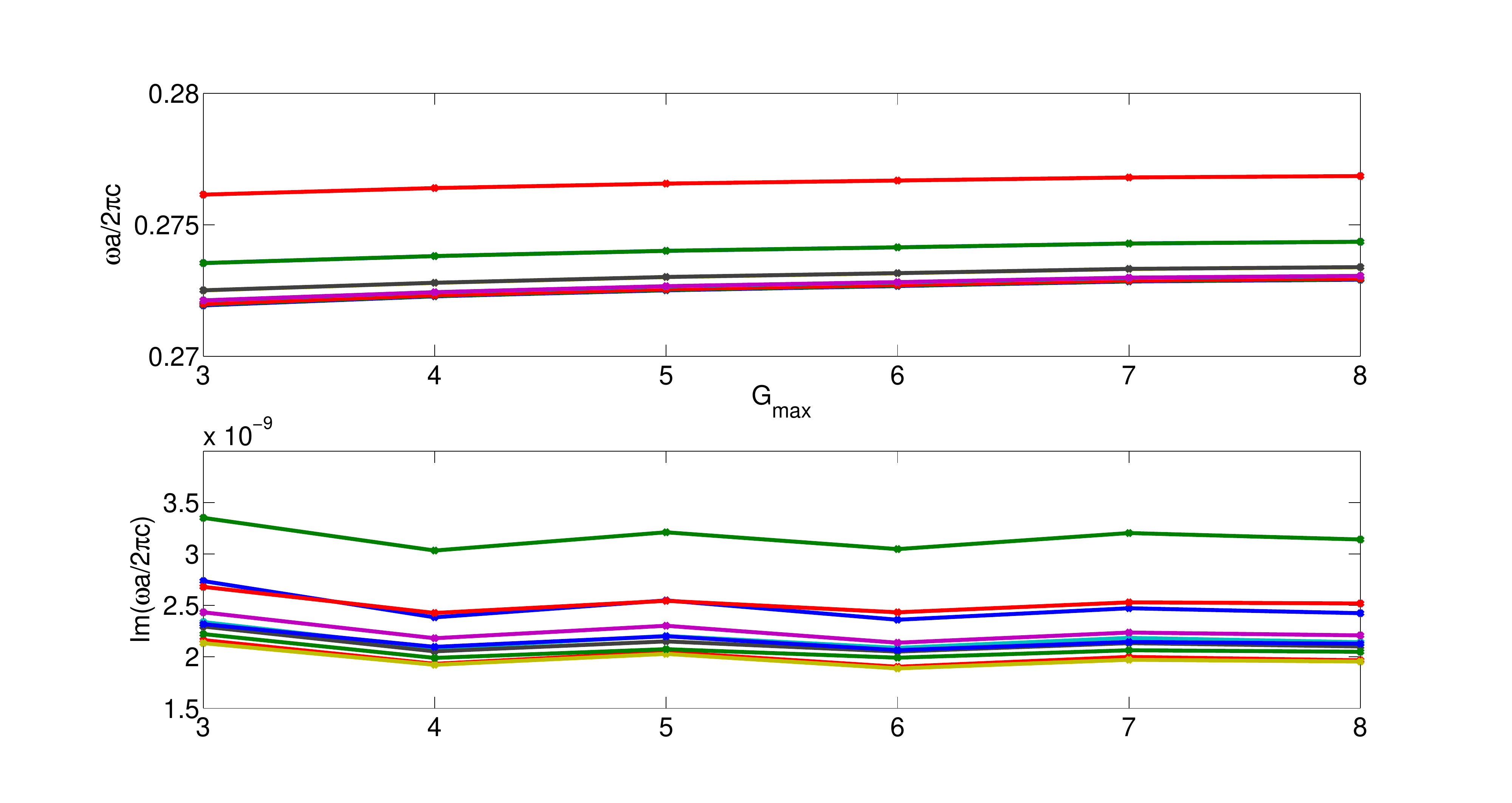}}
 \subfloat[]{\includegraphics[trim = 1in 0.8in 1in 1in, type=pdf,ext=.pdf,read=.pdf,width = 0.49\textwidth]{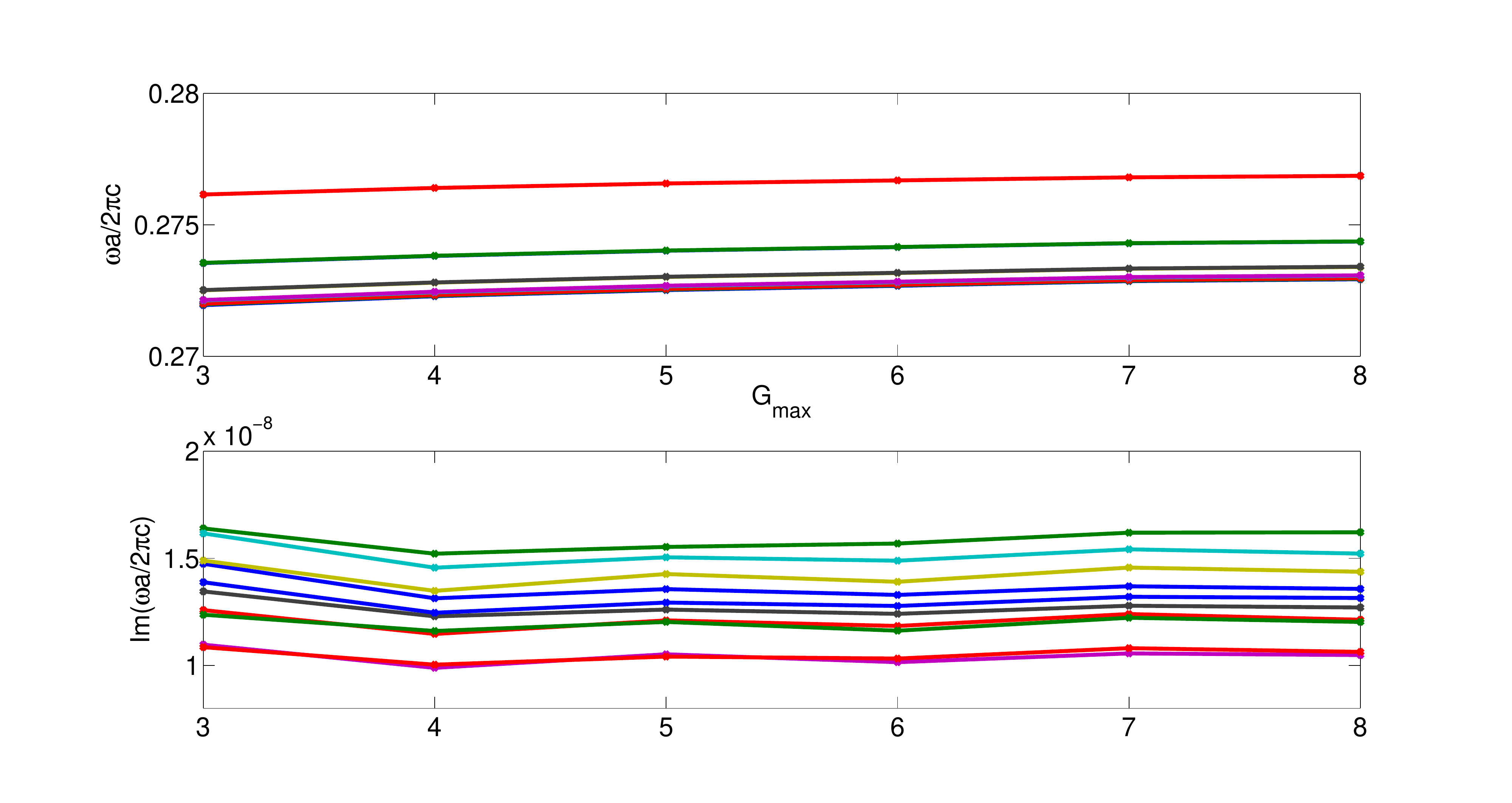}}
 \caption{(a): dependence on $G_{max}$ of the energies (up) and radiative rates (down) of the 10 lowest modes of a 32-supercell W1 guide with $\sigma = 0.002a$, $\delta = 0.005(2\pi)$ and $m_{max}=0$. (b): same for $m_{max} = 10$.}
 \label{app3}
 \end{center}
\end{figure}

\begin{figure}
\begin{center}
 \subfloat[]{\includegraphics[trim = 1in 0.8in 1in 1in, type=pdf,ext=.pdf,read=.pdf,width = 0.49\textwidth]{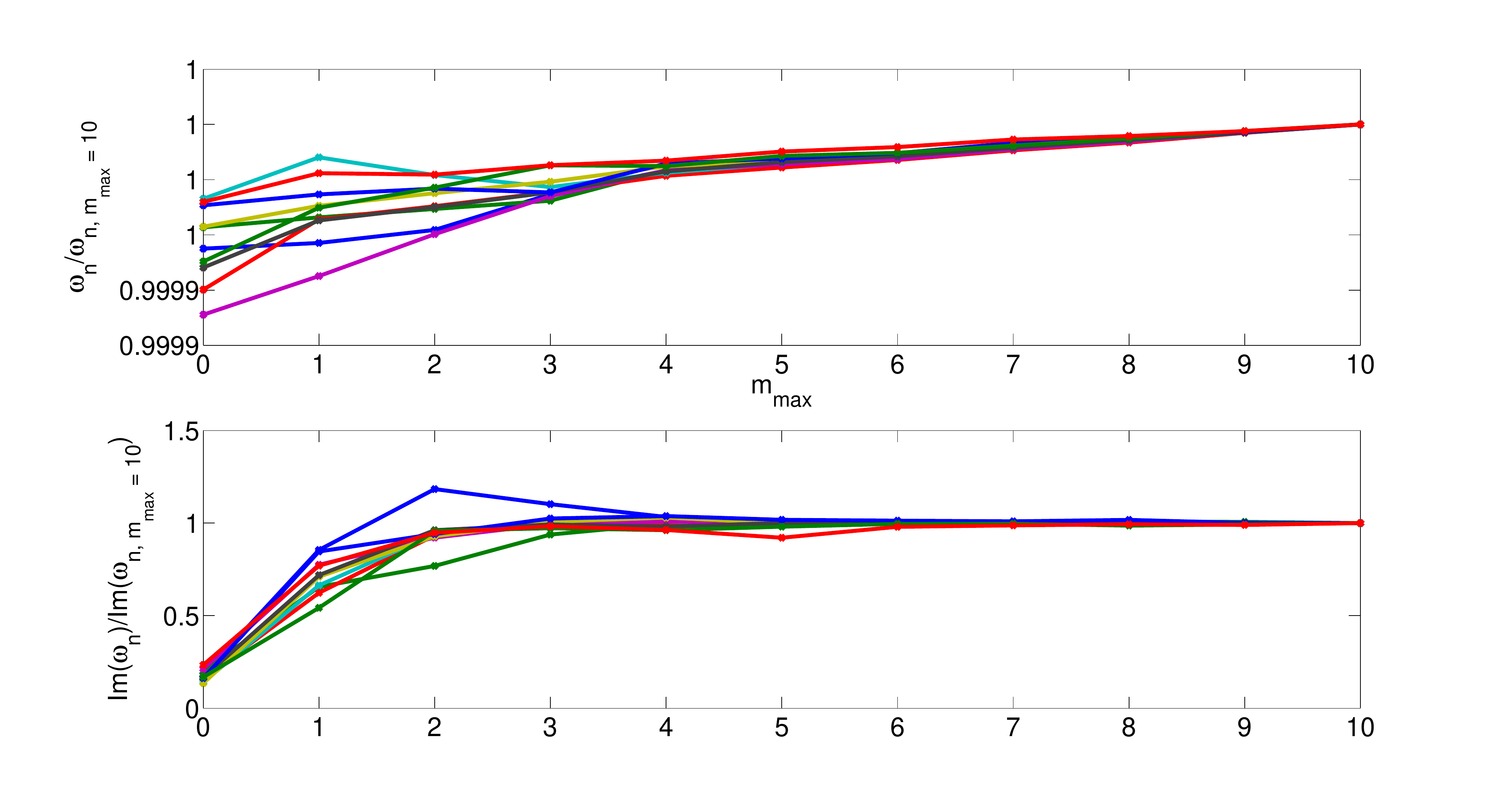}}
 \subfloat[]{\includegraphics[trim = 1in 0.8in 1in 1in, type=pdf,ext=.pdf,read=.pdf,width = 0.49\textwidth]{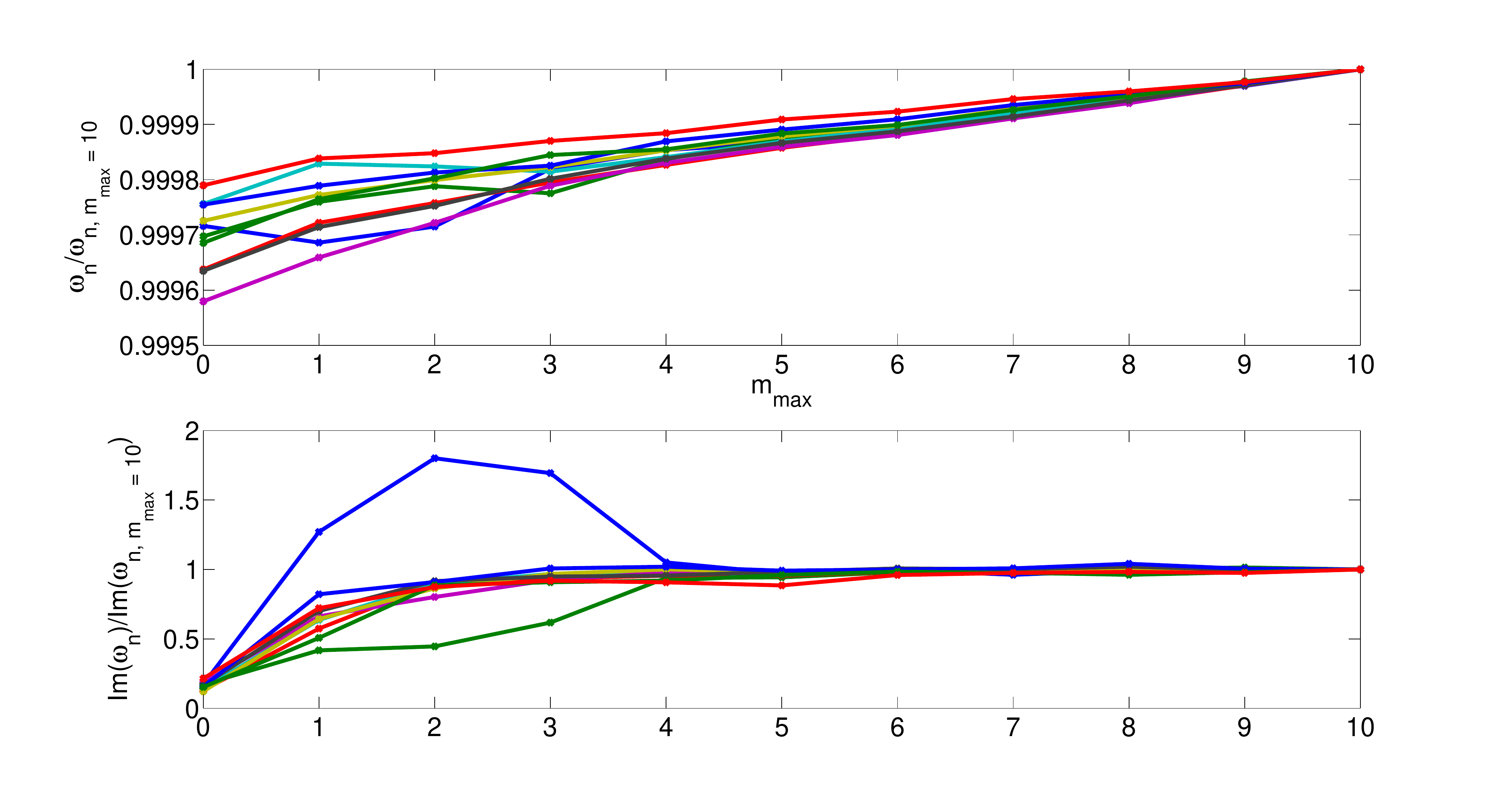}}
 \caption{(a): dependence on $m_{max}$ of the energies (up) and radiative rates (down) of the 10 lowest modes of a 32-supercell W1 guide with $\sigma = 0.006a$, $\delta = 0.005(2\pi)$ and $G_{max}=3$. (b): same for $G_{max} = 7$.}
 \label{app4}
 \end{center}
\end{figure}

\begin{figure}
\begin{center}
 \subfloat[]{\includegraphics[trim = 1in 0.8in 1in 1in, type=pdf,ext=.pdf,read=.pdf,width = 0.49\textwidth]{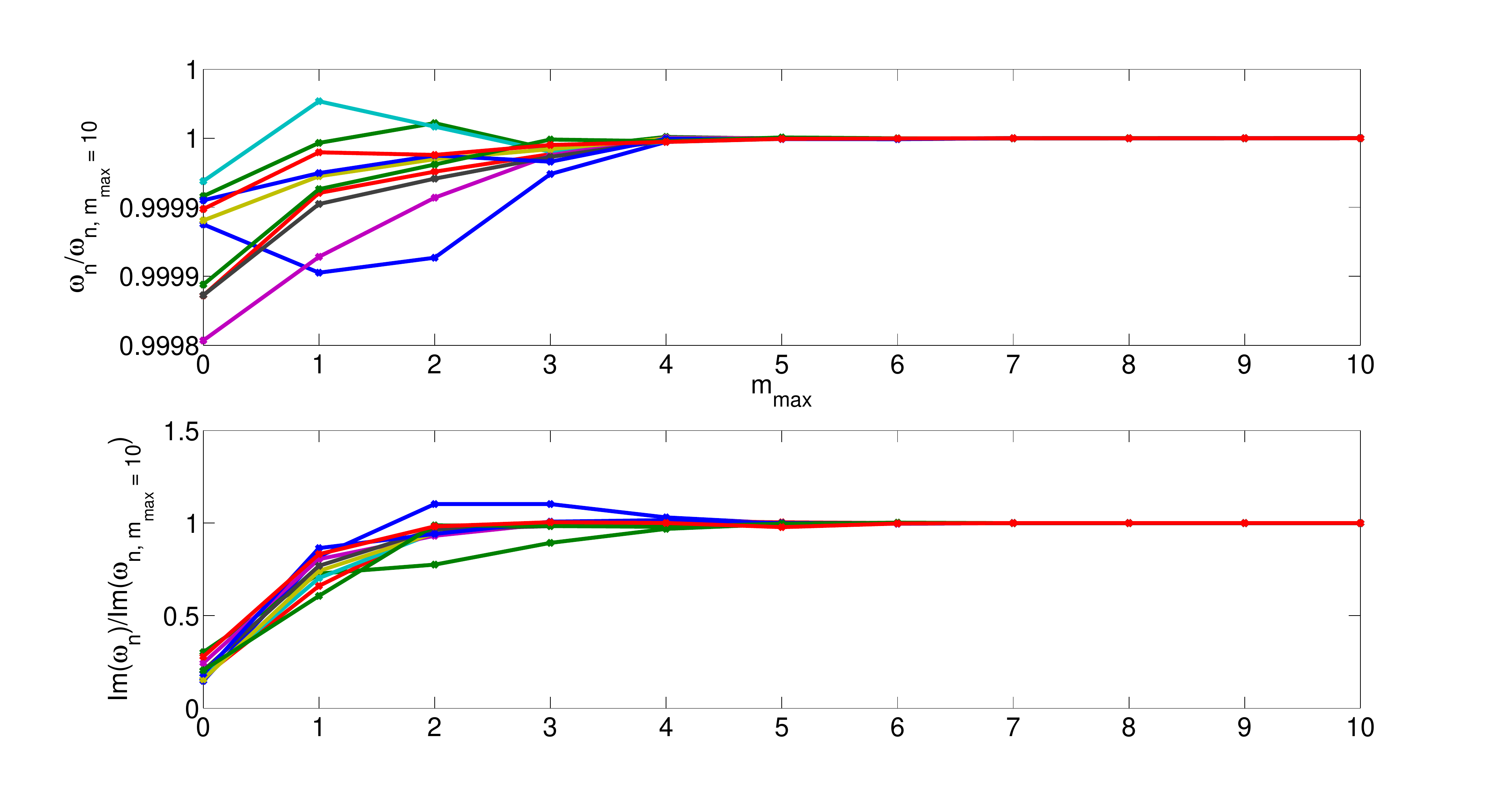}}
 \subfloat[]{\includegraphics[trim = 1in 0.8in 1in 1in, type=pdf,ext=.pdf,read=.pdf,width = 0.49\textwidth]{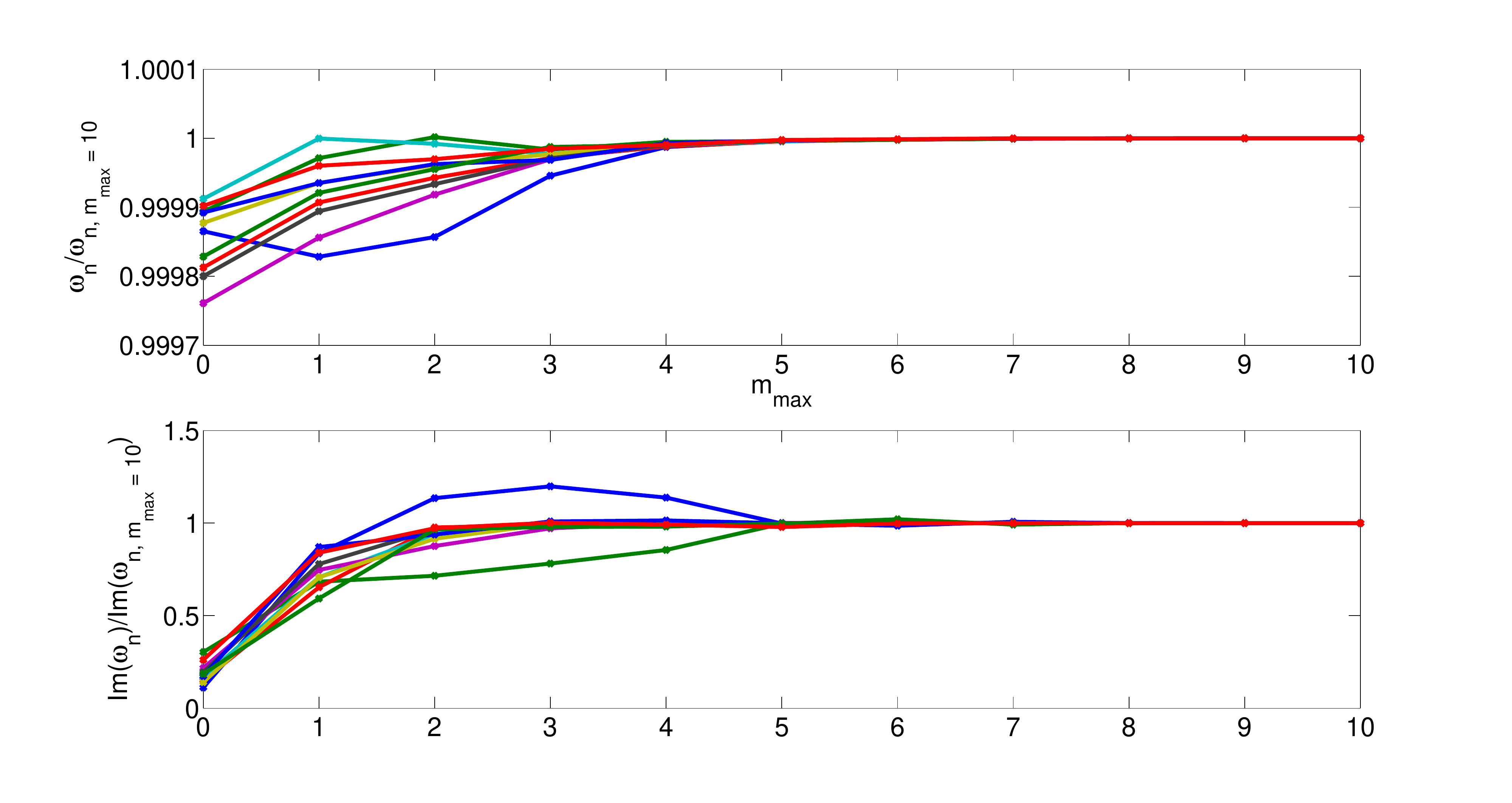}}
 \caption{(a): dependence on $m_{max}$ of the energies (up) and radiative rates (down) of the 10 lowest modes of a 32-supercell W1 guide with $\sigma = 0.002a$, $\delta = 0.06(2\pi)$ and $G_{max}=3$. (b): same for $G_{max} = 7$.}
 \label{app5}
 \end{center}
\end{figure}

\begin{figure}[ht]
\begin{center}
 \subfloat[]{\includegraphics[trim = 1in 0.8in 1in 1in, type=pdf,ext=.pdf,read=.pdf,width = 0.49\textwidth]{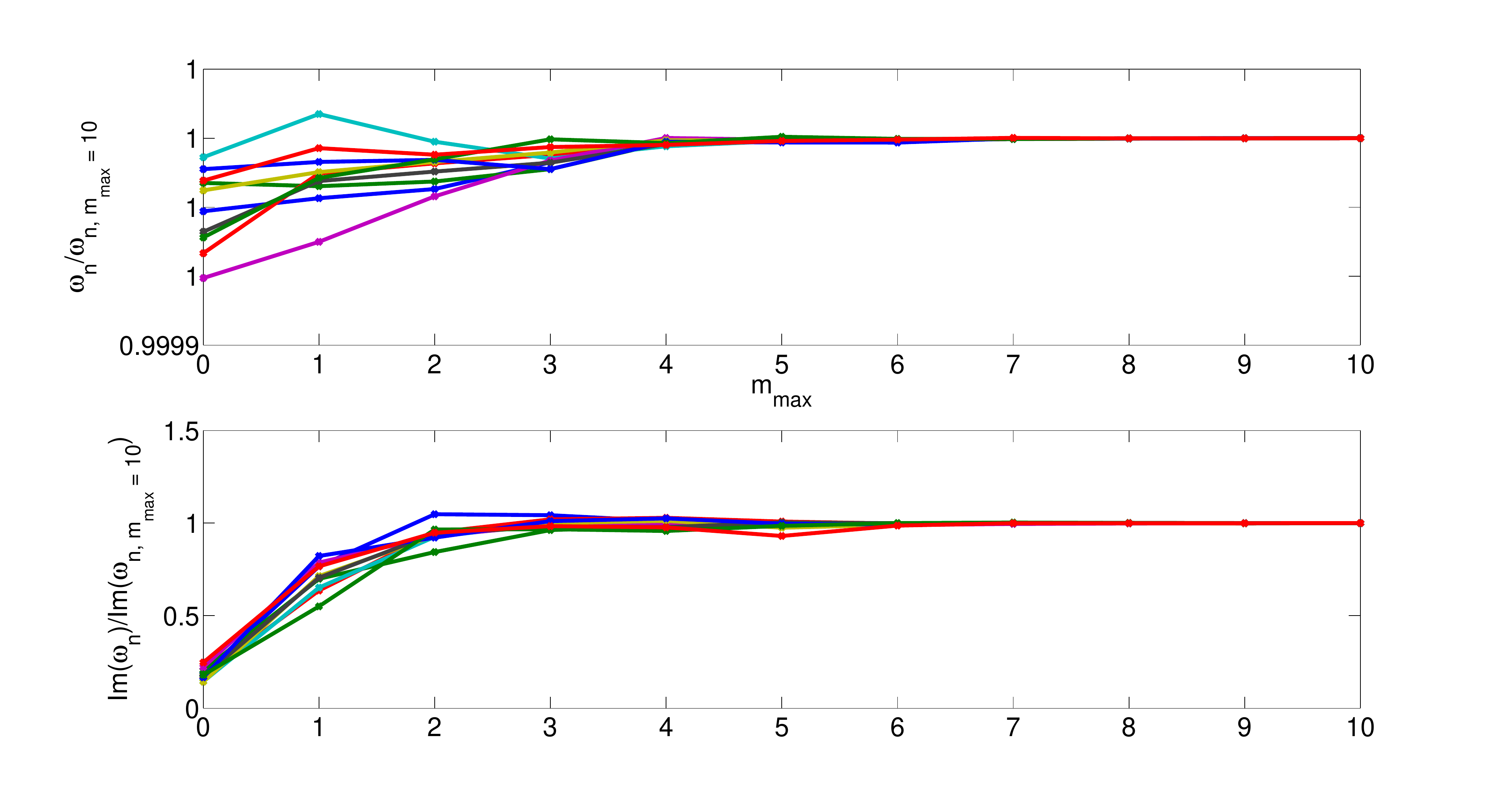}}
 \subfloat[]{\includegraphics[trim = 1in 0.8in 1in 1in, type=pdf,ext=.pdf,read=.pdf,width = 0.49\textwidth]{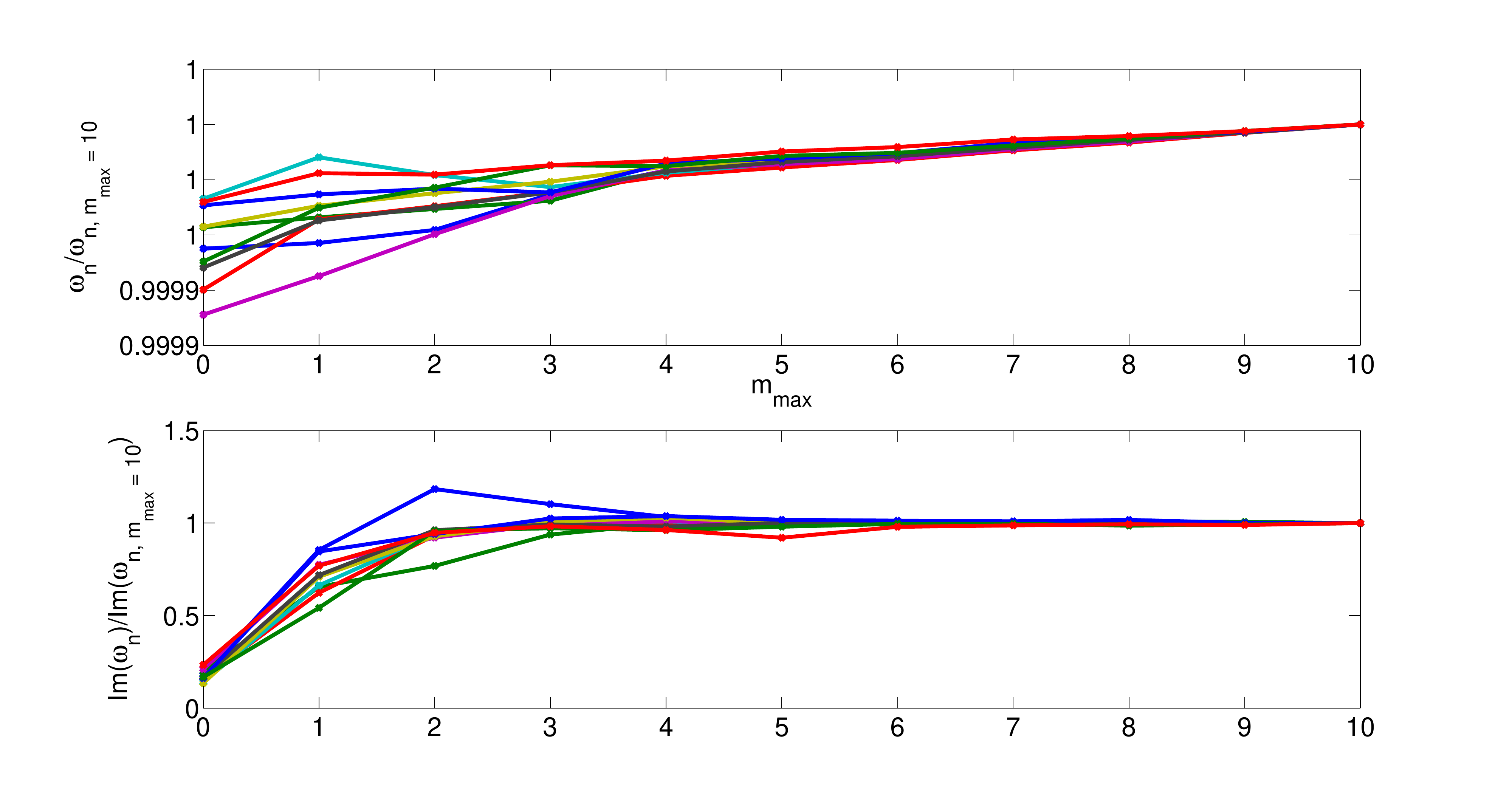}}
 \caption{(a): dependence on $m_{max}$ of the energies (up) and radiative rates (down) of the 10 lowest modes of a 32-supercell W1 guide with $\sigma = 0.002a$, $\delta = 0.005(2\pi)$ and $G_{max}=3$. (b): same for $G_{max} = 7$.}
 \label{app6}
 \end{center}
\end{figure}

In this appendix we present data related to the convergence of the BME vs. $G_{max}$ and $m_{max}$, similar to the data presented in sec. \ref{bmedisc}, but for the three remaining cases: $\sigma = 0.006a, \delta = 0.005(2\pi)$ (fig. \ref{app1} and \ref{app4}), $\sigma = 0.002a, \delta = 0.06(2\pi)$ (fig. \ref{app2} and \ref{app5}), and $\sigma = 0.002a, \delta = 0.005(2\pi)$ (fig. \ref{app3} and \ref{app6}). The discussion from sec. \ref{bmedisc} applies equally well (and sometimes better) in all cases. Most importantly, the fact that the moments up to $\approx m = 5$ have the main contribution to energies and loss rates is visible in all cases shown in fig. \ref{app4}, \ref{app5}, and \ref{app6}. 

\clearpage


\end{document}